\documentclass[12pt]{article}
\usepackage[dvips]{graphicx}
\makeatletter
\@addtoreset{equation}{section}

\makeatother
\begin{document}
\begin{titlepage}
\null
\begin{flushright}
UT-Komaba/02-06\\
hep-th/0206141
\end{flushright}
\begin{center}

\vspace{10mm}

  {\Large {\bf Various Wrapped Branes from Gauged Supergravities}}
\normalsize

\vspace{15mm}

  {\large Michihiro Naka}

\vspace{8mm}

  {\it Institute of Physics, University of Tokyo\\
Komaba, Meguro-ku, Tokyo 153-8902, Japan}\\
{hiro@hep1.c.u-tokyo.ac.jp}

\vspace{3mm}

June, 2002

\end{center}

\vspace{15mm}

\begin{abstract}

We study wrapped brane configurations
via possible maximally supersymmetric gauged supergravities.
First,
we construct various supersymmetric wrapped D3 brane configurations from
$D=5$ $N=8$ $SO(6)$ gauged supergravity. 
This procedure provides certain new examples of wrapped D3 branes around
supersymmetric cycles inside non-compact special holonomy manifolds.
We analyze their behaviors numerically
in order to discuss a correspondence to Higgs and Coulomb branches of 
sigma models on wrapped D3 branes.
We also realize supersymmetric
wrapped M2 branes from $D=4$ $N=8$ $SO(8)$ gauged supergravity.
Then, we study supersymmetric wrapped type IIB NS5 branes 
by $D=7$ $N=4$ $SO(4)$ gauged supergravity.
We show a method to derive them by using supersymmetric 
wrapped M5 branes in $D=7$ $N=4$ $SO(5)$ gauged supergravity. 
This method is based on a domain wall like reduction.
Solutions include  
NS5 branes wrapped around holomorphic
$CP^2$ inside non-compact Calabi-Yau threefold.
Their behavior shows a similar feature to that for NS5 branes wrapped around
holomorphic $CP^1$ inside non-compact $K3$ surface.
This construction also provides a check of preserved
supersymmetry for a solution interpreted within 
a string world-sheet theory introduced by Hori and Kapustin.
Finally, we find new non-supersymmetric solutions including 
AdS space-times in $D=6$ $N=2$ $SU(2)\times U(1)$ massive gauged supergravity.
These solutions can be interpreted as non-supersymmetric
wrapped D4-D8 configurations which
are dual to non-supersymmetric conformal field theories 
realized on wrapped D4 branes.
\end{abstract}

\end{titlepage}
\newpage
\section{Introduction}
\hspace{5mm}
In a context of string theory,
we have realized a correspondence between gravity and gauge theory.
We usually consider gauge theories which live on branes.
Then, strong coupling dynamics of gauge theories could be analyzed by 
classical supergravity solutions for brane configurations.
One important research direction is to 
give such examples and check a correspondence 
with less amount of supersymmetry than original 
AdS/CFT correspondence \cite{Maldacena}.
One of practical problems to overcome here is to construct explicit 
supergravity solutions for 
brane configurations with various background fluxes which realize 
given gauge theories.
There are several approaches to provide
such supergravity solutions based on AdS/CFT correspondence.
For example, see \cite{PoSt, HeKlOu, CvGiLuPo, CoGuWaZa} and so on.

In this article, we consider wrapped branes
introduced by Maldacena and Nunez \cite{MaNu1}.
Wrapped branes are objects which wrap around
supersymmetric cycles inside non-compact special holonomy manifolds.
Wrapped branes have been identified with supersymmetric
magnetically charged solutions in various gauged supergravities.
Solutions in gauged supergravities 
give near horizon geometries of various branes.
They also have information on isometries of compactified spheres
via gauge fields and scalar fields in their field contents.
Wrapped branes correspond to solutions with asymptotic AdS space-time 
at boundary (UV) region.
They usually have singularities at interior (IR) region.
In some cases, we encounter supersymmetric AdS solutions at IR region.
Here, we have solutions interpolating two AdS space-times 
with different dimensions.
When we construct solutions for wrapped branes, 
we need to identify spin connections of metric
on wrapped cycles with particular gauge fields in gauged supergravities
in order to ensure supersymmetry for wrapped brane solutions. 
Then, 
worldvolume theories on wrapped branes are twisted supersymmetric 
theories \cite{BeSaVa}.
It follows that supergravity solutions serve as dual gravity systems to  
low energy dynamics of worldvolume theories on wrapped branes.
They give examples for less supersymmetric AdS/CFT correspondence.
This construction in \cite{MaNu1} has been extended in various ways
\cite{MaNu2, AcGaKi, NiOz, GaKiWa, 
NuPaScTr, Gomishol, EdNu, ScTr, MaNa, GaKiPaWa1, Hernandez, 
GaKiMaWa1, BiCoZa, GoMa, GaKi, GoRu, GaKiMaWa2,
Gomis, DiEnImLo, ApBiCoPeZa, AhScSo, HeSf, GaKiPaWa2,
GuNuSc, GaMaPaWa, HeSf2, DiLeMe, BrGoMaRa}.

In early literatures, 
wrapped brane solutions have been constructed from
various truncated gauged supergravities in each dimension.
As for wrapped M5 branes \cite{GaKiWa},
it is systematic to handle possible wrapped branes at once within
maximally supersymmetric gauged supergravity.
Here, we consider wrapped D3 branes.
Several wrapped D3 branes have been discussed in \cite{MaNu1, NiOz}
by using truncated five-dimensional gauged supergravities.
We wish to derive possible wrapped D3 branes in a uniform manner.
We write down the solutions by using maximally supersymmetric
$D=5$ $N=8$ $SO(6)$ gauged supergravity. 
Our present treatment includes wrapped D3 branes
around special Lagrangian three cycles
inside Calabi-Yau threefolds, 
and holomorphic two cycles inside Calabi-Yau fourfolds.
We find that the former has no AdS solution at IR region, and 
the latter has AdS solution when two cycles have negative curvature.
Then, we analyze their behaviors numerically 
with naive interpretations on dual worldvolume theories.
We will also embed wrapped M2 branes \cite{GaKiPaWa1} into 
maximally supersymmetric $D=4$ $N=8$ $SO(8)$ gauged supergravity.
These prescriptions should also give a starting point to discuss possible
resolutions of IR singularities for wrapped branes. 
As a side remark, we will observe a universal form of 
BPS equations for wrapped D3, M2 and M5 branes.

Another wrapped branes arise from NS5 branes
which give dilatonic backgrounds with less supersymmetry.
Several supergravity solutions have been constructed in
\cite{MaNu2, MaNa, GaKiMaWa1, BiCoZa, GoRu, GaKiMaWa2}.
However, preserved supersymmetry has not 
been understood in seven-dimensional supergravity. 
We will write down BPS equations for all possible wrapped NS5 branes
via $D=7$ $N=4$ $SO(4)$ gauged supergravity.
We find that it is useful to derive these BPS equations
by reducing BPS equations for wrapped M5 branes obtained from
$D=7$ $N=4$ $SO(5)$ gauged supergravity.
We pursue this procedure by borrowing a domain wall like
reduction \cite{CvLiLuPo}.
This method is equivalent to a direct derivation from the $SO(4)$ supergravity.
Then, we concentrate on supersymmetric type IIB
  NS5 branes wrapped around
holomorphic $CP^2$ inside Calabi-Yau threefold.
We give their ten-dimensional solution which 
has a curvature singularity at IR region.
The solution is related to a string world-sheet description
of wrapped NS5 branes introduced by Hori and Kapustin \cite{HoKa}.
Present derivation gives an explicit check of $1/8$ preserved
supersymmetry of the wrapped NS5 brane solution.
We also notice that the supergravity
solution has a similar structure with
that of NS5 branes wrapped around holomorphic $CP^1$ inside $K3$ surface. 

It is natural to study behaviors of wrapped branes without supersymmetry.
Here, we wish to consider non-supersymmetric solutions with ansatz 
which is the same as supersymmetric configurations.
We start with second-order differential equations of motion 
obtained from
Lagrangian in gauged supergravities. 
We have no technology to find exact
solutions for these differential equations.
Nevertheless, we are able to make an exact statement on 
solutions if we assume that solutions include AdS space-times.
We find such non-supersymmetric solutions for wrapped D4-D8 systems
in massive IIA supergravity by generalizing supersymmetric results
\cite{NuPaScTr}.
An interest on these solutions lies on a possible role as dual configurations
for non-supersymmetric CFTs realized on wrapped D4 branes.
Similar results have been obtained for wrapped M5 branes in \cite{GaKiPaWa2}.
We have checked that similar solutions do not exist for wrapped D3
and M2 branes.

There are a lot of directions to study further by
using present wrapped brane solutions.
An obvious direction is to consider Penrose limit 
of supergravity solutions for various wrapped branes along 
recent discussions \cite{BlFiHuPa1, BeMaNa, BlFiHuPa2}.
A main interest here will be to understand a detailed 
structure on preserved supersymmetry found in various simple examples.
We will return to address this analysis elsewhere.

The organization of this paper is the following.
In section 2, we derive solutions of wrapped D3 branes by using $D=5$
$N=8$ $SO(6)$ gauged supergravity.
We solve BPS equations numerically with small comments
about dual worldvolume theories.
We also include known results for reader's convenience.
In section 3, we realize 
wrapped M2 branes in $D=4$ $N=8$ $SO(8)$ gauged supergravity.
In section 4, we discuss wrapped type IIB NS5 branes.
We show a derivation of BPS equations 
for the NS5 branes by using wrapped M5 branes.
Then we discuss a ten-dimensional solution for 
wrapped NS5 branes around holomorphic $CP^2$ inside
Calabi-Yau threefolds.
In section 5, we discover non-supersymmetric solutions
with AdS space-time for D4-D8 wrapped systems via
$D=6$ $SU(2)\times U(1)$ massive gauged supergravity.
Section 6 includes summary and discussions. 

\section{Wrapped D3 branes}
\hspace{5mm}
We start with supersymmetric wrapped D3 branes.
We derive wrapped D3 branes in type IIB supergravity 
from five-dimensional BPS solutions
in $D=5$ $N=8$ $SO(6)$ gauged supergravity \cite{GuRoWa}.
Five-dimensional BPS solutions with required amount of supersymmetry
are constructed by noting projections on spinors \cite{GaLaWe}.
Then, we embed solutions in five dimensions
into those in ten dimensions \cite{CvLuPoSaTr1}.
In resulting solutions in type IIB supergravity,
only metric and self-dual five-form field strength are excited.
We discuss numerical evaluation of BPS solutions to see behaviors of
ten-dimensional metrics.
We also include a little observations from worldvolume theories on the 
wrapped D3 branes.

\subsection{$D=5$ $N=8$ $SO(6)$ gauged supergravity}
\hspace{5mm}
Five-dimensional $N=8$ gauged supergravity \cite{GuRoWa}
consists of a graviton, 8 gravitinos, 12
tensor fields, 15 vector gauge fields, 48 gauginos and 42 scalars. 
Ungauged $N=8$ supergravity has global $E_{6(6)}$ symmetry and 
composite local $USp(8)$ symmetry. 
Scalar fields are parameterized by coset space $E_{6(6)}/USp(8)$.
This global $E_{6(6)}$ is translated into local $SO(6)$ and global
$SL(2,{\bf R})$ symmetry in gauged supergravity.
The bosonic fields have following representations in these symmetries 
\begin{eqnarray}
\begin{array}{|c|c|c|c|c|c|}\hline
& e^m_{\mu} & 
B_{\mu\nu}^{I\alpha} &
A_{\mu IJ} &
V^{IJ ab} &
V_{I\alpha}^{ab}\\\hline
USp(8) &&&& 27 & 27\\\hline
SO(6) &&6 &15 &15 & 6\\ \hline
SL(2,{\bf R})&&2 &&&2 \\\hline
\end{array}
\end{eqnarray}
Here, we denote indices of $USp(8), SO(6)$ and $SL(2,{\bf R})$
by $a,b=1,\dots,8$,
$I,J=1,\dots,6$, and $\alpha,\beta=1,2$.
The 42 scalar fields lie in $SO(6)$
representations, ${\bf 1}\oplus {\bf 1}\oplus
{\bf 10}\oplus \overline{{\bf 10}}\oplus {\bf 20}'$.
We consider only the scalars in ${\bf 20}'$ representations.
We ignore a contribution of $SL(2,{\bf R})$
part of scalar fields in ${\bf 1}\oplus {\bf 1}$ representations.
This means that we consider constant values of the dilaton
and axion fields in ten dimensional type IIB supergravity.
We also set values of $B$ fields to be zero.
Now, let us proceed to bosonic part of the Lagrangian.
The Lagrangian has following form
\begin{eqnarray}
L=\sqrt{-g}\left[
-\frac{1}{4}R+\frac{1}{24}P_{\mu abcd}P^{\mu abcd}
-\frac{1}{8}F_{\mu\nu ab}F^{\mu\nu ab}
+g^2\left(\frac{6}{45^2}(T_{ab})^2-\frac{1}{96}(A_{abcd})^2\right)
\right],
\end{eqnarray}
where $g$ is gauge coupling constant for $SO(6)$ gauge fields,
and a signature of metric is $\left(+----\right)$.
The first three terms are kinetic terms for
metric, scalar fields and gauge fields.
The last term is a $SO(6)\times SL(2,{\bf R})$ invariant
potential for scalar fields.
Scalar functions in this potential will be defined later.
We have omitted $F^2A, FA^3, A^5$ terms for gauge fields.
These terms are irrelevant to our treatment of wrapped D3 branes.
Supersymmetry transformations of fermionic gravitinos and gauginos are
\begin{eqnarray}
\label{susyd3}
\delta\psi_{\mu a}&=&
\partial_{\mu} \epsilon_a
+\frac{1}{4}\omega_{\mu}^{\rho\sigma}\gamma_{\rho\sigma}\epsilon_a
-Q_{\mu a}^{b}\epsilon^b
-\frac{2}{15}gT_{ab}\gamma_{\mu}\epsilon_b
-\frac{1}{6}
F_{\nu\rho\;ab}
\left(
\gamma^{\nu\rho}\gamma_{\mu}+2\gamma^{\nu}\delta_{\mu}^{\rho}
\right)\epsilon^b,\nonumber\\
\delta\chi_{abc}&=&\sqrt{2}\gamma^{\mu}P_{\mu abcd}\epsilon^{d}
-\frac{1}{\sqrt{2}}gA_{dabc}\epsilon^{d}
-\frac{3}{2\sqrt{2}}\gamma^{\mu\nu}
F_{\mu\nu\;[ab}\epsilon_{c]|},
\end{eqnarray} 
where $[\dots]|$ 
denotes skew symmetrization with all symplectic traces removed.
We denote five-dimensional gamma matrices by $\gamma_{\mu}$.
These matrices satisfy anti-commutation relation
$\{\gamma_{\mu}, \gamma_{\nu}\}=2g_{\mu\nu}$
with a five-dimensional metric tensor $g_{\mu\nu}$.

Let us introduce a definition of each field.             
Coset elements for scalar fields $V^{IJab},V_{I\alpha}^{\quad ab}$
and their inverse elements $\widetilde{V}_{abIJ},
\widetilde{V}_{ab}^{\quad I\alpha}$ are represented by
\begin{eqnarray}
V^{IJab}=\frac{1}{4}
\left(\Gamma_{KL}\right)^{ab}S_K^IS_L^J,&&\qquad
V_{I\alpha}^{\quad ab}
=\frac{1}{\sqrt{2}}                   
\left(\Gamma_{K\beta}\right)^{ab}
S_I^KS_{\alpha}^{'\beta},\\
\widetilde{V}_{abIJ}=
\frac{1}{4}
\left(\Gamma_{KL}\right)^{ab}
\left(S^{-1}\right)_I^K
\left(S^{-1}\right)_J^L,&&\qquad
\widetilde{V}_{ab}^{\quad I\alpha}
=-\frac{1}{2\sqrt{2}}
\left(\Gamma_{K\beta}\right)^{ab}
\left(S^{-1}\right)_K^I
\left(S'^{-1}\right)_{\beta}^{\alpha}.\nonumber
\end{eqnarray}
Here, $\Gamma_I$ are $8\times 8$ gamma matrices
satisfying $\{\Gamma_I, \Gamma_J \}=2\delta_{IJ}$.
We also introduce gamma matrices
$\Gamma_{I\alpha}=\left(\Gamma_I, i\Gamma_I\Gamma_0\right)$, $\alpha=1,2$.
We parameterize
a subsector $SO(6)\times SL(2,{\bf R})$ of scalar manifolds
by matrices $S_I^J \in SO(6)$ and
$S_{\alpha}^{'\beta}\in SL(2,{\bf R})$.
The matrix $S'$ will not give a contribution into wrapped D3 branes.
We define $USp(8)$ gauge fields $Q_{\mu a}^b$ by 
\begin{eqnarray}
Q_{\mu a}^{b}=
-\frac{1}{3}\left[
\widetilde{V}^{bc IJ}\partial_{\mu}V_{IJ ac}
+gA_{\mu IL}\delta^{JL}
\left(
2V_{ac}^{IK}\widetilde{V}^{bc}_{JK}-V_{J\alpha ac}\widetilde{V}^{bc I\alpha}
\right)
\right].
\end{eqnarray}
The coset elements $V^{IJab}$ play an role to transform 
$SO(6)$ gauge field strength 
into $USp(8)$ composite local gauge field strength,
$F_{\mu\nu}^{ab}=V^{IJab}F_{\mu\nu IJ}$.
The term which enters in scalar kinetic term is given by
\begin{eqnarray}
P_{\mu}^{abcd}=\widetilde{V}^{ab IJ}\partial_{\mu}V_{IJ}^{\;\;\;cd}.
\end{eqnarray}
The functions of scalar fields are defined by 
\begin{eqnarray}
T_{ab}
&=&-\frac{15}{4}\Omega^{cd}W_{acbd},\qquad\qquad
W_{abcd}=\epsilon^{\alpha\beta}\delta^{IJ}V_{I\alpha ab}
V_{J\beta cd},\\
A_{abcd}&=&-3\left[
W_{a[bcd]}+\frac{1}{6}\left(
\Omega_{bc}\Omega^{ef}W_{a[efd]}
+\Omega_{cd}\Omega^{ef}W_{a[efb]}
+\Omega_{db}\Omega^{ef}W_{a[efc]}
\right)
\right],\nonumber
\end{eqnarray}
where $[\dots ]$ denotes anti-symmetrization of indices,
and 
$\epsilon_{\alpha\beta}=\epsilon^{\alpha\beta}=-\epsilon^{\beta\alpha}$,
$\epsilon_{12}=1$.
Here, $\Omega_{ab}$ are anti-symmetric matrices satisfying 
$\Omega_{ab}=\Omega^{bc}=\delta_a^c,\;
\Omega_{ab}\Omega^{ab}=8$.  
These matrices are realized as
$\Omega^{ab}=-\Omega_{ab}
=-i\left(\Gamma_0\right)^{ab}$
with $8\times 8$
$SO(7)$ gamma matrices $\Gamma_i \left(i=0,\dots,6 \right)$.

Solutions in the gauged supergravity are 
arranged into solutions in type IIB supergravity \cite{CvLuPoSaTr1}.
The resulting ten-dimensional metric $ds_{10}^2$ is  
\begin{eqnarray}
\label{upliftd3}
ds_{10}^2&=&
\Delta^{\frac{1}{2}}ds_{5}^2
-\frac{1}{g^2\Delta^{\frac{1}{2}}}T_{IJ}^{-1}
D\mu^I D\mu^J
,\\
\Delta&=&T_{IJ}\mu^I\mu^J,\qquad\quad
D\mu^I=d\mu^I+gA^{IJ}\mu^J.\nonumber
\end{eqnarray}
Here, $ds_5^2$ is metric in the five-dimensional theory.
The internal unit five-sphere $S^5$ is parameterized by 
$\left(\mu^1,\dots,\mu^6\right)$ satisfying $\sum_{I=1}^6\mu^I\mu^I=1$.
Generally, this five-sphere is squashed by five-dimensional 
gauge and scalar fields.
In our application, scalar function $T_{IJ}$
will be given by $T_{IJ}=S_I^KS_K^L\delta_{LJ}$.
Self-dual Ramond-Ramond 5-form field strength ${\cal F}_5$ is 
provided by 
\begin{eqnarray}
{\cal F}_5&=&F_5+*_{10}F_5,\\
F_5&=&-g U {\rm Vol}_5 
+\frac{1}{g}\left(
T_{IJ}^{-1}*DT_{JK}
\right)\wedge
\left(\mu^KD\mu^I\right)
-\frac{1}{2g^2}T_{IK}^{-1}T_{JL}^{-1}
*F_2^{IJ}\wedge D\mu^K \wedge D\mu^{L},\nonumber\\
F^{IJ}&=&dA^{IJ}+gA^{IK}\wedge A^{KJ},\nonumber\\
DT_{IJ}&=&dT_{IJ}+gA^{IK}T_{KJ}+gA^{JK}T_{IK},\qquad
U=2T_{IJ}T_{JK}\mu^I\mu^K-\Delta T_{II},\nonumber
\end{eqnarray}
where ${\rm Vol}_5$ is a volume form in five-dimensions,
and $*$ and $*_{10}$ denote Hodge duals in five and ten dimensions.
Other fields in type IIB supergravity are not excited.
This shows us that resulting ten-dimensional solutions 
represent wrapped D3 branes.

\subsection{BPS equations}
\hspace{5mm}
Let us consider BPS configurations of wrapped D3 branes.
Possible supersymmetric wrapped cycles are 
two cycles inside $K3$ surfaces, Calabi-Yau threefolds,
and Calabi-Yau fourfolds, and also
three cycles inside Calabi-Yau threefolds and $G_2$ holonomy manifolds.
We will show ansatz on metric, gauge fields and 
scalar fields in the gauged supergravity.
These are chosen to be consistent with 
a required twisting procedure on curved worldvolume,
and projections on spinors in order to
preserve required amount of supersymmetry \cite{GaLaWe}.
Then, we derive first order BPS equations for all wrapped D3 branes
from supersymmetry variations of fermions.
Here we implicitly assume that this gives the same result based on
second order equations of motion from the Lagrangian.

We use a metric
for D3 branes wrapped around supersymmetric $d$-cycles 
$\Sigma_{d}$ with $d=2, 3$
\begin{eqnarray}
ds_5^2=e^{2f(r)}\left(d\xi_{4-d}^2-dr^2\right)-e^{2g(r)}
d\widetilde{s}_{\Sigma_d}^2,
\end{eqnarray}
where $f(r), g(r)$ are radial functions to be determined.
Here, coordinates $\xi^i, i=0\dots 3-d$ denotes
unwrapped directions of D3 brane.
We denote metric on the directions 
by $d\xi_{4-d}^2=ds^2\left(R^{1,3-d}\right)$. 
Metrics $d\widetilde{s}_{\Sigma_d}^2$ on supersymmetric cycles satisfy
Einstein condition.
These constant curvature
metrics are normalized to have unit cosmological constants $\ell=\pm 1$.
Ansatz on other fields are specified individually.
We identify $SO(6)$ gauge fields with spin connections of 
metrics on supersymmetric cycles.
We decompose $SO(6)$ into their subgroup $SO(p)\times SO(q)$
$(p+q=6)$ to excite only gauge fields in $SO(p)$.
This decomposition divides six transverse directions of D3 brane
into two classes of directions.
The first $p$ directions are
within non-compact special holonomy manifolds including 
supersymmetric cycles.
The second $q$ directions are transverse to them.
In the following, we show this process
on twisting conditions on gauge fields and spinor fields.
Then, we provide required ansatz on scalar fields.

We begin with D3 branes wrapped around
holomorphic two-cycles inside Calabi-Yau two, three and fourfolds.
We use a following five-dimensional metric 
\begin{eqnarray}
ds_5^2=e^{2f(r)}\left(d\xi_2^2-dr^2\right)-e^{2g(r)}
d\widetilde{s}_{\Sigma_2}^2.
\end{eqnarray}
Let us introduce orthonormal basis of this metric
\begin{eqnarray}
e^0=e^{f(r)}d\xi_0,\quad
e^1=e^{f(r)}dx_1
\quad
e^2=e^{f(r)}dr,\quad
e^{3}=e^{g(r)}\widetilde{e}^1,\quad
e^{4}=e^{g(r)}\widetilde{e}^2,
\end{eqnarray}
where $\widetilde{e}^1, \widetilde{e}^2$ are orthonormal basis for 
a metric on supersymmetric cycle $d\widetilde{s}_{\Sigma_2}$.

Now, we specify ansatz on $SO(6)$ gauge fields 
which are consistent with projections on spinor fields $\epsilon_a$.
We will omit indices $a$ for spinor fields in the following.  
Wrapped D3 branes around holomorphic two cycles
inside non-compact $K3$ surfaces preserve eight supercharges.
This $1/4$ supersymmetric configuration is realized by an ansatz
on gauge fields and spinor fields
\begin{eqnarray}
&&\gamma_{34}\epsilon=\Gamma_{12}\epsilon, 
\qquad \gamma_2\epsilon=\epsilon,\\
&&
F_{34}^{12}=-\frac{\ell}{g}e^{-2g(r)}.\nonumber
\end{eqnarray}
We set other components of $SO(6)$ gauge fields to be zero.
Here, $SO(6)$ symmetry breaks into $SO(2)\times SO(4)$.
Then, we identify $SO(2)=U(1)$ subgroup with a structure group $U(1)$ of 
spin connections for holomorphic two-cycles.

Let us turn to wrapped D3 branes around holomorphic two cycles
inside non-compact Calabi-Yau threefolds.
We decompose $SO(6)$ symmetry into $SO(4)\times SO(2)$.
We identify a diagonal $U(1)$ group of $U(2) \subset SO(4)$
with a structure group $U(1)$ of spin connections for 
holomorphic two-cycles.
For these $1/8$ supersymmetric configurations, we use following ansatz
on non-zero components of $SO(6)$ gauge fields with projections on
spinor fields
\begin{eqnarray}
&&\gamma_{34}\epsilon=\Gamma_{12}\epsilon
=\Gamma_{34}\epsilon, \qquad \gamma_2\epsilon=\epsilon,\\
&&
F_{34}^{12}=F_{34}^{34}=-\frac{\ell}{2g}e^{-2g(r)}.\nonumber
\end{eqnarray}

We also have wrapped D3 branes around holomorphic two cycles
inside non-compact Calabi-Yau fourfolds.
Here, a diagonal $U(1)$ group of $U(3)$ within $SO(6)$ symmetry
is identified with a structure group $U(1)$ of
spin connections for holomorphic two-cycles.
Then, we impose following ansatz in order to realize
the $1/16$ supersymmetric configurations
\begin{eqnarray}
&&\gamma_{34}\epsilon=\Gamma_{12}\epsilon
=\Gamma_{34}\epsilon=\Gamma_{56}\epsilon, 
\qquad \gamma_2\epsilon=\epsilon,\\
&&
F_{34}^{12}=F_{34}^{34}=F_{34}^{56}=-\frac{\ell}{3g}e^{-2g(r)}.\nonumber
\end{eqnarray}
We choose other components of $SO(6)$ gauge fields to be zero.

We turn to BPS configurations with supersymmetric three cycles.
We use five-dimensional metric defined by
\begin{eqnarray}
ds_5^2=e^{2f(r)}\left(d\xi^2-dr^2\right)-
e^{2g(r)}d\widetilde{s}_{\Sigma_3}^2.
\end{eqnarray}
Then, let us introduce orthonormal basis of this metric
\begin{eqnarray}
e^0=e^{f(r)}d\xi,
\quad
e^1=e^{f(r)}dr,
\quad
e^2=e^{g(r)}\widetilde{e}^1,
\quad
e^3=e^{g(r)}\widetilde{e}^2,
\quad
e^4=e^{g(r)}\widetilde{e}^3,
\end{eqnarray}
where $\widetilde{e}^{a} (a=1,2,3)$ are orthonormal basis for
Einstein metrics $d\widetilde{s}_{\Sigma_3}^2$ on supersymmetric three
cycles under consideration.

We consider D3 branes wrapped around special Lagrangian three cycles
inside Calabi-Yau threefolds.
A structure group of spin connections for three cycles are $SO(3)$.
This group is identified with $SO(3)$
the subgroup of decomposition $SO(6)\to SO(3)\times SO(3)$.
Then, these $1/8$ supersymmetric configurations are specified by ansatz on  
non-zero $SO(6)$ gauge field strength and projections on spinor fields
\begin{eqnarray}
&&\gamma_{23}\epsilon=\Gamma_{12}\epsilon,
\quad
\gamma_{34}\epsilon=\Gamma_{23}\epsilon,
\quad
\gamma_{42}\epsilon=\Gamma_{31}\epsilon,
\quad \gamma_1\epsilon=\epsilon,\\
&&
F^{12}_{23}=F^{23}_{34}=F^{31}_{42}=-\frac{\ell}{g}e^{-2g(r)}.\nonumber
\end{eqnarray}

We turn to wrapped D3 branes around
associative three cycles inside $G_2$ holonomy manifolds.
Here, we decompose $SO(6)$ symmetry into $SO(4)\times SO(2)$.
Then, we identify $SU(2)_L$ in $SO(4)\approx SU(2)_L\times SU(2)_R$
with a structure group of spin connections for the three cycles.
Ansatz on non-zero $SO(6)$ gauge field strength
for these $1/16$ supersymmetric configuration is specified as
\begin{eqnarray}
&&\gamma_{23}^+\epsilon=\Gamma_{14}\epsilon,
\quad
\gamma_{34}^+\epsilon=\Gamma_{24}\epsilon,
\quad
\gamma_{42}^+\epsilon=\Gamma_{34}\epsilon,
\quad \gamma_1\epsilon=\epsilon,\\
&&
F^{14}_{34}=F^{24}_{42}=F^{34}_{23}=-\frac{\ell}{2g}e^{-2g(r)},\nonumber
\end{eqnarray}
where a sign $+$ denotes self-dual parts of corresponding gamma matrices.

We proceed to scalar fields.
We turn on only one scalar field which breaks 
$SO(6)$ symmetry into the $SO(p)\times SO(q)$ with $p+q=6$.
We introduce this scalar field $\phi(r)$ as
\begin{eqnarray}
\label{scad3}
S_I^J=\left(e^{-q\phi(r)}\; {\bf 1}_{p\times p},
\; e^{p\phi(r)}\; {\bf 1}_{q\times q}\right),
\end{eqnarray}
where ${\bf 1}_{n\times n}$ denotes $n\times n$ unit matrix.
We set $\phi(r)=0$ in a case $p=6, q=0$ for two cycles inside
Calabi-Yau fourfolds.
Values of $p,q$ are read off by noting that $d+p$ is a real dimension
of non-compact special holonomy manifolds under consideration.

We wish to give comments on resulting fields based on above ansatz.
Let us decompose indices of $SO(6)$ symmetry $I,J=1,\dots,6$ into
$A,B=1,\dots,p$ for $SO(p)$ subgroup and $\widehat{A}, \widehat{B}=1,\dots,q$
for $SO(q)$ subgroup.
Then, we always have a relation
\begin{eqnarray}
F_{\mu\nu}^{AB}\Gamma_{AB}\gamma_{\mu\nu}\epsilon
=\frac{2d\ell}{g}e^{-2g(r)}\epsilon.
\end{eqnarray}
On the other hand, $USp(8)$ gauge fields are given by 
$Q_{\mu}^{ab}=-\frac{1}{4}gA_{\mu AB}
\left(\Gamma_{AB}\right)^{ab}$.
We omit expressions for 
$P_{\mu}^{abcd}$ and scalar functions $T_{ab}, A_{abcd}$ here.

Now, we are able to derive first order BPS equations 
for three radial functions $f(r), g(r)$ and $\phi(r)$ by handling 
supersymmetry transformations of fermions (\ref{susyd3})
\begin{eqnarray}
\label{bpsd3}
f'(r)e^{-f(r)}&=&-\frac{g}{12}\left(
pe^{-2q\phi(r)}+qe^{2p\phi(r)}
\right)
+\frac{d\ell}{6g}e^{2q\phi(r)-2g(r)},\nonumber\\
g'(r)e^{-f(r)}&=&-\frac{g}{12}\left(
pe^{-2q\phi(r)}+qe^{2p\phi(r)}
\right)
+\frac{(d-6)\ell}{6g}e^{2q\phi(r)-2g(r)},\\
\phi'(r)e^{-f(r)}&=&
-\frac{g}{12}\left(
e^{-2q\phi(r)}-e^{2p\phi(r)}
\right)+\frac{d\ell}{6pg}e^{2q\phi(r)-2g(r)},\nonumber
\end{eqnarray}
where $d$ is a real dimension of supersymmetric cycles, and 
$p+q=6$ which are chosen such that $d+p$ is a real dimension
of non-compact special holonomy manifolds.
We use a convention that
curvature $\ell$ of supersymmetric cycles are given by $\pm 1$.
We denote a derivative with respect to $r$ by $'$.
Note that these BPS equations have a similar form to those for wrapped
M5 and M2 branes \cite{GaKiWa, GaKiPaWa1}. 
 
\subsection{BPS solutions}
\hspace{5mm}
Let us discuss solutions for BPS equations (\ref{bpsd3}).
We have no exact solutions for them.
Thus, we will analyze behaviors of radial functions
$f(r),g(r)$ and $\phi(r)$ numerically.
Note that we always have a solution for spinor fields
\begin{eqnarray}
\epsilon(r)=e^{\frac{f(r)}{2}}\epsilon_{0},
\end{eqnarray}
where $\epsilon_{0}$ is certain constant spinor.
This solution is obtained from a radial component 
for supersymmetry variation of gravitinos.

We begin with specifying boundary behaviors of 
radial functions for each solution.
We impose that five-dimensional metric $ds_5^2$ at small $r\to0$
should behave asymptotically as follows
\begin{eqnarray}
ds_5^2=
\frac{2}{g^2r^2}\left(d\xi_{4-d}^2-dr^2-d\widetilde{s}_{\Sigma_d}^2\right).
\end{eqnarray}
This metric is like AdS$_5$ space-time except that a metric on $R^{1,3}$
is replaced by a metric on $R^{1,3-d}\times \Sigma_d$.
Then, we have following boundary behavior of $f(r), g(r)$ 
\begin{eqnarray}
\label{boundaryfg}
f(r)=g(r)=\log{\left(\frac{gr}{2}\right)},\qquad r\to 0.
\end{eqnarray}
We are able to determine asymptotic behavior of remaining
radial function $\phi(r)$ by substituting the asymptotic
behaviors of $f(r),g(r)$ into the third BPS equation 
including $\phi'(r)$ in (\ref{bpsd3}). 
Resulting behavior of $\phi(r)$ is 
\begin{eqnarray}
\phi(r)=\frac{d\ell}{12p}r^2 \log{r}+C r^2, \qquad r\to 0,
\end{eqnarray}
where $C$ is arbitrary constant.
This constant $C$ will be a single moduli which changes radial evolution of 
radial functions.
This asymptotic behavior of scalar function $\phi(r)$ 
may has an interpretation in AdS/CFT correspondence
\cite{MaNu1, BaKrLa}.
The leading term is interpreted 
as an insertion of certain boundary operator 
into worldvolume theory on wrapped D3 branes.
This operator is induced by curvature of wrapped cycles.
The subleading term corresponds to an expectation value of the
boundary operator.

Ten-dimensional metrics of wrapped D3 branes are given 
by embedding formulas (\ref{upliftd3}) 
\begin{eqnarray}
\label{d310me}
ds^2_{10}&=&
\Delta^{\frac{1}{2}}ds_{5}^2
-\frac{1}{g^2\Delta^{\frac{1}{2}}}T_{IJ}^{-1}
D\mu^I D\mu^J,\\
\Delta&=&e^{-2q\phi}\mu^A\mu^A
+e^{2p\phi}\mu^{\hat{A}}\mu^{\hat{A}},\qquad
D\mu^A=d\mu^A+gA^{AB}\mu^B,\nonumber
\end{eqnarray}
where $\mu^I$ are coordinates on $S^5$ satisfying $\sum_{I=1}^6\mu^I\mu^I=1$.
In general, we will have singularities 
in ten-dimensional solutions when we change $r$ to be large.
Here, we adopt a criteria for acceptable singularities in \cite{MaNu1}. 
Good singularities are singularities whose time $(00)$ component of embedded 
ten-dimensional metric $g_{00}^{D=10}$ goes to constant at singularities,
or changes in a bounded way near singularities.
On the other hand, bad singularities
are singularities whose $g_{00}^{D=10}$ component diverges at singularities
or is unbounded near singularities.
In practice, we are able to see the behaviors 
by numerically calculating dominant contribution 
$e^{2f(r)+p\phi(r)}$ when $\phi(r)$ diverges near singularities,
or $e^{2f(r)-q\phi(r)}$ when $\phi(r)$ becomes small near singularities. 
We expect that supergravity backgrounds with good singularities 
have alternative descriptions in order to describe 
physical phases in dual worldvolume theories on wrapped D3 branes.
As a side remark, one might try to turn on 
additional scalar fields in order to give marginal deformations \cite{MaNu1}
or change a feature of singularities \cite{BiCoZa}.
We do not include this generalization here.

Now, we analyze behaviors of each configuration numerically.
We begin with the simplest case without non-trivial scalar field $\phi(r)$.
In this example, we have $\phi(r)=0$ and
no moduli in solutions $f(r), g(r)$ for BPS equations.
We consider wrapped D3 branes around two cycles insides 
Calabi-Yau fourfolds ($CY_4$) for $d=2, p=6, q=0$.
We have following BPS equations
\begin{eqnarray}
f'(r)e^{-f(r)}=-\frac{g}{2}+\frac{\ell}{3g}e^{-2g(r)},\qquad
g'(r)e^{-f(r)}=-\frac{g}{2}-\frac{2\ell}{3g}e^{-2g(r)}.
\end{eqnarray}
We numerically study behaviors of $f(r), g(r)$ determined 
by these first order differential equations.
In fact, if we ignore the boundary behavior of $f(r), g(r)$ 
(\ref{boundaryfg}),
we are able to find an exact solution with AdS$_3$ space-time given by 
\begin{eqnarray}
\label{2cy4ads}
e^{f(r)}=\frac{4}{3g}\frac{1}{r},\qquad
e^{g(r)}=\frac{2}{\sqrt{3}g}, \qquad
\ell=-1.
\end{eqnarray}
\begin{figure}
\begin{center}
\includegraphics[height=4cm]{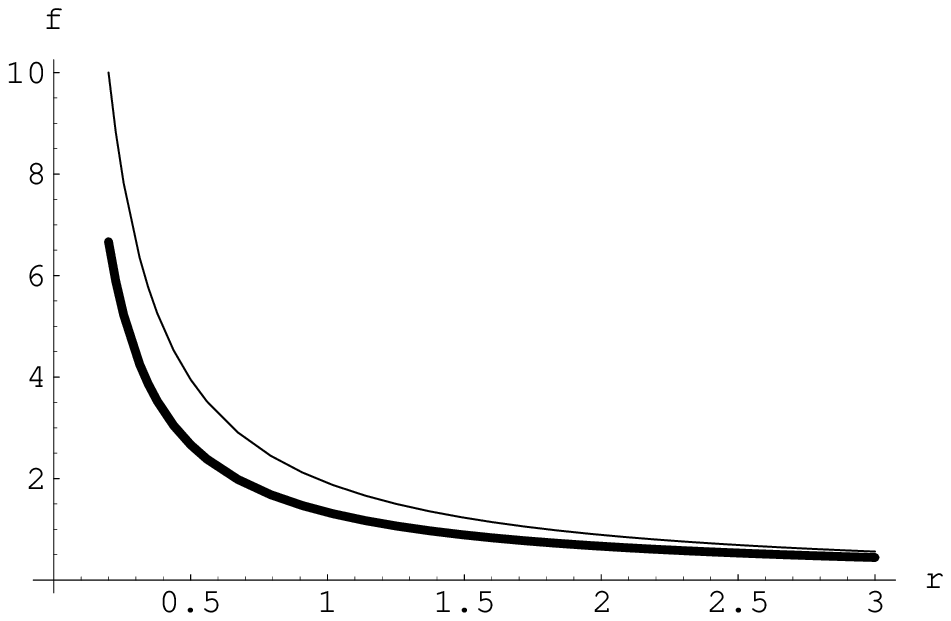}
\includegraphics[height=4cm]{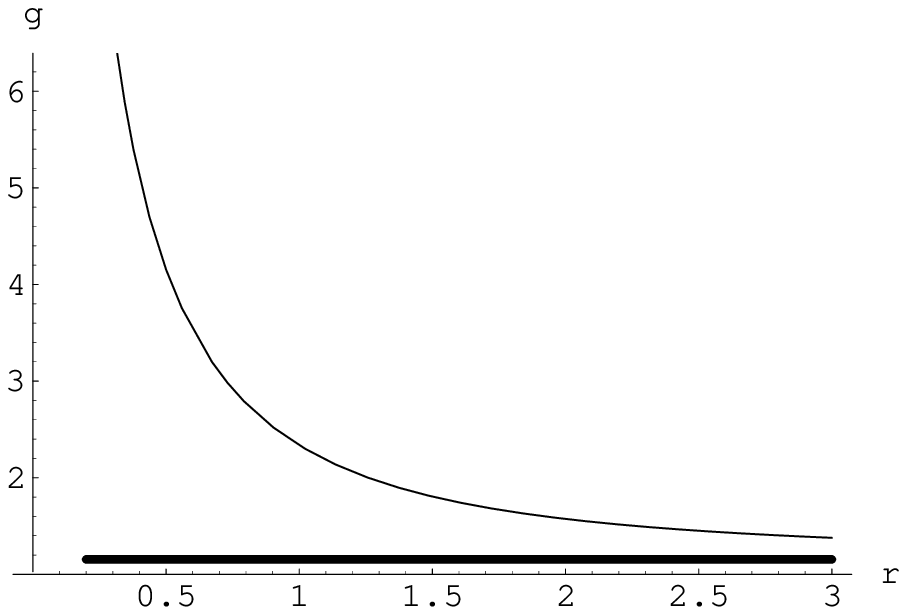}
\caption{Behavior of $e^{f(r)},e^{g(r)}$
for wrapped D3 branes around supersymmetric $\Sigma_{g\geq 2}$ in $CY_4$.
Numerical solution approaches to the AdS$_3$ solution (\ref{2cy4ads})
plotted by bold lines.}
\end{center}
\end{figure}
We may choose present holomorphic two cycles 
with constant negative curvature $\ell=-1$ to be
genus $g\geq 2$ Riemann surfaces $\Sigma_g$. 
We plot numerical evaluation of radial functions $f(r), g(r)$
by BPS equation with $\ell=-1$ in Figure 1.
Numerical solution approach this AdS$_3$ solution when $r$ becomes large.
Thus, we have obtained a solution which interpolates
almost AdS$_5$ (at small $r$) and AdS$_3$ space-times (at large $r$).
We expect that the AdS$_3$ solution is a dual supergravity background to 
$D=2$ $N=(1,1)$ superconformal field theory on wrapped D3 branes
at low energy with respect to the inverse size of $\Sigma_g$.
We expect that the numerical solution gives a dual background to
certain renormalization group flow from
twisted $D=4$ $N=4$ super Yang-Mills theory to the $D=2$ SCFT.
We might have $D=2$ $N=(1,1)$ sigma model on Higgs branch 
up to subtleties on vacuum expectation values of 
massless fields in two-dimensions \cite{MaNu1}.
This Higgs branch seems not to be visible in the supergravity solution.
We do not expect Coulomb branch here because we have no codimension 
both wrapped D3 branes and non-compact $CY_4$.
We are able to estimate a central charge of
the $D=2$ $N=(1,1)$ superconformal
field theory by assuming a AdS/CFT correspondence \cite{MaNu1} 
\begin{eqnarray}
c=\frac{3R_{AdS_3}}{2G_N^3}=
\frac{3\cdot\frac{4}{3g}}{2G_N^3}
=\frac{2}{g}\cdot\frac{{\rm Volume}(S^5)\cdot {\rm Volume}(\Sigma_g)}
{G_N^{10}}=
8\sqrt{2}\pi^5g^2N^2
\cdot {\rm Volume}(\Sigma_g),
\end{eqnarray}
where ten-dimensional Newton's constant is defined as
$G_N^{10}=8\pi^6g^2\alpha'^4$.
The radius of $S^5$ is given by $\sqrt{2}/g
=\sqrt{\alpha'}\left(4\pi g_{st}N\right)^{\frac{1}{4}}$
with the number of D3 branes $N$,
and a volume of unit five sphere is given by
${\rm Volume}(S^5)=\pi^3$.
It would be possible to derive this central charge from a precise
definition of dual superconformal field theory obtained by 
a dimensional reduction of twisted 
$D=4$ ${\cal N}=4$ $SU(N)$ super Yang-Mills theory \cite{VaWi}.
Let us proceed to D3 branes wrapped around holomorphic $CP^1$ 
with positive curvature $\ell=+1$ inside $CY_4$.
We do not have a AdS$_3$ solution for this configuration.
Numerical solution for radial functions $f(r), g(r)$ 
are plotted in Figure 2.
\begin{figure}
\begin{center}
\includegraphics[height=4cm]{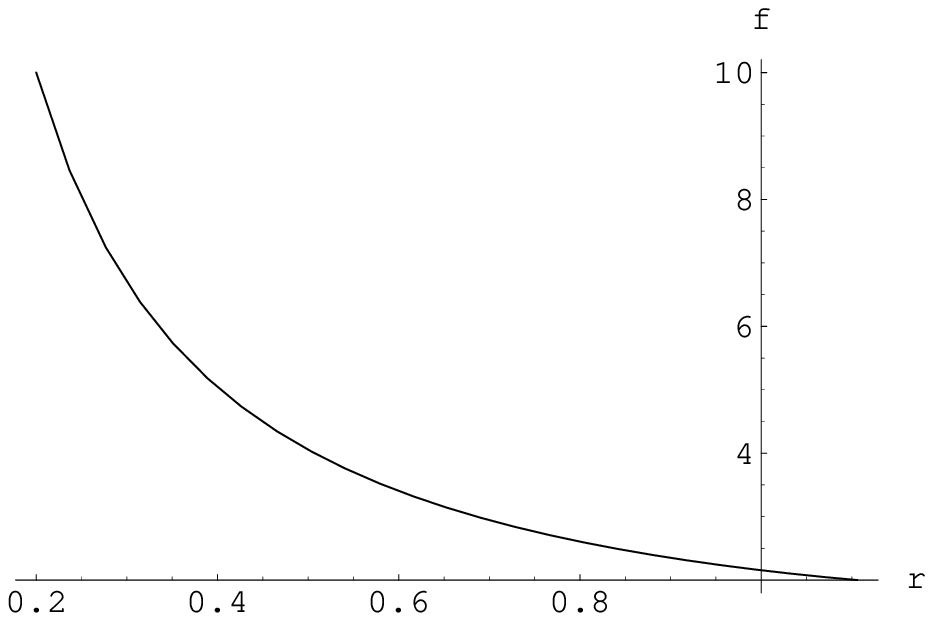}
\includegraphics[height=4cm]{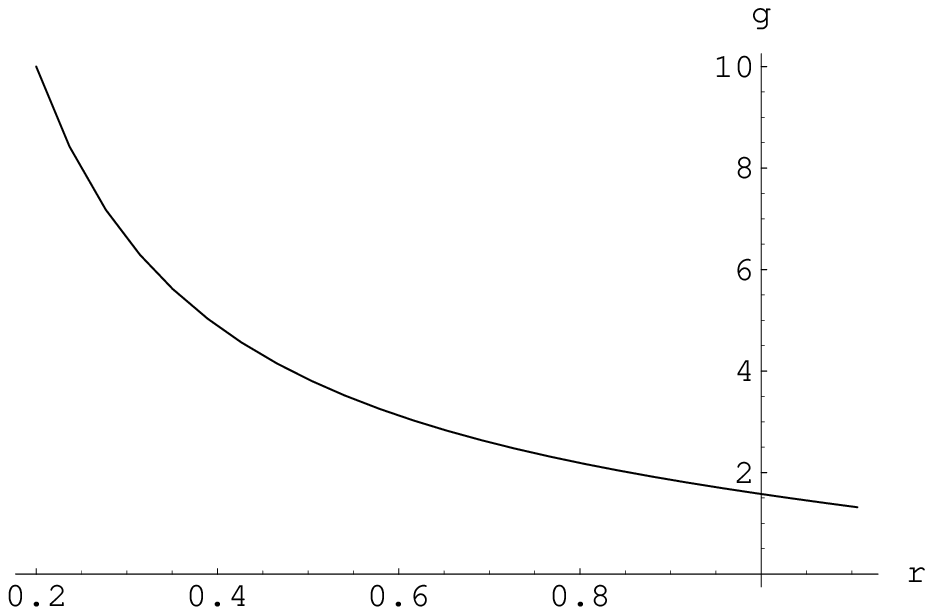}
\end{center}
\caption{Behavior of $e^{f(r)},e^{g(r)}$
for wrapped D3 branes on supersymmetric $CP^1$ in $CY_4$.}
\end{figure}
We see that $e^{f(r)}$ approaches to zero at a finite value
$r=r_0>0$.
This signals an appearance of
singularity in ten-dimensional metric (\ref{d310me}).
We easily see that the singularity is regarded as a good singularity.
Naively, we interpret that this solution near the singularity 
gives a supergravity background 
dual to $D=2$ $N=(1,1)$ sigma model on wrapped D3 branes.
We do not expect Higgs branch because we have no zero modes 
from $CP^1$ inside $CY_4$.
Also we seem to have no Coulomb branch
because we have no codimension 
both wrapped D3 branes and non-compact $CY_4$.
These observation give no sensible interpretation of
the supergravity solution from
a worldvolume theory on wrapped D3 branes.

Let us proceed to solutions of wrapped D3 branes 
around supersymmetric three cycles inside Calabi-Yau threefolds $(CY_3)$
for $d=p=q=3$.
We consider three cycles with negative $(\ell=-1)$
and positive $(\ell=+1)$ constant curvature  
as hyperbolic space $H^3$ and sphere $S^3$. 
We have no solutions with AdS$_2$ space-time from BPS equations (\ref{bpsd3})
for $\ell=\pm 1$.
First, we depict numerical evaluations of $f(r), g(r)$ and $\phi(r)$
for solutions with $\ell=-1$ in Figure 3.
We have chosen five values of moduli $C$ in order to
show a feature of solutions. 
\begin{figure}
\begin{center}
\includegraphics[height=4cm]{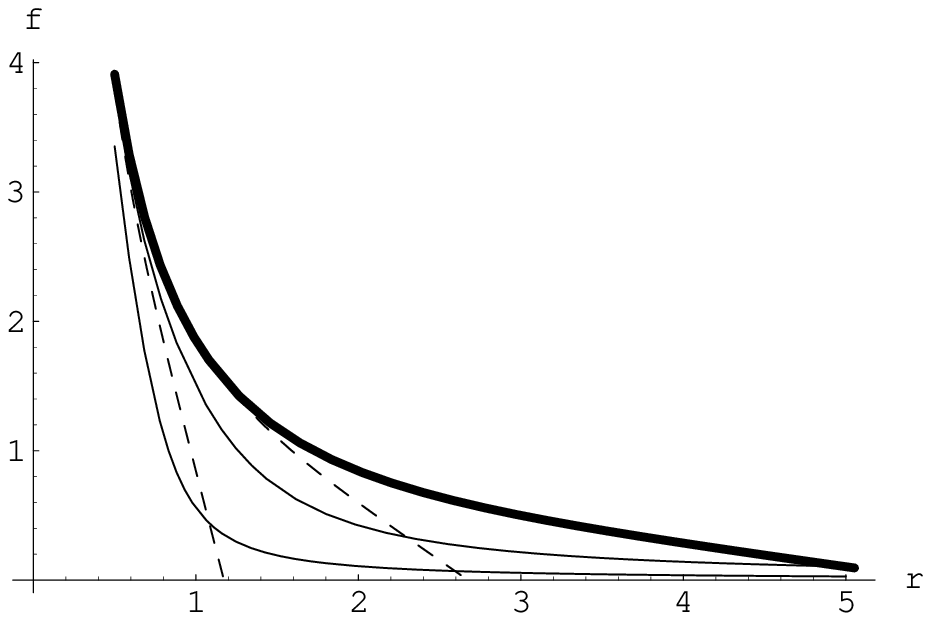}\quad
\includegraphics[height=4cm]{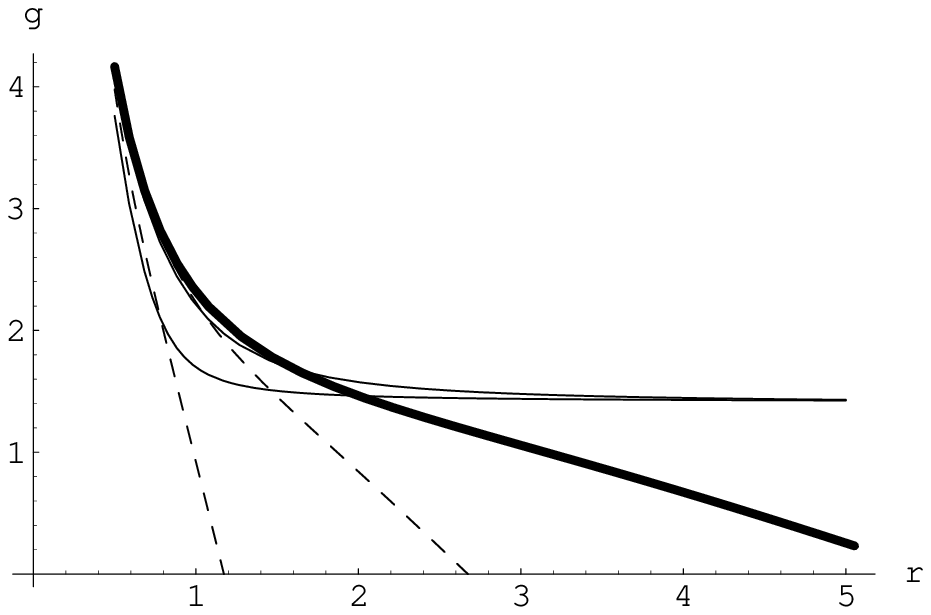}
\includegraphics[height=4cm]{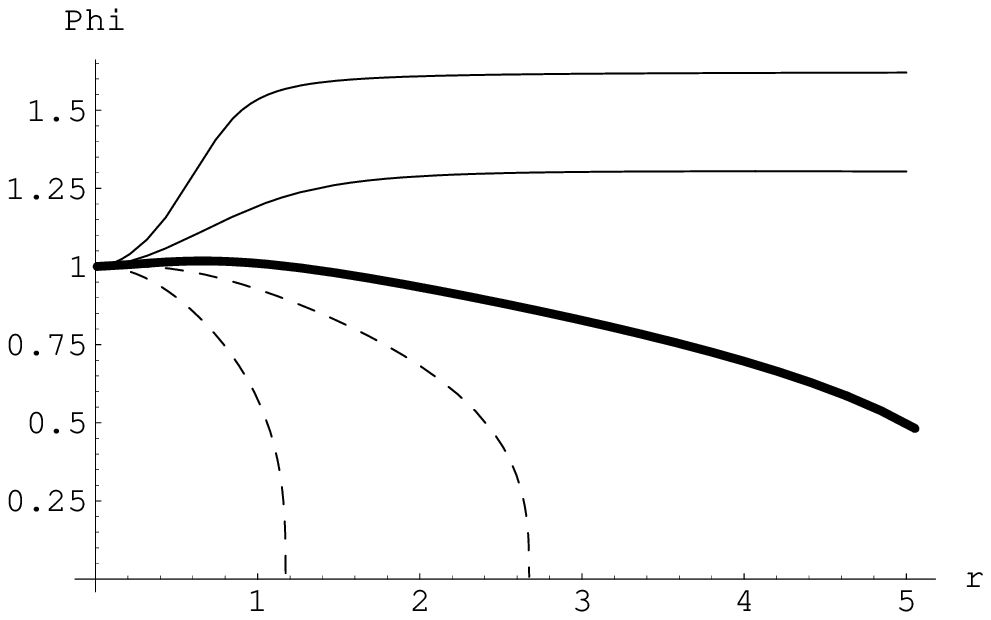}
\end{center}
\caption{Behavior of $e^{f(r)},e^{g(r)}$ and $e^{\phi(r)}$ for
wrapped D3 branes around supersymmetric $H^3$ in $CY_3$.
We denote plots of solutions with $C=0.75, 0.25$ by usual lines,
a plot with $C=0.01$ by a bold line, and plots $C=-0.1, -0.5$ by dotted lines.}
\end{figure}
Behaviors of solutions depending on a moduli $C$
change their feature around $C=0.01$.
We see that for solutions with $C=0.75$ and $0.25$,
$e^{f(r)}$ goes to zero at large $r$,
and $e^{g(r)}, e^{\phi(r)}$ approach to constants at large $r$. 
For plots with $C=-0.1$ and $-0.5$, $e^{2g(r)}$ goes to zero at first.
We can check that we always have good singularities for these five solutions. 
We expect that this is correct for solutions with arbitrary values of $C$.
Let us interpret this result from dual worldvolume theories 
on wrapped D3 branes.
We expect that the worldvolume theories are
$D=1$ supersymmetric quantum mechanics with eight supercharges 
at low energy limit.
These theories may be viewed as sigma models on Higgs or Coulomb
branches spanned by motions of wrapped D3 branes.
Naively, we expect that 
solutions with $C>0.01$ correspond to Higgs branches, and
solutions with $C<0.01$ correspond to Coulomb branches 
\cite{MaNu1,FrGuPiWa2}.
Present analysis indicates
that both Higgs and Coulomb branches can be captured in a supergravity limit.
Note that we are able to have Higgs branches spanned by zero modes on $H^3$.
Slight different behaviors of solutions should be interpreted 
in terms of a correspondence between Higgs and Coulomb branches.
Secondly, we turn to D3 branes wrapped around positive curvature cycle
$S^3$ inside $CY_3$.
We show numerical plots of three radial functions with particular
values of $C$ in Figure 4.
\begin{figure}
\begin{center}
\includegraphics[height=4cm]{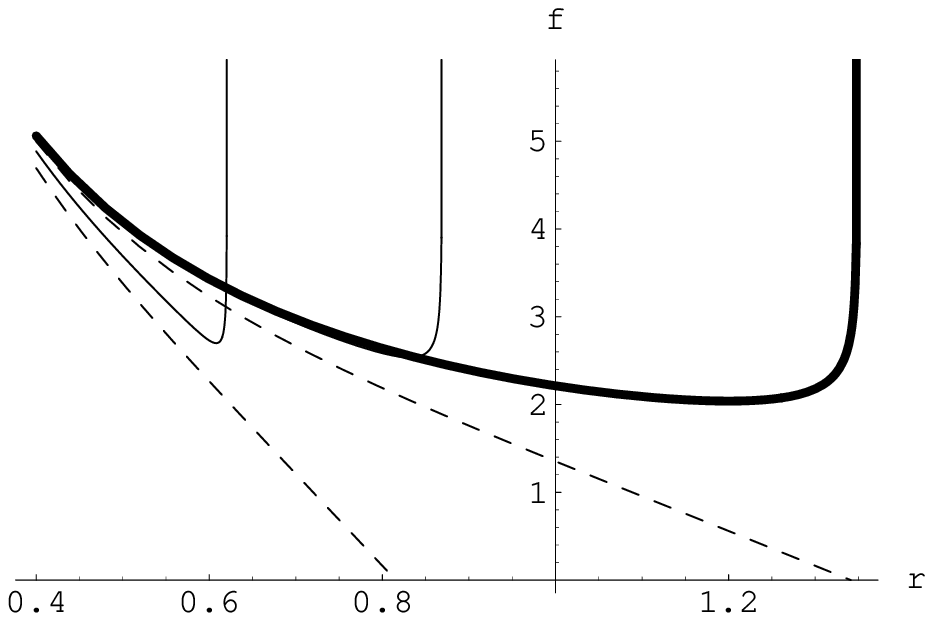}\quad
\includegraphics[height=4cm]{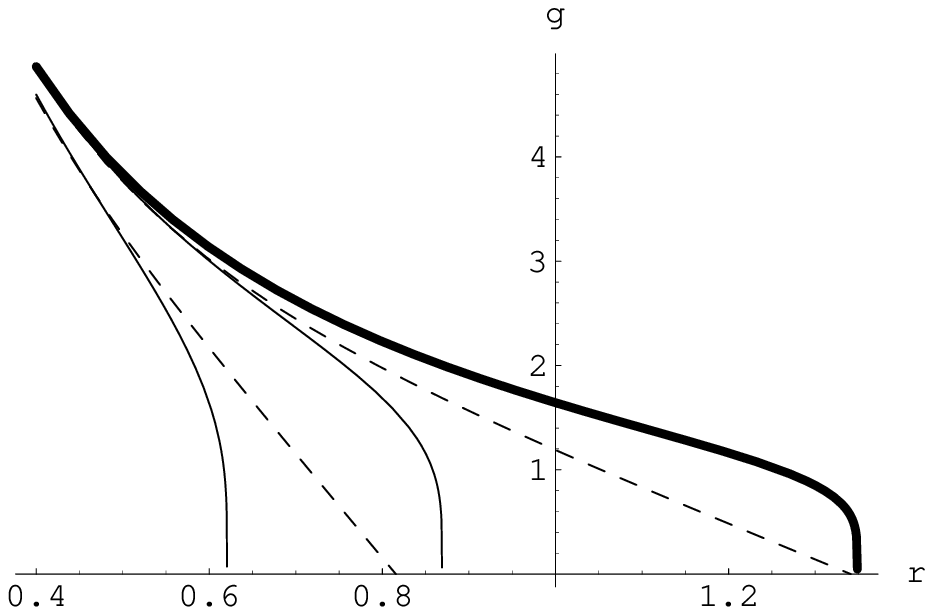}
\includegraphics[height=4cm]{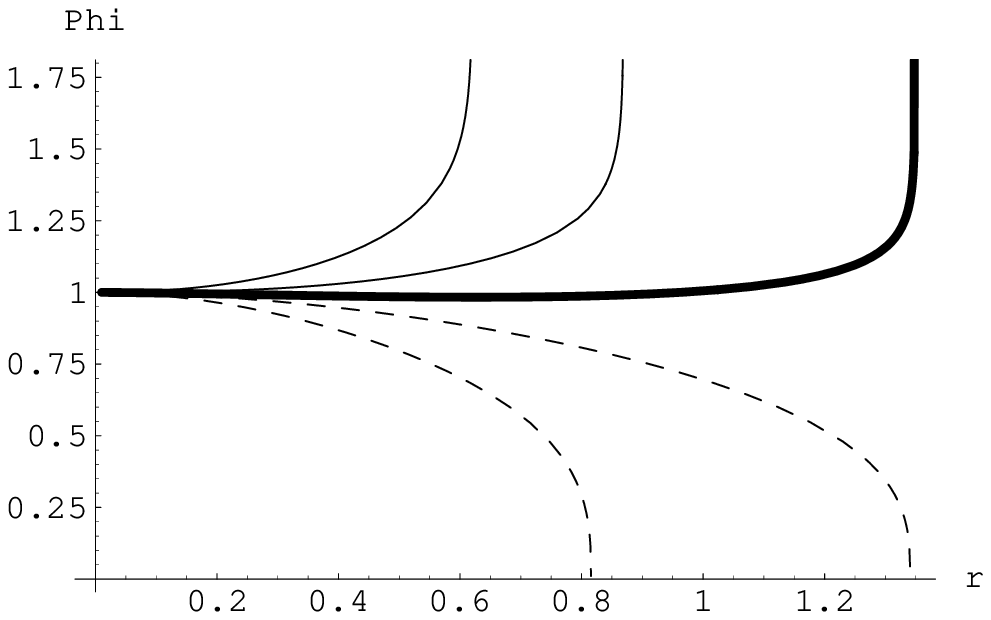}
\end{center}
\caption{Behavior of $e^{f(r)},e^{g(r)}$ and $e^{\phi(r)}$
for wrapped D3 branes around supersymmetric $S^3$ in $CY_3$.
We denote plots of solutions with $C=0.75, 0.25$ by usual lines,
a plot with $C=0$ by a bold line, and plots with 
$C=-0.25,-0.75$ by dotted lines.}
\end{figure}
In this case, we have bad singularities for $C=0,0.25$ and $0.75$
due to divergence of $e^{2\phi(r)}$ near singularities where
$e^{g(r)}=0$.
On the other hand, we have good singularities for $C=-0.25$ and $-0.75$.
These behaviors indicate that one-dimensional supersymmetric
quantum mechanics on wrapped D3 branes do not have Higgs branches.
This is natural because three sphere $S^3$ does not give rise to zero modes
which parameterize a moduli space of Higgs branch.
Coulomb branches are still captured by solutions with $C<0.01$.

Wrapped D3 branes around holomorphic two cycles inside
non-compact $K3$ manifolds $(d=2, p=2, q=4)$
show the same feature as above D3 branes wrapped 
around three cycles inside $CY_3$.
This gives a consistent result with \cite{MaNu1}. 
An interpretation of the solutions is similarly stated to above cases. 

We turn to wrapped D3 branes around 
associative three cycles inside non-compact $G_2$ holonomy manifolds
for $d=3, p=4, q=2$.
In this case, we have an exact solution with AdS$_2$ space-time \cite{NiOz}
\begin{eqnarray}
\label{d3g2ads}
e^{f(r)}=\frac{1}{2g}e^{4\phi}\frac{1}{r},\qquad
e^{g(r)}=\frac{1}{\sqrt{2}g}e^{4\phi},\qquad
e^{12\phi(r)}=4,\qquad \ell=-1.
\end{eqnarray}
We choose negatively curved cycle as three-dimensional
hyperbolic space $H^3$ here.
Numerical solution which approaches this AdS$_2$ solution at large $r$
is expected to give a critical solution for one parameter numerical solutions 
when we change values of moduli $C$.
We wish to fix a value of moduli $C$ for this asymptotic AdS$_2$ solution. 
Let us note a relation obtained from the BPS equation 
\begin{eqnarray}
e^{2g(r)+4\phi(r)}=e^{2g(r)-8\phi(r)}-\frac{3\ell}{g^2}
\left(g(r)+4\phi(r)\right)+C_0,
\end{eqnarray}
where $C_0$ is arbitrary constant.
This relation is not depend on $r$.
We are able to see that we have
$C_0=\frac{3}{2g^2}-\frac{5}{2g^2}\log{2}
+\frac{3}{g^2}\log{g}$
for the AdS$_2$ solution.
By substituting asymptotic behaviors of $g(r)$ and $\phi(r)$
at small $r$ into above relation,
we find that a value of moduli $C$ is
$C=\frac{1}{32}+\frac{1}{96}\log{2}$.
Let us show numerical solutions of radial functions 
with $\ell=-1$ for five specific values of $C$ in Figure 5.
\begin{figure}
\begin{center}
\includegraphics[height=4cm]{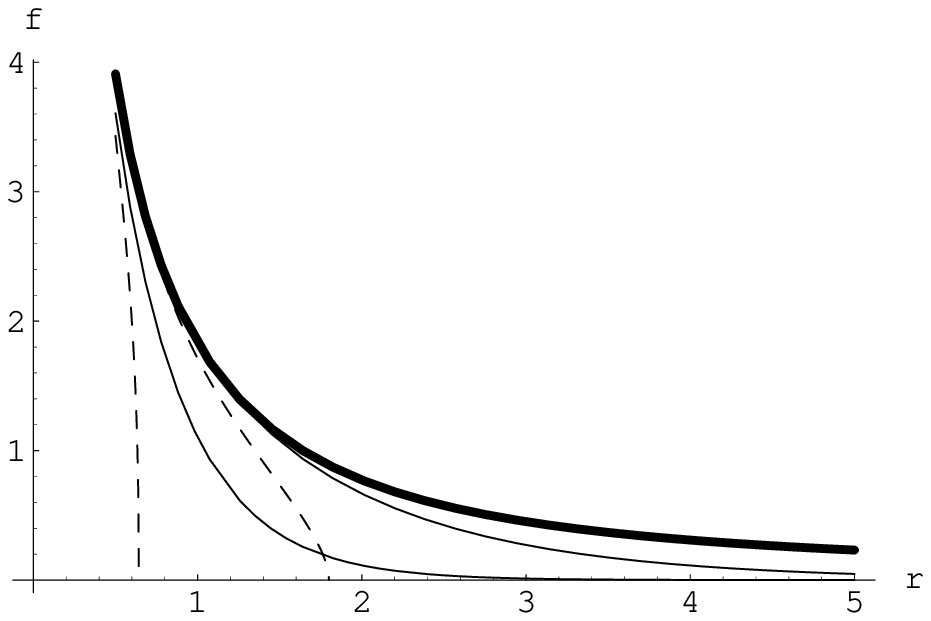}\quad
\includegraphics[height=4cm]{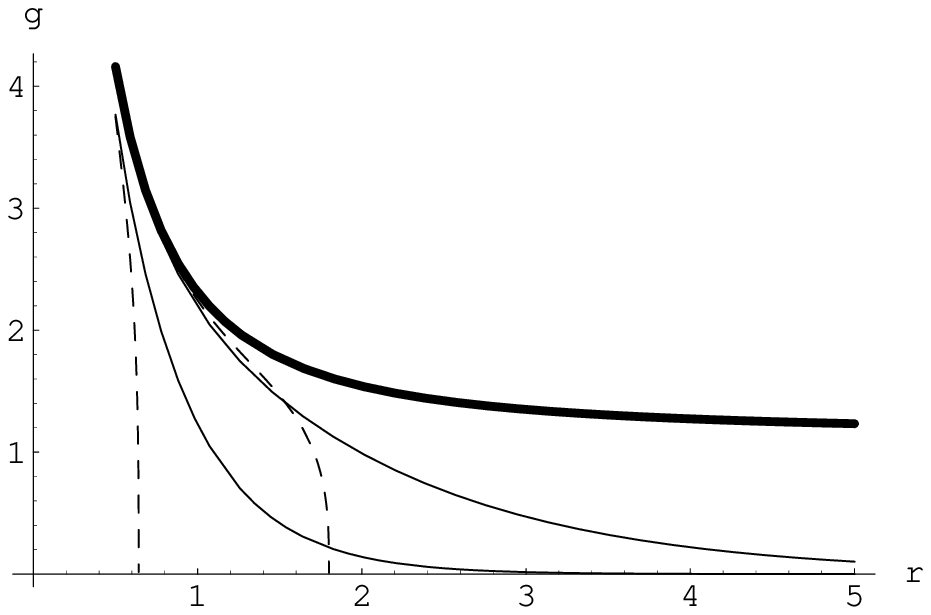}
\includegraphics[height=4cm]{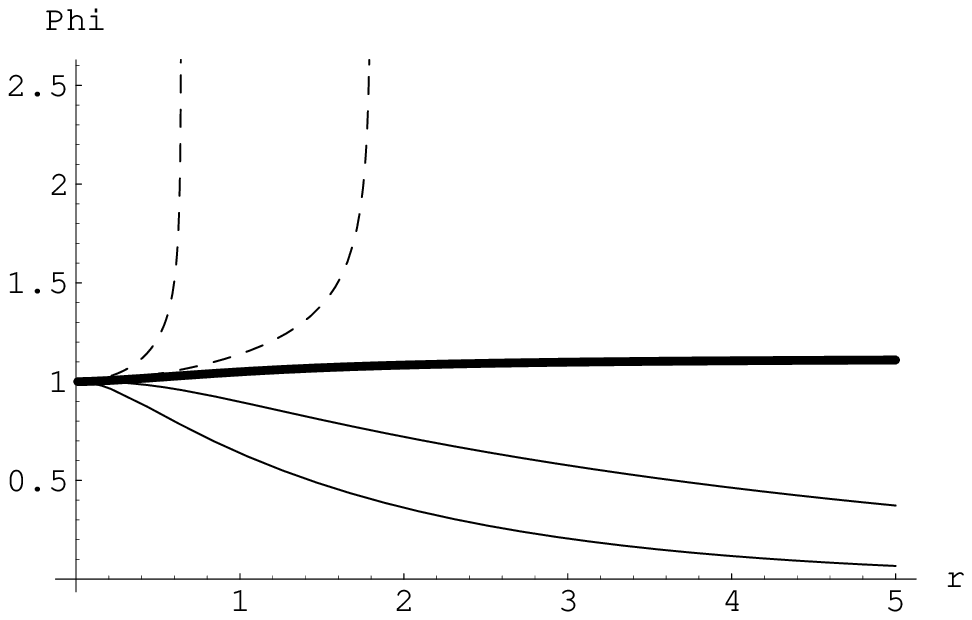}
\end{center}
\caption{Behavior of $e^{f(r)},e^{g(r)}$ and $e^{\phi(r)}$
for wrapped D3 branes around supersymmetric
$H^3$ in non-compact $G_2$ holonomy manifolds.
We denote plots of solutions with $C=-1, -0.15$ by usual lines,
a plot with $C=\frac{1}{32}+\frac{\log{2}}{96}$ by a bold line, and
plots with $C=0.1, 0.5$ by dotted lines.
Solution with  $C=\frac{1}{32}+\frac{\log{2}}{96}$ approaches to 
the AdS$_2$ solution (\ref{d3g2ads}) at large $r$.}
\end{figure}
It follows that we have good singularities 
for solutions with $C=-1$ and $-0.15$.
On the other hand, we have bad singularities for solutions with 
$C=0.1$ and $0.5$.
Let us interpret that the former
solutions reflect Higgs branches
of motions of wrapped D3 branes, and
the latter solutions reflect Coulomb branches \cite{NiOz}. 
This suggests that we have Coulomb and Higgs branches which
intersect a conformal fixed point given by
AdS$_2$ solution with $C=\frac{1}{32}+\frac{1}{96}\log{2}$.
These solutions are expected to give the dual background of the 
worldvolume $D=1$ supersymmetric quantum mechanics with four
supercharges on the wrapped D3 branes at low energies 
smaller than the inverse size of wrapped cycles.
The solutions suggest that
we loss an information of Coulomb branches and 
are able to capture only Higgs branches in the decoupling limit.
It is possible to have zero modes on $H^3$ in order to have Higgs branches.
The AdS$_2$ solution is expected to give a dual background of a
superconformal quantum mechanics on the wrapped D3 branes.
The corresponding central charge are evaluated from the AdS$_2$ solution
\begin{eqnarray} 
c\propto\frac{1}{G_N^2}
=\frac{{\rm Volume}(S^5)\cdot{\rm Volume}(H^3)}{G_N^{10}}
=4\sqrt{2}\pi^5g^3N^2\cdot {\rm Volume}(H^3),
\end{eqnarray}
where $G_N^{2}$ and $G_N^{10}$ are two- and ten-dimensional Newton's constants.
We denote the number of D3 branes by $N$.
Normalization of this central charge should be determined 
in an appropriate way.
Here, the radius of AdS$_2$ space-time does not appear in 
the expression for central charge of the one-dimensional conformal theory.
We proceed to wrapped D3 branes around positive curvature $(\ell=+1)$
cycles $S^3$.
We do not have AdS$_2$ solution to the BPS equations.
We show numerical evaluations for radial functions 
with five chosen values of moduli $C$ in Figure 6.
\begin{figure}
\begin{center}
\includegraphics[height=4cm]{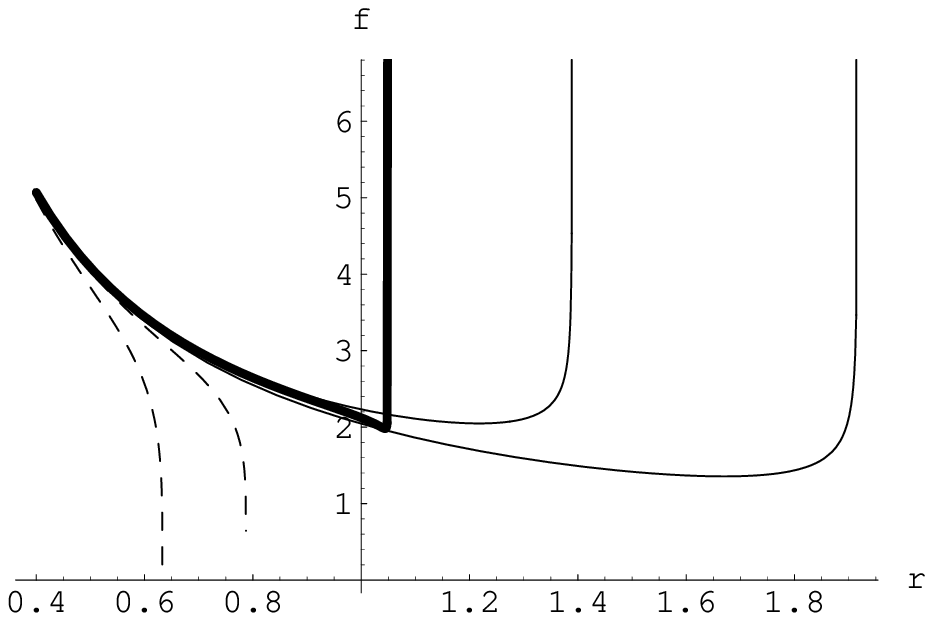}\quad
\includegraphics[height=4cm]{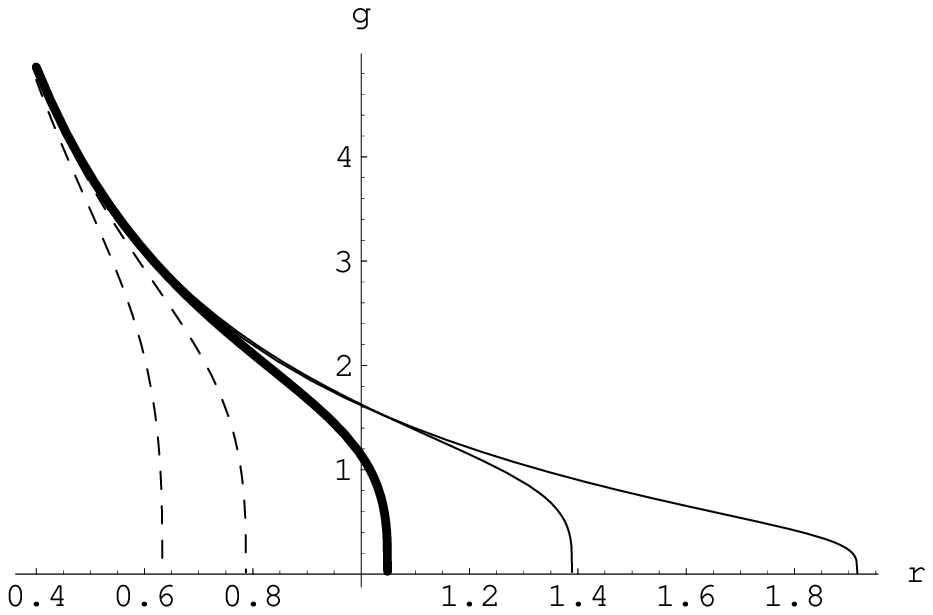}
\includegraphics[height=4cm]{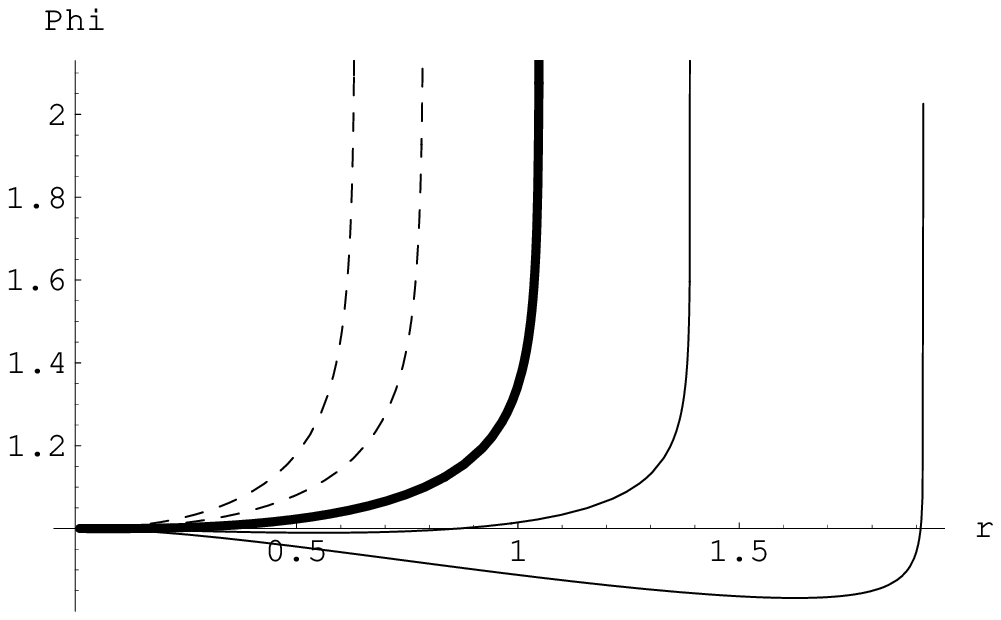}
\end{center}
\caption{Behavior of $e^{f(r)},e^{g(r)}$ and $e^{\phi(r)}$
for wrapped D3 branes around supersymmetric
$S^3$ in non-compact $G_2$ holonomy manifolds.
We denote plots of solutions with  $C=-0.15, 0$ by usual lines,
a plot with $C=0.125$ by a bold line, and 
plots with $C=0.3, 0.5$ by dotted lines.
Note that 
$e^{\phi(r)}$ always diverge near singularities where $e^{g(r)}=0$.}
\end{figure}
The numerical solutions always have bad singularities
due to a divergence of $e^{2\phi(r)}$ at singularities
where $e^{g(r)}$ goes to zero.
These bad singularities suggest that 
we have no physically sensible branches on worldvolume
theories in the supergravity limit.  
As in $H^3$ case, we expect 
that only Higgs branches are captured at low energy limit.
But, these Higgs branches are not allowed because $S^3$ gives no zero modes.
Thus, we have a sensible interpretation of 
solutions by dual worldvolume theories \cite{NiOz}.
Notice that we have chosen an identification between 
moduli $C$ and Higgs/Coulomb branches 
in order to match physically natural pictures.
It would be necessary to give an direct argument on this identification.

Numerical behavior of wrapped D3 branes around holomorphic two cycles inside
non-compact Calabi-Yau threefolds $(d=2, p=4, q=2)$
show the same feature as that of above D3
branes wrapped around three cycles inside $G_2$ holonomy manifolds.
An interpretation of the result is also stated in the same manner.
We have an exact AdS$_3$ solution for two cycles with negative curvature 
\cite{MaNu1}
\begin{eqnarray}
e^{f(r)}=\frac{1}{g}e^{4\phi}\frac{1}{r},\qquad
e^{g(r)}=\frac{1}{g}e^{4\phi},\qquad
e^{12\phi(r)}=2,\qquad \ell=-1.
\end{eqnarray}
Here, we determine a value of moduli $C$ for the
asymptotic AdS$_3$ solution.
We have a $r$-independent relation between $g(r)$ and $\phi(r)$
from the BPS equations
\begin{eqnarray}
e^{2g(r)+4\phi(r)}=e^{2g(r)-8\phi(r)}-\frac{2\ell}{g^2}
\left(g(r)+8\phi(r)\right)+C_0,
\end{eqnarray}
where $C_0$ is arbitrary constant.
We are able to see $C_0=\frac{-2}{g^2}\log{\frac{2}{g}}$ 
for the AdS$_3$ solution.
Then, we arrive at a value of moduli $C=1/48$ 
for the AdS$_3$ solution.
One parameter numerical solutions change their behaviors
around this value of moduli $C=1/48$.

It might appear different types of singularities 
by choosing values of moduli $C$ carefully.
Thus, for completeness, 
let us mention a general interpretation 
about behaviors of supergravity solutions.
Supersymmetric worldvolume theories on wrapped D3 branes have 
Higgs and Coulomb branches.
We expect a dynamics of Higgs branch 
from motions of wrapped D3 branes transverse to supersymmetric cycles
and tangent to non-compact special holonomy manifolds. 
On the other hand, 
we are able to have a dynamics of Coulomb branches
from motions which are transverse to all directions in
non-compact special holonomy manifolds.
If we have zero modes on wrapped supersymmetric cycles, 
we expect Higgs branches classically.
We should have at least good singularities in supergravity solutions
for each Higgs/Coulomb branch.
However, if we have only bad singularities, we manage to interpret that 
physical branches are not visible in a supergravity limit.
On the other hand,
if we have AdS solutions at large $r$,
it is natural to expect that corresponding conformal theories 
appear at an intersection of Higgs and Coulomb branches.
In any cases,
it would be certainly interesting 
to give an argument based on a precise study on 
dual worldvolume theories.
Then, we should start with Lagrangian of the ${\cal N}=4$ 
super Yang-Mills theory and its dimensional reductions.
It will be also desirable to give reasons why we have AdS$_{3}$ and
AdS$_2$ solutions only for several restricted configurations.

Alternatively, we are able to analyze solutions of BPS equations in the
same way as \cite{GaKiWa} on wrapped M5 branes.
We introduce new radial functions $h(r), x(r)$ and $F(r)$
\begin{eqnarray}
h(r)=e^{f(r)-2q\phi(r)},\qquad
x(r)=e^{2g(r)-4q\phi(r)},
\qquad
F(r)=x(r)^{\frac{q}{2}}e^{12\phi(r)}.
\end{eqnarray}
Then, we are able to derive ordinary differential equations
for $F$ and $x$ from BPS equations (\ref{bpsd3})
when we have non-trivial scalar function $\phi(r)$.
We write down these equations for each configuration
specified by $d$ and $p$
\begin{eqnarray}
\underline{d=2,p=2}\qquad
\frac{dF}{dx}=\frac{g^2F}{2g^2\left(2F\sqrt{x}-x\right)+2\ell},&&
\underline{d=2,p=4}\qquad
\frac{dF}{dx}=\frac{F}{x}\frac{g^2x+\ell}{g^2F+2\ell},\nonumber\\
\underline{d=3,p=3}\qquad
\frac{dF}{dx}=\frac{4g^2F}{3g^2\left(3x^{\frac{1}{3}}F-x\right)+6\ell},&&
\underline{d=3,p=4}\qquad
\frac{dF}{dx}=\frac{1}{2}\frac{F}{x}\frac{2g^2x+\ell}{g^2F+2\ell}.\nonumber\\
\end{eqnarray}
It is straightforward to characterize behaviors 
of ten-dimensional metrics (\ref{d310me}).
Here, one may wish to classify behaviors of wrapped D3 branes 
by combining with those of wrapped M5 and M2 branes.

\section{Wrapped M2 branes}
\hspace{5mm}
Here, we consider supersymmetric 
wrapped M2 branes obtained by truncated $D=4$ $U(1)^4$
gauged supergravity \cite{GaKiPaWa1}.
Our present aim is to embed these wrapped M2 brane configurations into 
$D=4$ $N=8$ $SO(8)$ gauged supergravity \cite{deNi}. 
This maximally supersymmetric gauged supergravity in four-dimensions
is believed to be a consistent 
$AdS_4\times S^7$ reduction of $D=11$ supergravity \cite{deNi2, deNi3, deNi4}. 
We correctly reproduce a result \cite{GaKiPaWa1}
by choosing suitable ansatz on scalar fields in $SO(8)$
gauged supergravity.

\subsection{$D=4$ $N=8$ $SO(8)$ gauged supergravity}
\hspace{5mm}
Let us introduce $D=4$ $N=8$ $SO(8)$ gauged supergravity \cite{deNi}. 
The field content consists of graviton $e^{\alpha}_{\mu}$,
$SO(8)$ gauge fields $A_{\mu}^{IJ}$, 
$E_7/SU(8)$ coset scalar fields ${\cal V}_{ij}^{IJ}$,
gravitinos $\psi_{\mu}^i$ and gauginos $\chi^{ijk}$.
We denote indices of $SU(8)$ and $SO(8)$ by
$i,j=1\dots 8$ and $I,J=1\dots 8$.
We have 70 real scalar fields which 
are described by 56-bein for the coset element
\begin{eqnarray}
\label{coset}
{\cal V}_{ij}^{IJ}=\left(
\begin{array}{cc}
u_{ij}^{IJ}& v_{ijKL}\\
v^{k\ell IJ}&u^{k\ell}_{KL}
\end{array}
\right).
\end{eqnarray}
Component $28 \times 28$ matrices $u_{ij}^{IJ},
u_{IJ}^{ij}$ and $v_{ijKL}, v^{ijIJ}$ with anti-symmetric indices 
are real matrices which behave under a complex conjugation as
$\left(u_{ij}^{IJ}\right)^*=u^{ij}_{IJ}$, 
$\left(v_{ij IJ}\right)^*=v^{ij IJ}$.
These matrices also satisfy following orthonormalization relations 
\begin{eqnarray}
\label{ortho}
u^{ij}_{IJ}u_{k\ell}^{IJ}-v^{ij IJ}v_{k\ell IJ}=\frac{1}{2}
\left(\delta^i_k\delta^j_{\ell}-\delta^i_{\ell}\delta^j_k\right),&&
u^{ij}_{IJ}v^{k\ell IJ}-v^{ij IJ}u^{k\ell}_{IJ}=0,\\
u^{ij}_{IJ}u_{ij}^{KL}-v_{ij IJ}v^{ij KL}=
\frac{1}{2}\left(\delta_I^K\delta_J^L
-\delta_J^K\delta_I^L\right),&&
u^{ij}_{IJ}v_{ij KL}-v_{ij IJ}u^{ij}_{KL}=0.\nonumber
\end{eqnarray}
On the other hand,
$SO(8)$ gauge field potential $A_{\mu}^{IJ}$ 
transforms under adjoint representation of $SO(8)$ group.
Corresponding gauge field strength is defined as 
$F_{\mu}^{IJ}=2\partial_{[\mu}A_{\nu]}^{IJ}
-2gA_{[\mu}^{IK}A_{\nu]}^{KJ}$
with $SO(8)$ gauge coupling constant $g$.
We introduce composite $SU(8)$ gauge field potential
${\cal B}_{\mu j}^i$ by
\begin{eqnarray}
{\cal B}_{\mu j}^i=
\frac{2}{3}\left(
D_{\mu}^{A}u_{ik}^{IJ}\cdot u_{IJ}^{kj}
-D_{\mu}^{A}v_{ik}^{IJ}\cdot v^{kj IJ}
\right),
\end{eqnarray}
where $D_{\mu}^A$ is a covariant derivative defined as
$D_{\mu}^Au_{ij}^{IJ}
=\partial_{\mu}u_{ij}^{IJ}-2gA_{\mu}^{K[I}u_{ij}^{J]K}$.
Let us also define a term which 
enters in a scalar kinetic term
within a Lagrangian, ${\cal A}_{\mu}^{ijk\ell}$ by
\begin{eqnarray}
{\cal A}_{\mu}^{ijk\ell}
=-2\sqrt{2}\left(
u^{ij}_{IJ}D_{\mu}v^{k\ell IJ}
-
v^{ij IJ}D_{\mu}u^{k\ell}_{IJ}
\right).
\end{eqnarray}
Here, we use a $SO(8)\times SU(8)$ covariant derivative $D_{\mu}$
which acts $u_{ij}^{IJ}$ as
\begin{eqnarray}
D_{\mu}u_{ij}^{IJ}=
\partial_{\mu}u_{ij}^{IJ}
+
{\cal B}_{\mu [i}^k u_{j] k}^{IJ}
-2gA_{\mu}^{K[I}u_{ij}^{J]K}.
\end{eqnarray}
Now, we write a bosonic part of the Lagrangian $L$ 
\begin{equation}
L=\sqrt{-g}\left[R-\frac{1}{48}\left({\cal A}_{\mu}^{ijk\ell}\right)^2
-\frac{1}{4}
F^+_{\mu\nu,IJ}\left[
2S^{IJ,KL}-\delta_{KL}^{IJ}
\right]F_{KL}^{+\mu\nu}
-g^2\left(
\frac{3}{2}|A^{ij}|^2-\frac{1}{12}|A_i^{\; jk\ell}|^2\right)\right],
\nonumber\\
\end{equation}
where we use a signature $( - + + + )$ of four-dimensional metric.
We denote self-dual part of $F_{\mu\nu}$ by $F^+_{\mu\nu}$.
Symbol $\delta^{IJ}_{KL}$ is equal to $\frac{1}{2}
\left(\delta_{K}^I\delta_L^J-\delta_K^J\delta_L^I\right)$.
Here, we also introduce scalar dependent matrices $S^{IJ,KL}$ by
$\left(u_{IJ}^{ij}+v_{ijIJ}\right)S^{IJ,KL}=u^{ij}_{KL}$.
Two scalar functions in a potential term of 
the Lagrangian are defined as  
\begin{eqnarray}
A^{ij}=\frac{4}{21}T_{k}^{\; ikj},\quad\quad
A_i^{\; jk\ell}=-\frac{4}{3}T_i^{\; [jk\ell]},
\end{eqnarray}
where a function $T_i^{\; jk\ell}$
with $SU(8)$ tensor indices is given by
\begin{eqnarray}
T_i^{\; jk\ell}=
\left(u^{k\ell}_{IJ}+v^{k\ell IJ}\right)
\left(u_{im}^{JK}u^{jm}_{KI}-v_{im JK}v^{jm KI}
\right).
\end{eqnarray}
Supersymmetry transformations of gravitinos $\psi_{\mu}^i$
and gauginos $\chi_{ijk}$ in bosonic backgrounds are given by                   \begin{eqnarray}
\label{d4susy}
\frac{1}{2}\delta\psi_{\mu}^i
&=&\partial_{\mu}\epsilon^i
+\frac{1}{4}\omega_{\mu}^{\rho\sigma}\gamma_{\rho\sigma}\epsilon^i
+\frac{1}{2}{\cal B}_{\mu\; j}^{\;\; i}\epsilon^j
+\frac{1}{\sqrt{2}}\left(
\frac{1}{4}\overline{F}_{\nu\lambda}^{-ij}\gamma^{\nu\lambda}
-gA^{ij}
\right)\gamma_{\mu}\epsilon_j\nonumber\\
\delta\chi^{ijk}&=&-\frac{1}{2}
{\cal A}_{\mu}^{ijk\ell}\gamma^{\mu}\epsilon_{\ell}
+\left(
\frac{3}{2}\gamma^{\mu\nu}\overline{F}_{\mu\nu}^{-[ij}\delta_{\ell}^{k]}
-2gA_{\ell}^{\; ijk}
\right)\epsilon^{\ell}.
\end{eqnarray}
Here, gauge fields strength $\overline{F}_{\mu\nu}^{ij}$
with $SU(8)$ indices $i,j$
are defined as 
$F_{\mu\nu}^{IJ}=\left(u_{ij}^{IJ}+v_{ij IJ}\right)
\overline{F}_{\mu\nu}^{ij}$,
and $\overline{F}_{\mu\nu}^{-ij}$ is an anti-self-dual part of 
the field strength $\overline{F}_{\mu\nu}^{ij}$.         
We also introduce $D=4$ gamma matrices $\gamma_{\mu}$ satisfying 
relations $\{\gamma_{\mu},\gamma_{\nu}\}=2g_{\mu\nu}$
with four-dimensional metric tensor $g_{\mu\nu}$.
We denote components of spin connection for four-dimensional metric
by $\omega_{\mu}^{\nu\rho}$.

Note that solutions in this gauged supergravity are 
arranged into solutions in eleven-dimensional supergravity 
by using an embedding formula \cite{deNi2, deNi3, deNi4}. 

\subsection{BPS equations}
\hspace{5mm}
Let us consider BPS equations in the gauged supergravity 
for possible configurations of wrapped M2 branes.
We are able to wrap M2 branes around holomorphic two cycles in
non-compact $K3$ surfaces (Calabi-Yau twofolds), 
Calabi-Yau three, four and fivefolds.
We show ansatz on four-dimensional metric, 
$SO(8)$ gauge fields and scalar fields.
We choose the ansatz consistent 
with twisting procedures on wrapped worldvolumes
and supersymmetric projections on spinor fields.
Then, we derive first order BPS equations for all the configurations
by handling supersymmetry variations of fermionic fields (\ref{d4susy}).

We begin with ansatz on four-dimensional metric 
with supersymmetric $2$-cycles $\Sigma_2$
\begin{eqnarray}
ds_4^2=e^{2f(r)}\left(-d\xi^2+dr^2\right)+e^{2g(r)}d\widetilde{s}^2_{\Sigma_2},
\end{eqnarray}
with two radial functions $f(r), g(r)$ to be determined.
Here, we write a metric
on supersymmetric two cycle by $d\widetilde{s}^2_{\Sigma_2}$.
We normalize this metric to satisfy 
Einstein condition with unit cosmological constants
$\widetilde{R}_{\mu\nu}=\ell \widetilde{g}_{\mu\nu}$ 
($\ell=\pm 1$).
We introduce orthonormal basis of the four-dimensional metric
\begin{eqnarray}
e^0=e^{f(r)}d\xi,
\quad
e^1=e^{f(r)}dr,
\quad
e^2=e^{g(r)}\widetilde{e}^1,
\quad
e^3=e^{g(r)}\widetilde{e}^2,
\end{eqnarray}
where $\widetilde{e}^1, \widetilde{e}^2$ are orthonormal basis of 
a metric $d\widetilde{s}^2_{\Sigma_2}$.
Then, we are able to write down projection conditions on spinor
fields and 
ansatz on gauge fields for each configuration
by following the same consideration as wrapped D3 branes in previous section.
Let us consider holomorphic
two cycles inside Calabi-Yau $n$-folds $(CY_n)$.
In a twisting procedure, we break $SO(8)$ symmetry 
into a subgroup $SO(2n-2)\times SO(10-2n)$.
The subgroup $SO(2n-2)$ represents a rotational symmetry
of directions which are
transverse to M2 branes and tangent to $CY_n$. 
Then, we identify a diagonal subgroup $U(1)$
of $U(n-1)\subset SO(2n-2)$ with a structure group
$U(1)$ of spin connections for holomorphic two cycles.
Following forms on $SO(8)$ gauge field strength
and projections on spinor fields realize the twisting
procedures and keep required amount of supersymmetry 
for configurations under consideration. 
We choose other components of $SO(8)$ gauge field strength to be zero.
\begin{eqnarray}
\underline{K3}
&&\gamma_{23}\epsilon=\Gamma_{12}\epsilon, \qquad \gamma_1\epsilon=\epsilon,\\
&&F_{23}^{12}=\frac{\ell}{g}e^{-2g(r)},\nonumber\\
\underline{CY_3}
&&\gamma_{23}\epsilon=\Gamma_{12}\epsilon
=\Gamma_{34}\epsilon, \qquad \gamma_1\epsilon=\epsilon,\\
&&F_{23}^{12}=F_{23}^{34}=\frac{\ell}{2g}e^{-2g(r)},\nonumber\\
\underline{CY_4}
&&\gamma_{23}\epsilon=\Gamma_{12}\epsilon
=\Gamma_{34}\epsilon=\Gamma_{56}\epsilon, \qquad \gamma_1\epsilon=\epsilon,\\
&&F_{23}^{12}=F_{23}^{34}=F_{23}^{56}=\frac{\ell}{3g}e^{-2g(r)},
\nonumber\\
\underline{CY_5}
&&\gamma_{23}\epsilon=\Gamma_{12}\epsilon
=\Gamma_{34}\epsilon
=\Gamma_{56}\epsilon=\Gamma_{78}\epsilon, \qquad \gamma_1\epsilon=\epsilon,\\
&&F_{23}^{12}=F_{23}^{34}=F_{23}^{56}
=F_{23}^{78}=\frac{\ell}{4g}e^{-2g(r)},\nonumber
\end{eqnarray}
where we introduce $8\times 8$ gamma matrices $\Gamma_I$ 
$(I=1\dots 8)$ satisfying an anti-commutation relation 
$\{\Gamma_{I}, \Gamma_{J}\}=2\delta_{IJ}$.
We have dropped $SU(8)$ indices $i$ for spinor fields here.
Amounts of supersymmetry preserved by
these configurations are $1/4,1/8,1/16$ and $1/16$ for $n=2,3,4$ and $5$.
Note that the projection condition on 
a configuration with $CY_5$ does not
lead further breaking of supersymmetry than $CY_4$.

Let us proceed to ansatz on scalar fields.
We decompose $SO(8)$ symmetry into $SO(p) \times SO(q)$
subgroup where $p+q=8$.
We also decompose $SO(8)$ indices $I,J=1\dots 8$
into $A,B=1\dots p$ for $SO(p)$ 
and $\widehat{A},\widehat{B}=1\dots q$ for $SO(q)$.
We should impose ansatz 
on scalar functions ${\cal V}_{ij}^{IJ}$ 
in order to realize required breaking of $SO(8)$ symmetry.
Here, we turn on only one scalar field $\phi(r)$ 
within coset elements of scalar functions (\ref{coset})
\begin{eqnarray}
&&u_{ij}^{AB}=\frac{1}{4}\cosh{\left[2q\phi(r)\right]}
\left(\Gamma_{AB}\right)_{ij},
\qquad\qquad\quad
v_{ij\; AB}=-\frac{1}{4}\sinh{\left[2q\phi(r)\right]}
\left(\Gamma_{AB}\right)_{ij},
\nonumber\\
&&u_{ij}^{A\widehat{B}}=
\frac{1}{4}
\cosh{\left[-(p-q)\phi(r)\right]}\left(\Gamma_{A\widehat{B}}\right)_{ij},
\qquad
v_{ij\; A\widehat{B}}=
-\frac{1}{4}
\sinh{\left[-(p-q)\phi(r)\right]}\left(\Gamma_{A\widehat{B}}\right)_{ij},
\nonumber\\
&&u_{ij}^{\widehat{A}\widehat{B}}=
\frac{1}{4}\cosh{\left[-2p\phi(r)\right]}
\left(\Gamma_{\widehat{A}\widehat{B}}\right)_{ij},
\qquad\qquad
v_{ij\; \widehat{A}\widehat{B}}=
-\frac{1}{4}\sinh{\left[-2p\phi(r)\right]}
\left(\Gamma_{\widehat{A}\widehat{B}}\right)_{ij}.
\nonumber
\end{eqnarray}
Other components can be read off by noting anti-symmetric indices.
These forms of scalar functions satisfy orthonormalization relation 
(\ref{ortho}).
We choose a value of $p,q$ such that $p+2$ gives a real dimension
$2n$ for $CY_n$. 
Note that we set $\phi(r)=0$ for two cycles inside $CY_5$ with 
$p=8, q=0$.
Let us briefly describe how to determine ansatz. 
We consider a quantity
\begin{eqnarray}
&&u_{ij}^{AB}+v_{ij AB}=\frac{1}{4}e^{-2q\phi(r)}\left(\Gamma_{AB}\right)_{ij}.
\end{eqnarray}
A factor $e^{-2q\phi(r)}$ in the right hand side indicates 
a breaking of $SO(8)$ symmetry into subgroup $SO(p)\times SO(q)$.
Here, we wish to introduce $SO(8)$ valued matrices $S_I^J$ such that 
$S_I^J=\left(e^{-q\phi(r)}\;{\bf 1}_{p\times p}, \; e^{p\phi(r)}\;
{\bf 1}_{q\times q}\right)$ where
${\bf 1}_{n\times n}$ is a $n\times n$ unit matrix.
For this parameterization, 
it would be helpful to notice a manipulation in \cite{deNi3}.
Then, the factor $e^{-2q\phi(r)}$ can be recovered from 
a combination $S_A^BS_C^D\delta_{BD}$.
Similar structure has been observed 
in $D=5$ $SO(6)$ gauged supergravity (\ref{scad3})
for wrapped D3 branes in previous section.

Now, we are able to write down resulting fields based on above ansatz.
We have $SU(8)$ gauge field with a form
${\cal B}_{\mu\; j}^i=
-\frac{1}{2}gA_{\mu}^{IJ}\left(\Gamma_{IJ}\right)_j^i$.
We also obtain a term which appears in kinetic term for scalars. 
We only need following form with a contraction by gamma matrices, 
$\left(\Gamma_{\widehat{1}\widehat{2}}\right)^{ij}
{\cal A}_{1}^{ijk\ell}=
-4\sqrt{2}p\phi'(r)e^{-f(r)}
\left(\Gamma_{\widehat{1}\widehat{2}}\right)^{k\ell}$.
Finally, we have an expression for $A^{ij}$ given by
$A^{ij}=\frac{1}{8}\left(pe^{-2q\phi(r)}+qe^{2p\phi(r)}
\right)\delta^{ij}$.
We omit a lengthy expression for $A_{i}^{jk\ell}$ here.

With these preparations, we are able to derive first order
BPS equations for three radial functions $f(r), g(r)$ and $\phi(r)$ 
from supersymmetry variations of fermions (\ref{d4susy})
\begin{eqnarray}
\label{bpsm2}
f'(r)e^{-f(r)}&=&
-\frac{g}{4\sqrt{2}}\left(pe^{-2q\phi(r)}+qe^{2p\phi(r)}
\right)
+\frac{d\ell}{4\sqrt{2}g}e^{2q\phi(r)-2g(r)},\nonumber\\
g'(r)e^{-f(r)}&=&
-\frac{g}{4\sqrt{2}}\left(pe^{-2q\phi(r)}+qe^{2p\phi(r)}
\right)
+\frac{(d-4)\ell}{4\sqrt{2}g}e^{2q\phi(r)-2g(r)},\\
\phi'(r)e^{-f(r)}&=&-\frac{g}{8\sqrt{2}}
\left(e^{-2q\phi(r)}-e^{2p\phi(r)}\right)
+\frac{d\ell}{8\sqrt{2}pg}e^{2q\phi(r)-2g(r)}.\nonumber
\end{eqnarray}
Here, $d=2$ is a real dimension of supersymmetric cycles and 
$\ell$ is given by $\ell=\pm 1$.
We choose $p,q$ $(p+q=8)$ such that $2d+p$ is a real dimension 
of non-compact $CY_n$ manifold under consideration.
We denote a derivative in terms of $r$ by $'$.
Note that we should set $\phi(r)=0$ for Calabi-Yau 5-fold case $d=2,p=8,q=0$.
BPS equations in \cite{GaKiPaWa1} can be recovered 
by a change of scalar function $\phi(r)$ into $\phi(r)/8$.
Thus, we have correctly reproduced wrapped M2 branes \cite{GaKiPaWa1}
from the $N=8$ $SO(8)$ gauged supergravity.

\subsection{BPS solutions}
\hspace{5mm}
Let us briefly discuss solutions for BPS equations (\ref{bpsm2}).
It is straightforward to repeat the same analysis as 
BPS solutions for wrapped D3 branes. 
The spinor field behaves $\epsilon(r)=e^{\frac{f(r)}{2}}\epsilon_{0}$
where $\epsilon_{0}$ is certain constant spinor.
Asymptotic behavior of four-dimensional metric $ds_4^2$ at small
radius $r\to 0$ is specified as
$ds_4^2=\frac{1}{2g^2r^2}\left(-d\xi+dr^2+d\widetilde{s}_{\Sigma_2}^2\right)$
which has AdS$_4$ like form. 
This determines consistent boundary behaviors of
radial functions $f(r)$ and $g(r)$ with the BPS equations (\ref{bpsm2}) to be 
\begin{eqnarray}
f(r)=g(r)=-\log{\left(\sqrt{2}gr\right)},\qquad r\to 0.
\end{eqnarray}
Then, we should choose asymptotic behavior of scalar field $\phi(r)$ as
\begin{eqnarray}
\phi(r)=Cr+\frac{d\ell}{8p}r^2, \qquad r\to 0,
\end{eqnarray} 
where $C$ is arbitrary constant and becomes a single moduli in 
resulting one-parameter eleven-dimensional solutions.

In general, numerical evaluations of three radial functions 
show an appearance of singularities in BPS solutions.
We are able to specify good/bad singularities \cite{MaNu1}
by noting eleven-dimensional metrics
\begin{eqnarray}
ds_{11}^2=\Delta^{\frac{2}{3}}ds_{6}^2+\dots,
\qquad \Delta=e^{-2q\phi(r)}\mu^A\mu^A+e^{2p\phi(r)}\mu^{\widehat{A}}
\mu^{\widehat{A}},
\end{eqnarray}
where $\mu^I$ are coordinates on seven-sphere
$\mu^A\mu^A+\mu^{\widehat{A}}\mu^{\widehat{A}}=1$.
We see that this analysis reproduces a feature of behaviors 
analyzed in \cite{GaKiPaWa1}.
Here, let us concentrate on solutions which include AdS$_2$ space-times.
We find two such solutions by assuming that radial functions
$g(r)$ and $\phi(r)$ are constant in BPS equations (\ref{bpsm2}). 
The first solution arises
in M2 branes wrapped around holomorphic two cycle inside 
non-compact $CY_4$ $(d=2, p=6, q=2)$.
We have following values when we obtain the AdS$_2$ solution  
\begin{eqnarray}
\label{m2ads4}
e^{f(r)}=\frac{1}{3\sqrt{6}g}e^{4\phi}\frac{1}{r},\qquad
e^{g(r)}=\frac{1}{\sqrt{6}g}e^{4\phi},\qquad
e^{16\phi(r)}=3,\qquad \ell=-1,
\end{eqnarray}
Supersymmetric two cycle has negative curvature in this solution.
The second case arises in
M2 branes wrapped around holomorphic two cycle inside 
$CY_5$ $(d=2, p=8, q=0)$.
In this case, we turn on no scalar fields, $\phi(r)=0$.
The solution is specified by 
\begin{eqnarray}
\label{m2ads5}
e^{f(r)}=\frac{1}{2\sqrt{2}g}\frac{1}{r},\qquad\;\quad
e^{g(r)}=\frac{1}{2g},\quad\qquad\;\;\;
e^{16\phi(r)}=1,\quad\quad\;\; \ell=-1.
\end{eqnarray}
Again, supersymmetric two cycle has negative curvature.
We choose these holomorphic two cycles as genus $g\geq 2$ 
Riemann surfaces $\Sigma_g$. 
These AdS$_2$ solutions are expected to give dual supergravity backgrounds 
to $D=1$ superconformal quantum mechanics on the wrapped M2 branes.
Their central charges can be estimated from the AdS$_2$ solutions
\begin{eqnarray}
c\propto \frac{1}{G_N^2}=\frac{{\rm Volume}\left(S^7\right)
\cdot {\rm Volume}\left(\Sigma_g\right)}{G_N^{11}}=
\frac{4\sqrt{2}}{3}g^2N^{\frac{3}{2}}\cdot{\rm Vol}(\Sigma_g),
\end{eqnarray}
where $G_N^2$ and $G_N^{11}$ 
are two- and eleven-dimensional Newton's constants.
Eleven-dimensional Newton's constant is
defined by $G_N^{11}=16\pi^7\ell_p^9$ where $\ell_p$ is Planck length in 
eleven dimensions. 
The radius of squashed $S^7$ is given by 
$\sqrt{2}/g=\ell_p\left(2^5\pi^2N\right)^{\frac{1}{6}}$
where $N$ is the number of M2 branes.
Note that in near horizon limit of flat M2 branes,
the radius of $S^7$ is twice as that of AdS$_4$ \cite{Maldacena}. 
A volume of unit seven-sphere is known as $\pi^4/3$.
A volume of genus $g\geq 2$ Riemann surfaces can be given by 
\begin{eqnarray}
{\rm Vol}(\Sigma_g)=4\pi\left(g-1\right)e^{2g(r)},
\end{eqnarray}
where $e^{2g(r)}$ are shown in (\ref{m2ads4}) and (\ref{m2ads5})
for each solution \cite{MaNu1}. 
The radius of AdS$_2$ space-times does not enter the expression
of central charge.
Normalization of central charge should be determined in an appropriate way.
In principle, the central charge should be recovered
from an analysis of one-dimensional superconformal quantum mechanics
with two supercharges on the wrapped M2 branes.
Note that we would be able to specify worldvolume theories on wrapped branes
when we take a limit to obtain 
wrapped D2 branes whose metrics contain no AdS space-times.

We close this section with a comment on another method
to analyze BPS equations \cite{GaKiWa}.
We introduce new radial variables $h(r), x(r)$ and $F(r)$ from
$f(r), g(r)$ and $\phi(r)$
\begin{eqnarray}
h(r)=e^{f(r)-2q\phi(r)},\qquad
x(r)=e^{2g(r)-4q\phi(r)},\qquad F(r)=x(r)^{\frac{2}{q}}e^{16\phi(r)}.
\end{eqnarray}
Then, we are able to obtain ordinary differential equations
for these functions $F, x$ for each case specified by $d=2$ and $p, q$
with non-trivial radial function $\phi(r)$ for a scalar field
\begin{eqnarray}
\underline{p=2,q=6}\qquad
\frac{dF}{dx}=\frac{1}{3}\frac{F}{x}\frac{2g^2x-\ell}
{g^2\left(3x^{\frac{2}{3}}F-x\right)+2\ell},
&&
\underline{p=4,q=4}\qquad
\frac{dF}{dx}=
\frac{g^2F}{2g^2\sqrt{x}F+\ell},\nonumber\\
\underline{p=6,q=2}~~~~~~\qquad
\frac{dF}{dx}=\frac{F}{x}
\frac{g^2x+\ell}{3g^2\left(F+x\right)+2\ell}.&&
\end{eqnarray}
Then, it is straightforward to show numerical behaviors of 
eleven-dimensional metrics with radial evolutions for $h(r)$.

\section{Wrapped NS5 branes}
\hspace{5mm}
In this section, we study supersymmetric wrapped type IIB NS5 branes
around various supersymmetric cycles.
Here, we wish to understand these BPS configurations
from BPS solutions in $D=7$ $N=4$ $SO(4)$ gauged supergravity \cite{SaSe}
and their embedding ansatz into solutions in
type IIB supergravity \cite{CvLuPo1}.
We give an explanation for this by a method based on 
a domain-wall like reduction \cite{CvLiLuPo} from
$D=7$ $N=4$ $SO(5)$ gauged supergravity
to $D=7$ $N=4$ $SO(4)$ gauged supergravity.
Then, we obtain BPS equations for all possible wrapped NS5 branes 
from those for wrapped M5 branes \cite{AcGaKi, GaKiWa, GaKi}.
New examples are wrapped NS5 branes wrapped 
around various supersymmetric four cycles inside non-compact
Calabi-Yau three $(CY_3)$, fourfold $(CY_4)$, 
$G_2$ holonomy manifolds, $Spin(7)$
holonomy manifolds and HyperK\"ahler manifolds $(HK_2)$.
Then, we concentrate on
ten-dimensional BPS solutions for wrapped type IIB NS5 branes around
holomorphic $CP^2$ cycle in non-compact $CY_3$.
Their behavior turns out to be quite similar to
that of wrapped NS5 branes around holomorphic 
$CP^1$ cycle in non-compact $K3$ surfaces. 
Hori and Kapustin \cite{HoKa} have introduced related 
solutions for wrapped NS5 branes 
in a string world-sheet perspective.
In a supergravity limit, our solution is mapped into their solution
by certain change of constant parameters.
A calculation here gives a check of preserved supersymmetry 
for these configurations in terms of seven-dimensional gauged supergravity.
Finally, let us note that we use mostly plus signature 
of metric in this section.

\subsection{BPS equations}
\hspace{5mm}
Near horizon space-time
of flat NS5 branes is known to be a linear dilaton background
$R^{1,5}\times R\times S^3$.
$SO(4)$ isometry of this three sphere $S^3$ corresponds to 
R symmetry of little string theories living on the NS5 branes.  
We expect that this space-time is realized as a solution 
in $D=7$ $N=4$ $SO(4)$ gauged supergravity \cite{SaSe}.
We assume that this gauged supergravity is a consistent reduction of
type IIB supergravity.
This reduction has been explicitly shown in a bosonic sector 
of both supergravities \cite{CvLuPo1}.
We are also able to consider supersymmetric wrapped 
type IIB NS5 branes around various two-, three- and four-dimensional 
supersymmetric cycles in the $SO(4)$ gauged supergravity.
Let us outline a construction for them.
We start with a field content of the $SO(4)$ gauged supergravity.
Bosonic fields consist of seven-dimensional metric, $SO(4)$ gauge fields, 
$SL(4,{\bf R})/SO(4)_c$ coset scalar fields and three-form fields.
We denote indices of gauge group $SO(4)$ by $A, B=1 \dots 4$,
and indices of local gauge $SO(4)_c$ by $a, b=1 \dots 4$.
We need to specify ansatz on these fields for each configuration
under consideration.
First, we adopt ansatz on seven-dimensional metric 
for wrapped NS5 branes around supersymmetric $d$-cycle $\Sigma_d$
\begin{eqnarray}
\label{ns5me}
ds_7^2=e^{2f(r)}\left(d\xi_i^2+dr^2\right)+
e^{2g(r)}d\widetilde{s}_{\Sigma_d}^2,
\end{eqnarray}
where $d\xi_i^2$ is a metric on Minkowski space $R^{1,5-d}$ and 
$d\widetilde{s}_{\Sigma_d}^2$ is a metric on supersymmetric $d$-cycle.
Secondly, $SO(4)$ gauge fields are specified in a  
consistent way with required projections on spinor fields \cite{GaLaWe}.
They can be read off from ansatz on $SO(5)$ gauge fields
for wrapped M5 branes wrapped
around the same supersymmetric cycles \cite{AcGaKi, GaKiWa, GaKi}.
Third, we turn on one scalar field $\phi(r)$
so that scalar function $T_{AB}=\delta^{ab}
\left(V^{-1}\right)_a^A\left(V^{-1}\right)_b^B$ defined by
$SL(4,{\bf R})/SO(4)_c$ coset elements $V_A^a$
has a following form 
\begin{eqnarray}
T_{AB}=e^{f(r)}\left(e^{-2q\phi(r)}\;
{\bf 1}_{p\times p},\; e^{2p\phi(r)}\; {\bf 1}_{q\times q}\right),
\end{eqnarray}
where ${\bf 1}_{n\times n}$ is a $n\times n$ unit matrix. 
This ansatz breaks $SO(4)$ symmetry into $SO(p)\times SO(q)$ with
$p+q=4$
where $d+p$ coincides with a real dimension of underlying 
non-compact special holonomy manifolds.
A factor $e^{f(r)}$ in the right hand side 
is a matter of convention to match previous results eventually. 
Finally, for cases with four cycles in eight-manifolds, 
we should give non-zero values to certain components of three-form fields.
These ansatz are determined by a consistency of equation of motions
for three-form fields in the gauged supergravity.
We omit details of ansatz on three-form fields here
because we provide another method in the following.
Then, from supersymmetry transformations of fermions in the gauged 
supergravity,
we are able to obtain BPS equations for every wrapped NS5 brane.
We will also obtain ten-dimensional type IIB solutions by an embedding 
procedure in \cite{CvLuPo1}
(If we wish to obtain IIA NS5 branes, we 
use an embedding ansatz in \cite{CvLuPoSaTr2} from the $SO(4)$
gauged supergravity into type IIA supergravity).
In this section, we wish to present a simple procedure
in order to derive BPS equations for wrapped NS5 branes 
as a suitable limit of those for wrapped M5 branes.
We are able to check that this procedure gives the same BPS equations
as those from a direct procedure mentioned above.
A present derivation
gives some understanding on a connection between 
BPS equations of wrapped M5 and NS5 branes.
Let us recall that various supersymmetric
wrapped branes always have similar BPS equations 
as we have seen for wrapped D3, M2 and M5 branes.
This situation extends to dilatonic wrapped NS5 branes
due to a similar structure of supersymmetry variations of fermions 
in various gauged supergravities.

Wrapped M5 branes can be constructed from BPS solutions in
$D=7$ $N=4$ $SO(5)$ gauged supergravity \cite{GaKiWa}.
Let us introduce relevant information about this gauged supergravity 
\cite{PePiva}.
This gauged supergravity has $SO(5)$ gauge group and $SO(5)_c$
local composite gauge group.
We denote indices of $SO(5)$ group by $I,J=1 \dots 5$,
and indices of $SO(5)_c$ group by $i,j=1 \dots 5$. 
Bosonic fields consist of seven-dimensional metric, $SO(5)$ gauge potential,
scalar fields and three-form tensor fields.
14 scalar fields $V_I^i$ are defined in a coset space $SL(5,{\bf R})/SO(5)_c$.
These coset elements transform as $5$ representation
under both $SO(5)$ and $SO(5)_c$ groups.
We define a term which gives scalar kinetic term $P_{\mu\; ij}$,
and $SO(5)_c$ composite gauge potential $Q_{\mu\; ij}$ by symmetric and 
anti-symmetric parts of following contribution
\begin{equation}
\left(V^{-1}\right)_i^I\left(\delta_I^J\partial_{\mu}+gA_{\mu\;I}^{\; J}\right)
V_J^k\delta_{kj}=
P_{\mu\; (ij)}+Q_{\mu\; [ij]},
\end{equation}
where $A_{\mu \; J}^{\; I}$ are $SO(5)$ 
gauge potentials whose field strength are given by
$F^{\;\;\; I}_{\mu\nu\; J}$.
We denote $SO(5)$ gauge coupling constant by $m$.
We also need scalar functions $T_{ij}$ and $T$ 
which are defined by coset elements $V_I^i$ as
\begin{equation}
T_{ij}=\left(V^{-1}\right)_i^I
\left(V^{-1}\right)_j^J
\delta_{IJ},\qquad
T=\delta^{ij}T_{ij}.
\end{equation}                   
We omit a contribution of three-form fields here for simplicity.
This contribution becomes relevant only for a supersymmetric
four cycles in non-compact eight-dimensional special holonomy manifolds
\cite{GaKiWa, GaKi}.
Finally, we write down supersymmetry transformations of gravitinos $\psi_{\mu}$
and gauginos $\lambda_i$ by
\begin{eqnarray}
\label{susyd7}
\delta\psi_{\mu}&=&                                             
\Biggl[
\partial_{\mu}+\frac{1}{4}\omega_{\mu}^{\nu\rho}\gamma_{\nu\rho}
+\frac{1}{4}Q_{\mu\; ij}\Gamma^{ij}
+\frac{1}{20}mT\gamma_{\mu}
-\frac{1}{40}\left(\gamma_{\mu}^{\;\nu\rho}
-8\delta_{\mu}^{\nu}\gamma^{\rho}\right)
\Gamma^{ij}V_I^iV_J^jF_{\nu\rho}^{IJ}
\Biggr]\epsilon,\\
\delta\lambda_i&=&
\Biggl[
\frac{1}{2}\gamma^{\mu}P_{\mu\; ij}\Gamma^j
+\frac{1}{2}m\left(T_{ij}-\frac{1}{5}T\delta_{ij}\right)\Gamma^j
+\frac{1}{16}\gamma^{\mu\nu}
\left(\Gamma_{k\ell}\Gamma_i-\frac{1}{5}\Gamma_i\Gamma_{k\ell}\right)
V_K^kV_L^{\ell}F_{\mu\nu}^{KL}
\Biggr]\epsilon,\nonumber
\end{eqnarray}      
where we use $D=7$ gamma matrices $\gamma^{\mu}$
satisfying anti-commuting relation $\{\gamma^{\mu},\gamma^{\nu}\}=2g^{\mu\nu}$
with seven-dimensional metric tensor $g_{\mu\nu}$,
and $SO(5)_c$ gamma matrices $\Gamma^i$ satisfying
anti-commutation relation $\{\Gamma^i,\Gamma^j\}=2\delta^{ij}$.

First order BPS equations for wrapped M5 branes are derived 
from supersymmetry variations of fermions (\ref{susyd7})
\cite{AcGaKi, GaKiWa, GaKi}.
We use ansatz on seven-dimensional
metric with a supersymmetric curved $d$-cycle $\Sigma_d$
\begin{eqnarray}
ds^2_7=e^{2f(r)}\left(d\xi_{6-d}^2+dr^2\right)+e^{2g(r)}
d\widetilde{s}_{\Sigma_d}^2,
\end{eqnarray}
where $d\xi_{6-d}^2$ is a metric on Minkowski space-time $R^{1, 6-d}$, and
$d\widetilde{s}_{\Sigma_d}^2$ is a metric on a $d$-cycle
with normalized cosmological constant $\ell=\pm 1$.
Ansatz on scalar fields breaks $SO(5)$ symmetry 
into $SO(p)\times SO(q)$ with $p+q=5$
\begin{eqnarray}
V_I^i=\left(e^{q\phi(r)}\;
{\bf 1}_{p\times p}, \; e^{-p\phi(r)}\;{\bf 1}_{q\times q}\right),
\end{eqnarray}
with unit $n\times n$ matrices ${\bf 1}_{n\times n}$.
Here, $d+p$ is a real dimension of non-compact special holonomy
manifolds under consideration.
Present discussion includes configurations with
$(d,p)=(2,2), (2,4), (3,6), (3,7), (4,6)$ and $(4,7)$.
Ansatz on gauge fields can be specified as in \cite{AcGaKi, GaKiWa, GaKi}.
Resulting first order BPS equations for three radial functions 
$f(r), g(r)$ and $\phi(r)$ are written down as
\begin{eqnarray}
f'(r)e^{-f(r)}&=&-\frac{m}{10}\left(
pe^{-2q\phi(r)}+qe^{2p\phi(r)}
\right)+\frac{d\ell}{20m}e^{2q\phi(r)-2g(r)},\nonumber\\
g'(r)e^{-f(r)}&=&-\frac{m}{10}\left(
pe^{-2q\phi(r)}+qe^{2p\phi(r)}
\right)+\frac{(d-10)\ell}{20m}e^{2q\phi(r)-2g(r)},\\
\phi'(r)e^{-f(r)}&=&-\frac{m}{5}
\left(e^{-2q\phi(r)}-e^{2p\phi(r)}
\right)+\frac{d\ell}{10pm}e^{2q\phi(r)-2g(r)}.\nonumber
\end{eqnarray}

We are able to uplift solutions obtained from above BPS equations
in the seven-dimensional $SO(5)$ gauged supergravity 
into supersymmetric solutions of wrapped M5 branes 
in eleven-dimensional supergravity 
via an embedding formula \cite{NaVava1, NaVava2}.
In general, resulting eleven-dimensional metric $ds_{11}^2$ is given by
\begin{eqnarray}
\label{m5met}
ds_{11}^2&=&\Delta^{-\frac{2}{5}}ds_7^2
+\frac{1}{m^2}\Delta^{\frac{4}{5}}\left(
e^{2q\phi} DY^{\alpha_p}DY^{\alpha_p}+e^{-2p\phi}dY^{\alpha_q}dY^{\alpha_q}
\right),\nonumber\\
&=&\Delta^{-\frac{2}{5}}\left(ds_7^2+
\frac{1}{m^2}\Delta^{\frac{6}{5}}T^{-1}_{IJ}DY^IDY^J\right),
\end{eqnarray}
where a covariant derivative $D$ and a function $\Delta$ are defined by
\begin{eqnarray}
&&DY^I=dY^I+2mA^{I}_JY^J, \qquad Y^IY^I=1,\\
&&\Delta^{-\frac{6}{5}}=e^{-2q\phi}Y^{\alpha_p}Y^{\alpha_p}
+e^{2p\phi}Y^{\alpha_q}Y^{\alpha_q}.\nonumber
\end{eqnarray}
Here, we denote coordinates parameterizing four sphere $S^4$ by $Y^I$.
We decompose indices of $SO(5)$ $I=1\dots 5$ into 
$\alpha_p=1\dots p$ of $SO(p)$ and $\alpha_q=p+1\dots 5$ of $SO(q)$. 
The inverse of a scalar function 
$T_{IJ}=\left(V^{-1}\right)_i^I
\left(V^{-1}\right)_j^J\delta^{ij}$ in ten-dimensional metric is given by 
\begin{eqnarray}
T_{IJ}^{-1}=\left(e^{2q\phi(r)}\; {\bf 1}_{p\times p}, 
\; e^{-2p\phi(r)}\; {\bf 1}_{q\times q}\right).
\end{eqnarray}

Now, we wish to explain how to obtain BPS equations for wrapped NS5 branes 
as a certain limit of those for wrapped M5 branes.
In this limit, $SO(5)$ symmetry group should be broken into $SO(4)$ symmetry.
In other words, this limit is realized by deforming 
$S^4$ into $S^3\times R$ in a space-time metric.
In the following, we provide this limit by handling
eleven-dimensional metrics for wrapped M5 branes.
This manipulation is based on a technique in \cite{CvLiLuPo}.
Then, we obtain a map between parameters and bosonic fields
in the $SO(5)$ and $SO(4)$ gauged supergravities.
This map gives a simple derivation of
BPS equations for wrapped NS5 branes by those for wrapped M5 branes.

Let us begin with a scalar function $T_{IJ}$ in the $SO(5)$ 
gauged supergravity.
We introduce a radial scalar function $\psi(r)$ in order to extract 
a $SO(4)$ scalar function $T_{AB}$ $\left(A,B=1\dots 4\right)$ 
in the $SO(4)$ gauged supergravity
\begin{eqnarray}
T^{-1}_{IJ}=
\left(e^{2\psi(r)}\;T_{AB}^{-1},\; e^{-8\psi(r)}\right).
\end{eqnarray}
We relate this parameterization with
a deformation of four sphere $S^4$ into $S^3\times R$ in a space-time metric.
We define squashed four sphere $S^4$ with real coordinates $\mu^I$
$(I=1\dots 5)$ 
\begin{eqnarray}
T_{IJ}\mu^I\mu^J={\rm constant}.
\end{eqnarray}
Here, we consider a limit 
such that $\mu^5\to\infty$ and $T_{55}^{-1}\to\infty$.
Then, scalar function $\psi(r)$ diverges $\psi(r) \to -\infty$.
We divide this function $\psi(r)$ into
a finite scalar function $\rho(r)$ and an infinite contribution $\psi_0$ 
\begin{eqnarray}
\psi(r)=\rho(r)+\psi_0, \qquad \psi_0\to -\infty.
\end{eqnarray}
Then, we proceed to consider the limit 
in a eleven-dimensional metric (\ref{m5met}).
We choose an non-compact direction to be $Y^5$ direction
required for the deformation into $S^3\times R$.
First, we rewrite an element in eleven-dimensional metric (\ref{m5met}) as 
\begin{eqnarray}
\frac{1}{m^2}T^{-1}_{IJ}DY^IDY^J&=&
\frac{1}{m^2}
\left(
e^{2\left(\rho+\psi_0\right)}
T_{AB}^{-1}DY^ADY^B
+
e^{-8\left(\rho+\psi_0\right)}dY^5dY^5
\right),\nonumber\\
&=&
\frac{1}{m^2}e^{2\psi_0}
\left(e^{2\rho} T_{AB}^{-1}DY^ADY^B+e^{-8\rho}dY^{5'}dY^{5'}\right),
\end{eqnarray}
where we introduce another scaled coordinate $Y^{5'}$ from $Y^5$ 
\begin{eqnarray}
Y^{5'}=e^{-5\psi_0}Y^5.
\end{eqnarray}
Four sphere $S^4$ defined by $Y^AY^A+e^{10\psi_0}Y^{5'}Y^{5'}=1$
correctly reduces three sphere $S^3$ defined 
by $Y^AY^A=1$ in the limit $\psi_0 \to -\infty$.
Secondly, we consider how 
another element $\Delta$ in eleven-dimensional metric
behaves in the limit $\psi_0\to-\infty$
\begin{eqnarray}
\Delta^{-\frac{6}{5}}&=&T_{IJ}Y^IY^J=
e^{-2\left(\rho+\psi_0\right)}T_{AB}Y^AY^B
+
e^{8\left(\rho+\psi_0\right)}e^{10\psi_0}
Y^{5'}Y^{5'}\nonumber\\
&\to&e^{-2\left(\rho+\psi_0\right)}T_{AB}Y^AY^B
=e^{-2\left(\rho+\psi_0\right)}\Delta'.
\end{eqnarray}
Here, we have introduced a quantity $\Delta'=T_{AB}Y^AY^B$.
This quantity should appear when we embed solutions
in the $SO(4)$ gauged supergravity into those in ten-dimensional 
supergravities.
Thus, we obtain following
the eleven-dimensional metric (\ref{m5met}) 
in the limit $\psi_0\to-\infty$ which decompactifies one direction
\begin{eqnarray}
ds_{11}^2=e^{-\frac{2}{3}\left(\rho+\psi_0\right)}
\Delta^{'\frac{1}{3}}\left[
ds_7^2+\frac{1}{m^2}e^{2\left(\rho+2\psi_0\right)}\Delta^{'-1}
\left(e^{2\rho} T_{AB}^{-1}DY^ADY^B
+e^{-8\rho} dY^{5'}dY^{5'}\right)\right].
\end{eqnarray}
Remarkably, we are able to factor out terms with $\psi_0$ in this metric
as follows
\begin{eqnarray}
\label{d10me}
e^{\frac{2}{3}\psi_0}ds_{11}^2=
e^{-\frac{2}{3}\rho}\Delta^{'\frac{1}{3}}
\left(ds_7^2+\frac{1}{m^{'2}}
e^{4\rho}\Delta^{'-1}T_{AB}^{-1}DY^ADY^B\right)+
e^{-\frac{20}{3}\rho}\Delta^{'-\frac{2}{3}}
\frac{1}{m^{'2}}dY^{5'}dY^{5'},
\end{eqnarray}
where we use a parameter $m'$ defined by $m'=me^{-2\psi_0}$.
It is known that equations of motion in 
eleven-dimensional supergravity
have a scaling symmetry under which
eleven-dimensional metric tensor $g_{\mu\nu}$
and three-form fields $A_{\mu\nu\rho}$ 
are multiplied by a constant factor 
\begin{eqnarray}
g_{\mu\nu}\to k^2g_{\mu\nu},\quad
A_{\mu\nu\rho}\to k^3A_{\mu\nu\rho},
\end{eqnarray}
where $k$ is an arbitrary constant \cite{CrLuPoSt}.
This means that we are able to ignore 
a contribution of decompactification $\psi_0\to-\infty$ 
in eleven-dimensional metric consistently.
Then, we are able to extract ten-dimensional metrics for wrapped NS5 branes.
Here, we do not include an analysis on
eleven-dimensional four-form field strength 
in the decompactification limit
although we would be able to do it in a straightforward manner.

Now, we are able to
write down a map between scalar, gauge fields and parameters in
the $SO(5)$ and $SO(4)$ gauged supergravity in the following way
\begin{eqnarray}
\label{so5so4map}
m&=&e^{2\psi_0}m',\nonumber\\
V_I^i&=&e^{\rho+\psi_0}V_A^a,\\
F_{\mu\nu}^{ij}&=&e^{2\rho}F_{\mu\nu}^{ab},
\nonumber\\
\left(V^{-1}\right)_i^I\nabla_{\mu}V_{I}^j
&=&
\left(
\left(V^{-1}\right)_a^A\nabla_{\mu}V_{A}^b
+\partial_{\mu}\rho {\bf 1}_{4\times 4},\;
-4\partial_{\mu}\rho\right),\nonumber
\end{eqnarray}
where the second and third relations should be read with
a restriction of indices.
Note that we use indices 
$I,J=1\dots 5$ for $SO(5)$, $i,j=1\dots 5$ for $SO(5)_c$,
$A,B=1 \dots 4$ for $SO(4)$, and $a,b=1\dots 4$ for $SO(4)_c$.
We also have a scaling relation  
$A_{\mu I}^{\; J}=e^{-2\psi_0} A_{\mu A}^{\; B}$
among gauge potentials in both the gauged supergravities.

In above construction, we 
are able to relate a scalar field $\rho(r)$ 
with the ten-dimensional type IIA dilaton field $\Phi$ as follows 
\begin{eqnarray}
e^{2\Phi}=e^{-10\rho}\Delta^{'-1},
\end{eqnarray}
by using a standard reduction formula
$ds_{11}^2=e^{-\frac{2}{3}\Phi}ds_{10(st)}^2+
e^{\frac{4}{3}\Phi}dx_{11}^2=
e^{-\frac{1}{6}\Phi}ds_{10(EIN)}^2+e^{\frac{4}{3}\Phi}dx_{11}^2$
with ten-dimensional metrics $ds_{10(st)}^2$ and $ds_{10(EIN)}^2$
in string and Einstein frames
and a coordinate in the eleventh direction $x_{11}$.
This will be relevant when we consider wrapped 
type IIA NS5 brane solutions via an embedding formula 
from the $SO(4)$ gauged supergravity 
into type IIA supergravity \cite{CvLuPoSaTr2}.

With these ingredients, we are able to find
supersymmetry variations of fermions for $SO(4)$ gauged supergravity
and BPS equations of wrapped NS5 branes. 
Here, we consider ansatz on bosonic
fields in $SO(4)$ gauged supergravity 
which are determined in the way as wrapped M5 branes. 
BPS equations for wrapped NS5 branes are 
given by a slightly modified form of those for wrapped M5 branes.
The BPS equations for radial functions $f(r), g(r)$ in a metric and
$\phi(r)$ in a scalar function are given by 
\begin{eqnarray}
\label{ns5bps1}
f'(r)e^{-f(r)}&=&-\frac{m'}{10}
\left(pe^{-2q\phi(r)}+qe^{2p\phi(r)}\right)e^{-2\rho(r)}
+\frac{d\ell}{20m'}e^{2\rho(r)
-2g(r)+2q\phi(r)},\nonumber\\
g'(r)e^{-f(r)}&=&-\frac{m'}{10}
\left(pe^{-2q\phi(r)}+qe^{2p\phi(r)}\right)e^{-2\rho(r)}
+\frac{\left(d-10\right)\ell}{20m'}
e^{2\rho(r)-2g+2q\phi(r)},\\
\phi'(r)e^{-f(r)}&=&-\frac{m'}{4}
\left(e^{-2q\phi(r)}-e^{2p\phi(r)}\right)e^{-2\rho(r)}
+\frac{d\ell}{8pm'}
e^{2\rho(r)-2g(r)+2q\phi(r)},\nonumber
\end{eqnarray}
where $p+q=4$.
We set $\phi(r)=0$ when we consider configurations for $p=4, q=0$.
Here, $p, q$ are determined such that $d+p$ is equal to a real dimension of 
non-compact special holonomy manifolds including supersymmetric cycles
under consideration.
We have a factor $e^{2\rho(r)}$ in above BPS equations
by a map (\ref{so5so4map}). 
We also have a rescaling factor $4/5$ in the last differential
equations for $\phi(r)$ when 
we make a reduction for supersymmetry variations of fermions
in the $SO(5)$ gauged supergravity into the $SO(4)$ supergravity. 

Let us rewrite the above BPS equations (\ref{ns5bps1})
in a convenient form.
First, we define a gauge coupling constant $g$
in $SO(4)$ gauged supergravity by 
\begin{eqnarray}
g=m'. 
\end{eqnarray}
Secondly, we choose a function $\rho(r)$ to be
\begin{eqnarray}
\rho(r)=\frac{f(r)}{2},
\end{eqnarray}
in order to obtain a unit warp factor
in unwrapped NS5 branes' worldvolume directions of
ten-dimensional string frame metric for wrapped NS5 branes.
This normalization is understood by combining (\ref{ns5me}) and (\ref{d10me}). 
Then, we have following first order BPS equations
\begin{eqnarray}
f'(r)&=&-\frac{g}{10}\left(pe^{-2q\phi(r)}+qe^{2p\phi(r)}\right)
+\frac{d\ell}{20g}e^{2f(r)-2g(r)+2q\phi(r)}\nonumber,\\
g'(r)&=&-\frac{g}{10}\left(pe^{-2q\phi(r)}+qe^{2p\phi(r)}\right)
+\frac{\left(d-10\right)\ell}{20g}e^{2f(r)-2g(r)+2q\phi(r)},\\
\phi'(r)&=&-\frac{g}{4}\left(e^{-2q\phi(r)}-e^{2p\phi(r)}\right)
+\frac{d\ell}{8pg}e^{2f(r)-2g(r)+2q\phi(r)}.\nonumber
\end{eqnarray}
Now, we write down this BPS equations by introducing a superpotential $W$ 
\begin{eqnarray}
W=g\left(pe^{x(r)}+qe^{-\frac{p}{q}x(r)}\right)
+\frac{d\ell}{2g}e^{-2h(r)-x(r)},
\end{eqnarray}
where new radial functions $A(r),h(r)$ and $x(r)$ are defined by
\begin{eqnarray}
\label{change}
A(r)=\frac{5-d}{2}f(r)+\frac{d}{2}g(r),
\quad\qquad
h(r)=g(r)-f(r),\quad\qquad
x(r)=-2q\phi(r).
\end{eqnarray}
BPS equations are given by following differential equations
\begin{eqnarray}
\label{ns5bps}
A'(r)=-\frac{1}{4}W,\quad\qquad
h'(r)=\frac{1}{2d}\frac{\partial W}{\partial h},\quad\qquad
x'(r)=\frac{q}{2p}\frac{\partial W}{\partial x}.
\end{eqnarray}
Here, let us specify possible configurations of wrapped NS5 branes
specified by values of $d$ and $p,q$.
\begin{eqnarray}
\begin{array}{|c|c|c|c|}\hline
d&p&q&\\\hline
2&2&2& 2 \subset K3\\\hline
4&2&2&4 \subset CY_3\\\hline
\end{array}
\qquad
\begin{array}{|c|c|c|c|}\hline
d&p&q&\\\hline
3&3&1& 3 \subset CY_3\\\hline
4&3&1& 4 \subset G_2\\\hline
\end{array}
\qquad
\begin{array}{|c|c|c|c|}\hline
d&p&q&\\\hline
2&4&0& 2 \subset CY_3\\\hline
3&4&0& 3 \subset G_2\\\hline
\end{array}
\end{eqnarray}
Here, for example, we denote 
wrapped NS5 branes around supersymmetric 2 cycles in non-compact $K3$
surfaces by $2\subset K3$.
It turns out that behaviors of resulting ten-dimensional wrapped 
NS5 brane solutions show a similar feature 
for cases with the same $p,q$. 
We will show this situation by constructing explicit ten-dimensional solutions 
for $p=q=2$ cases in next subsection.
The other two case are also checked straightforwardly 
by using numerical evaluations of radial functions.

Finally, we mention NS5 branes wrapped around
supersymmetric four cycles in eight-dimensional special holonomy manifolds.
We begin with wrapped M5 branes around these cycles.
In these cases, we should turn on three-form fields in the $SO(5)$
gauged supergravity \cite{GaKiWa, GaKi}.
BPS equations for these wrapped M5 branes are given by
\begin{eqnarray}
f'(r)e^{-f(r)}&=&-\frac{m}{10}\left(
4e^{-2\phi(r)}+e^{8\phi(r)}\right)
+\frac{\ell}{5m}e^{2\phi(r)-2g(r)}
-\frac{a}{20m^3}e^{-4\phi(r)-4g(r)},\nonumber\\
g'(r)e^{-f(r)}&=&-\frac{m}{10}\left(
4e^{-2\phi(r)}+e^{8\phi(r)}\right)
-\frac{3\ell}{10m}e^{2\phi(r)-2g(r)}
-\frac{a}{30m^3}e^{-4\phi(r)-4g(r)},\\
\phi'(r)e^{-f(r)}&=&
-\frac{m}{5}\left(e^{-2\phi(r)}-e^{8\phi(r)}\right)
+\frac{\ell}{10m}e^{2\phi(r)-2g(r)}+
\frac{a}{60m^3}e^{-4\phi(r)-4\phi(r)},\nonumber
\end{eqnarray}
where we choose a constant $a=1/2, 3/8,
1/4$ and $1/8$ for complex Lagrangian four cycles in 
HyperK\"ahler manifolds $(HK_2)$, K\"ahler four cycles
in $CY_4$, special Lagrangian
four cycles in $CY_4$,
and Cayley four cycles in $Spin(7)$ holonomy manifolds. 
It is straightforward to repeat the same reduction procedure as
a precious prescription without three-form fields.
Seven-dimensional ansatz for the wrapped NS5 branes has no
contribution from a scalar field.
Here, we have two radial functions in a metric (\ref{ns5me})
arranged as 
\begin{eqnarray}
A(r)=\frac{1}{2}f(r)+2g(r),\qquad h(r)=g(r)-f(r).
\end{eqnarray}
Let us write down first order BPS equations  
\begin{eqnarray}
\label{ns5bps4}
A'(r)=-\frac{1}{4}W,\qquad h'(r)=\frac{1}{8}\frac{\partial W}{\partial h},
\end{eqnarray}
with a superpotential defined by
\begin{eqnarray}
W=4g+\frac{2\ell}{g}e^{-2h(r)}-\frac{4a}{3g^3}e^{-4h(r)}.
\end{eqnarray}
BPS solutions in ten-dimensions from these BPS equations
show singular behaviors as in a case $d=2, p=4, q=0$ 
such as $2\subset CY_3$ \cite{MaNu2} without 
three-form fields in the $SO(4)$ gauged supergravity.

\subsection{NS5 branes wrapped around 
holomorphic $CP^2$ in $CY_3$}
\hspace{5mm}
Now, we are able to construct type IIB solutions for any
wrapped NS5-branes from BPS equations (\ref{ns5bps}) and (\ref{ns5bps4}) 
and an embedding formula \cite{CvLuPo1}. 
They are supersymmetric solutions
already in the seven-dimensional $SO(4)$ gauged supergravity.
Here, we concentrate on a solution for wrapped type IIB
NS5 branes around holomorphic $CP^2$ inside Calabi-Yau threefold  
${\cal O}_{CP^2}(-3)$ obtained by a crepant resolution of orbifold
singularity $C^3/Z_3$.
Especially, we give a map to a solution provided by \cite{HoKa}
as a supergravity approximation of exact string world-sheet description 
for the wrapped NS5 branes.   
Also we briefly give comments on other configurations.

Let us start again with specific ansatz on bosonic fields 
in $D=7$ $SO(4)$ gauged supergravity \cite{SaSe}.
We choose usual ansatz on seven-dimensional metric as follows
\begin{eqnarray}
\label{ansatzcp2ns5}
ds^2_7=e^{2f(r)}\left(d\xi_2^2+dr^2\right)+e^{2g(r)}d\widetilde{s}_{CP^2}^2,
\end{eqnarray}
where $d\xi_2^2=-d\xi_0^2+d\xi_1^2$ is a metric on $R^{1,1}$
in unwrapped directions of NS5 branes.
Here, we use a Fubini-Study metric with real coordinates
for supersymmetric $CP^2$ cycle
\begin{eqnarray}
d\widetilde{s}_{CP^2}^2=\frac{1}{\left(1+\frac{u^2}{6}\right)^2}
\left[du^2+\frac{1}{4}u^2\left(
d\widetilde{\psi}+\cos{\widetilde{\theta}}d\widetilde{\phi}\right)^2
+\frac{u^2}{4}\left(d\widetilde{\theta}^2+\sin^2
{\widetilde{\theta}}d\widetilde{\phi}^2\right)\right].
\end{eqnarray}
Here $u$ is a radial coordinate, 
and ranges of angular coordinates $\widetilde{\psi},
\widetilde{\theta}$ and $\widetilde{\phi}$
are those for usual Euler angles
$0\leq\widetilde{\psi}<4\pi, 
0\leq\widetilde{\theta}<2\pi$ and  
$0\leq\widetilde{\phi}<2\pi$.
This metric on $CP^2$ has a unit cosmological constant $\ell=+1$.
Let us introduce orthonormal basis of this metric
\begin{eqnarray}
\widetilde{e}^0=\frac{du}{\left(1+\frac{u^2}{6}\right)},\quad
\widetilde{e}^1=\frac{u\left(
d\widetilde{\psi}+\cos{\widetilde{\theta}}d\widetilde{\phi}\right)
}{2\left(1+\frac{u^2}{6}\right)},\quad
\widetilde{e}^2=\frac{ud\widetilde{\theta}}
{2\left(1+\frac{u^2}{6}\right)},\quad
\widetilde{e}^3=\frac{u\sin{\widetilde{\theta}}d\widetilde{\phi}}
{2\left(1+\frac{u^2}{6}\right)}.
\end{eqnarray}
Then, we define K\"ahler form $\widetilde{J}$
and K\"ahler potential $\widetilde{A}$ for $CP^2$ with this
metric by 
\begin{eqnarray}
\widetilde{J}=\widetilde{e}^0\wedge 
\widetilde{e}^1+
\widetilde{e}^2\wedge \widetilde{e}^3=d\widetilde{A},
\qquad
\widetilde{A}=\frac{u^2}{4\left(1+\frac{u^2}{6}\right)}\left(
d\widetilde{\psi}+\cos{\widetilde{\theta}}d\widetilde{\phi}\right).
\end{eqnarray}
Let us return to a seven-dimensional metric (\ref{ansatzcp2ns5}) and
we denote orthonormal basis for the metric
(\ref{ansatzcp2ns5}) by
\begin{eqnarray}
e^0=e^{f(r)}d\xi_0, \quad
e^1=e^{f(r)}d\xi_1,\quad
e^2=e^{f(r)}dr,\quad 
e^{3+a}=e^{g(r)}\widetilde{e}^a \quad (a=0,1,2,3).
\end{eqnarray}
Then, we are able to specify projections on spinor fields and 
compatible ansatz on non-vanishing components of gauge field
strength $F_{\mu\nu}^{AB}$
\begin{eqnarray}
&&\gamma_{12}\epsilon=\gamma_{34}\epsilon=
\Gamma^{12}\epsilon,\qquad
\gamma_2\epsilon=\epsilon,\\
&&F^{12}_{34}=F^{12}_{56}=\frac{1}{2g}e^{-2g(r)}.\nonumber
\end{eqnarray}
Here, $\gamma_{\mu}$ are $D=7$ gamma matrices satisfying
anti-commutation relation $\{\gamma_{\mu}, \gamma_{\nu}\}=2g_{\mu\nu}$
with seven-dimensional metric tensor $g_{\mu\nu}$.
We also have $SO(4)_c$ gamma matrices $\Gamma^a$
satisfying $\{\Gamma^a,\Gamma^b\}=2\delta^{ab}$.
We denote indices of $SO(4)$ by $A,B=1\dots 4$.
We denote $SO(4)$ gauge coupling constant by $g$.  
Solutions satisfying this projection 
preserve four supercharges in the $SO(4)$ gauged and type IIB supergravity.
On the other hand, 
spin connection of $CP^2$ cycle is a $U(2)= U(1)\times SU(2)$ connection.
Then, we identify $U(1)$ subgroup of the spin connection with 
$U(1)=SO(2)$ subgroup of $SO(4)$ symmetry in the seven-dimensional 
$SO(4)$ gauged supergravity.
This requires that ansatz on scalar fields is chosen to break $SO(4)$
symmetry into $SO(2)\times SO(2)$ subgroup.
We are able to realize this breaking of $SO(4)$ symmetry
by turning on a single radial function $\phi(r)$
such that scalar function $T_{AB}$ has following form
\begin{eqnarray}
T_{AB}=e^{f(r)}\left(e^{-4\phi(r)}\; {\bf 1}_{2\times 2}, \;
e^{4\phi(r)}\;
{\bf 1}_{2\times 2}\right),
\end{eqnarray}
where ${\bf 1}_{2\times 2}$ is a unit $2\times 2$ matrix.
Here, an appearance of $f(r)$ in the right hand side
is a matter of convention
in order to simplify resulting ten-dimensional string frame metrics
for the wrapped NS5 branes.

We have BPS equations
for $f(r), g(r)$ and $\phi(r)$ 
(\ref{ns5bps}) with $d=4, p=q=2$ and $\ell=+1$ with the ansatz
on bosonic fields.
Let us write down solutions for the BPS equations.
We start with introducing
another radial functions defined by (\ref{change})
\begin{eqnarray}
x(r)=-4\phi(r),\qquad h(r)=g(r)-f(r),\qquad A(r)=\frac{3}{2}f(r)+g(r).
\end{eqnarray}
We also introduce another radial variable $z$ defined as
\begin{eqnarray}
z=e^{2h(r)}.
\end{eqnarray}
Then, the BPS equations are rewritten into following first order
differential equations
\begin{eqnarray}
\dot{A}(z)&=&g^2\frac{e^{2x(z)}+1}{2}+\frac{1}{2z},\nonumber\\
\dot{h}(z)&=&\frac{1}{2z},\\
\dot{x}(z)&=&-g^2\left(e^{2x(z)}-1\right)+\frac{1}{z},\nonumber
\end{eqnarray}
where $\dot{}$ denotes a derivative in terms of $z$.
These first order BPS equations have following 
solutions for radial functions $A(z), x(z)$
\begin{eqnarray}
\label{xbeh}
e^{-2x(z)}&=&1-\frac{1}{g^2z}+\frac{1+ke^{-2g^2z}}{2g^4z^2},\\
e^{2A(z)+x(z)}&=&z^2e^{2g^2z},\nonumber
\end{eqnarray}
where $k$ is only integration constant.
Other possible integration constant can be absorbed 
by a change of coordinates in seven-dimensional metric 
(\ref{ansatzcp2ns5}).
Behaviors of $e^{-2x(z)}$ for specific values of $k$
are given in Figure 7.
\begin{figure}
\begin{center}
\includegraphics[height=6cm]{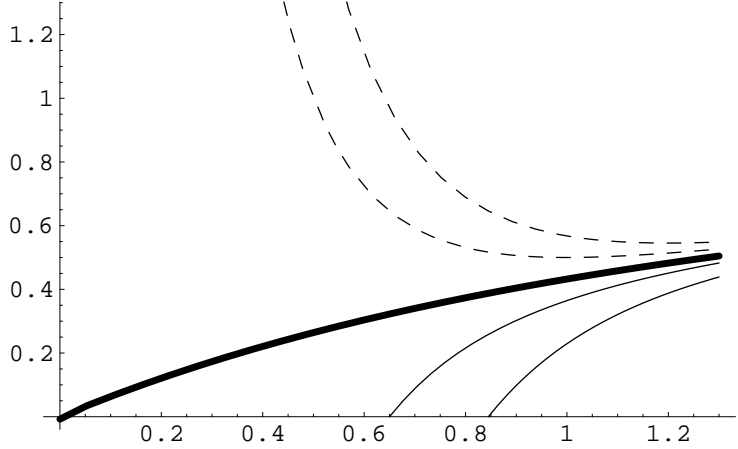}
\end{center}
\caption{Behaviors of $e^{-2x(z)}$ for $k=-4,-2,-1,0, 1$.
Bold line shows a behavior for $e^{-2x(z)}$ for $k=-1$.
We denote plots for $k=-4, -2$ by usual lines,
and plots for $k=-1, 0$ by dotted lines.
There are cut off at finite $z=z_0$ for
plots for $k=-4, -2$. We set $g=1$.}
\end{figure}
We have three different behaviors depending on a parameter $k$.
We always have singular behaviors at certain values of $z$. 
We have a singularity which $e^{-2x(z)}$ diverges at $z=0$
for behaviors with $k>-1$.  
For a behavior with $k=-1$, $e^{-2x(z)}$ vanishes at $z=0$.
There also exist singularities where
$e^{-2z(z)}$ vanishes at a finite positive value $z=z_0$
for behaviors with $k<-1$.

Now, we write down solutions in ten-dimensional type IIB
supergravity by using an formula \cite{CvLuPo1}.
Note that when we use the formula,
we choose coordinates which parameterize squashed three sphere $S^3$ as  
\begin{eqnarray}
\mu_1=\cos{\theta}\cos{\phi_1},\quad
\mu_2=\cos{\theta}\sin{\phi_1},\quad
\mu_3=\sin{\theta}\cos{\phi_2},\quad
\mu_4=\sin{\theta}\sin{\phi_2},
\end{eqnarray}
where $\mu_i \; (i=1\dots 4)$ satisfy
$\sum_{i=1}^4\mu_i^2=1$.
The ranges of angle variables $\theta$ and $\phi_1, \phi_2$ are
given by $0\leq \theta < \frac{\pi}{2}$ and  
$0\leq \phi_1, \phi_2 < 2\pi$.
This parameterization matches a present procedure 
that we break $SO(4)$ symmetry in gauged supergravity 
into $SO(2)\times SO(2)$ subgroup.
By an application 
of formulas in \cite{CvLuPo1} into the
ansatz on seven-dimensional bosonic fields, 
we obtain ten-dimensional metric, dilaton field and 
NS three-form field strength in type IIB supergravity.
Other fields in type IIB supergravity are zero
for the wrapped NS5 branes.
A ten-dimensional metric in string frame is given by
\begin{equation}
ds_{10(st)}^2=d\xi_{2}^2+g^2e^{2x}dz^2
+zd\widetilde{s}_{CP^2}^2+
\frac{d\theta^2}{g^2}+\frac{e^{-x(z)}}{g^2\Omega(z)}\cos^2{\theta}\left(
d\phi_1+\widetilde{A}\right)^2+\frac{e^{x(z)}}{g^2\Omega(z)}
\sin^2{\theta}d\phi_2^2,
\end{equation}
where radial function $x(z)$ is provided by (\ref{xbeh}), 
and we have introduced a function $\Omega$ as
\begin{eqnarray}
\Omega(z)=e^{x(z)}\cos^2{\theta}+e^{-x(z)}\sin^2{\theta}.
\end{eqnarray}
We omit dependence of $\Omega$ on $\theta$ to write $\Omega(z)$
because this dependence is not important here.
The dilaton field has following form
\begin{eqnarray}
e^{-2\Phi+2\Phi_0}=e^{2g^2z}\left[1-\sin^2{\theta}\frac{1}{g^2z}
\left(1-\frac{1+ke^{-2g^2z}}{2g^2z}\right)\right],
\end{eqnarray}
where $\Phi_0$ is the expectation value of dilaton field.
Then, NS three-form field strength is 
\begin{equation}
H_{3}^{NS}=
\frac{2\sin{\theta}\cos{\theta}}{g^2\Omega(z)^2}
\left(\sin{\theta}\cos{\theta}\dot{x}(z)dz-d\theta\right)
\wedge\left(d\phi_1+\widetilde{A}\right)\wedge d\phi_2
+\frac{e^{-x(z)}\sin^2{\theta}}{g^2\Omega(z)}
d\widetilde{A}\wedge d\phi_2.
\end{equation}
Notice that
moduli parameters of this solution are integration constant $k$
and the expectation value of dilaton field $\Phi_0$. 

Let us see asymptotic behaviors of the solution. 
We begin with a behavior at $z\to\infty$.
In this limit, we easily see $e^{-2x(z)} \to 1$ from (\ref{xbeh}).
Then, the ten-dimensional metric and dilaton field become 
\begin{eqnarray}
ds^2_{10(st)}&=&d\xi_{2}^2+g^2dz^2
+z d\widetilde{s}_{CP^2}^2+\frac{1}{g^2}d\theta^2
+\frac{1}{g^2}\cos^2{\theta}\left(d\phi_1+\widetilde{A}\right)^2
+\frac{1}{g^2}\sin^2{\theta}d\phi_2^2,\nonumber\\
e^{-2\Phi+2\Phi_0}&=&e^{2g^2z}.
\end{eqnarray}
This metric has a similar form to that for a linear dilaton background
arising from flat NS5 branes. 
Here, we have a worldvolume $R^{1,1}\times CP^2$ instead of 
flat Minkowski space-time $R^{1,5}$, 
and an effect of twisting procedure on a worldvolume. 
A divergent size of wrapped cycle $CP^2$ at $z\to\infty$
indicates that we have an asymptotically free theory
as a dual two-dimensional theory on the wrapped NS5 branes. 
Note that we are able to relate $SO(4)$ gauge coupling 
in the seven-dimensional gauged supergravity and 
the number of wrapped NS5 branes $N$ as follows
\begin{eqnarray}
\frac{1}{g^2}=N,
\end{eqnarray}
from the amount of NS three-form flux over the
transverse squashed three sphere.

It turns out that all the solutions specified by a parameter $k$
show a singular behavior at small $z$ region.
The function $e^{-2x(z)}$ goes to zero or infinity at certain value 
of radial variable $z$ depending on the parameter $k$.
For $k<-1$, allowed range of radial direction $z$ is restricted to
$z_0\leq z<\infty$ where $z_0$ is a value to give $e^{-2x(z_0)}=0$.
For a solution with $k=-1$, $e^{-2x(z)}$ vanishes at $z=z_0=0$.
For $k>-1$, radial variable is defined on $0\leq z<\infty$.
The function $e^{-2x(z)}$ diverges at $z\to 0$.
Let us note that time component $g_{00}$ of 
ten-dimensional Einstein frame metric is given by $e^{-\frac{\Phi}{2}}$.
Solutions with $k\leq-1$ are singular at $z=z_0$ and $\theta=\pi/2$.
We also have an orbifold singularity where $z=0$ and $\theta=0$ for $k=-1$.
The $g_{00}$ component in these cases goes to zero near the singularities. 
On the other hand, solutions with $k>-1$ are singular at $z\to 0$
and generic values of $\theta$. 
The corresponding $g_{00}$ components diverge near the singularities.
If we assume a validity of criteria about singularities in \cite{MaNu1},
we expect that solutions with $k\leq-1$ may give 
supergravity  backgrounds which are dual to 
two-dimensional theories with four supercharges
on the wrapped NS5 branes.
We show plots for behaviors of dilaton field for $\theta=\pi/2$ in Figure 8.
\begin{figure}
\begin{center}
\includegraphics[height=6cm]{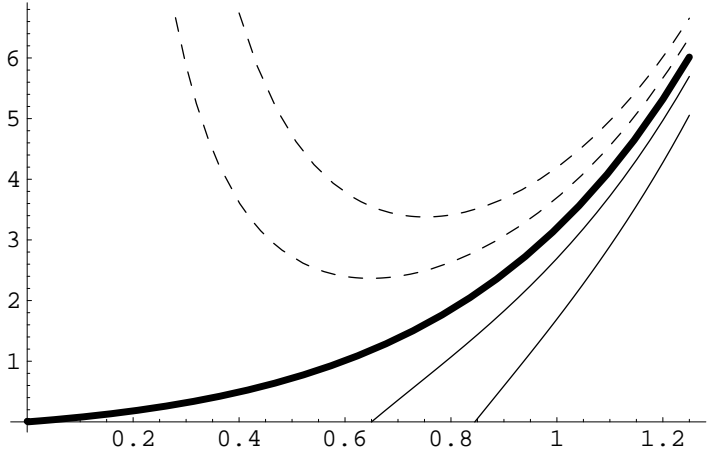}
\end{center}
\caption{Behaviors of $e^{-2\Phi(z)}$ with $\theta=\pi/2$
for $k=-4,-2,-1,0$ and $1$.
Bold line shows a behavior for $k=-1$.
We denote plots for $k=-4, -2$ by usual lines,
and plots for $k=-1, 0$ by dotted lines.
Note that $e^{-2\Phi(z)}$ for $k=-4, -2$ go to zero
at finite $z=z_0$.}
\end{figure}

Let us see the behavior near the singularity $z=z_0$
and $\theta=\pi/2$ for $k \leq -1$ more properly.
We linearly approximate $e^{-2x(z)}$ as $2g^2(z-z_0)$.
Then, we introduce following variables $y$ and $\psi$ 
from $z$ and $\theta$ 
\begin{eqnarray}
\sqrt{2g^2\left(z-z_0\right)}=\sqrt{2gy}\;\sin{\frac{\psi}{2}},\qquad
\theta-\frac{\pi}{2}=\sqrt{2gy}\;\cos{\frac{\psi}{2}},
\end{eqnarray}
where these new variables behave $y\to0$ and $\psi\to 0$ near the singularity. 
Then, the ten-dimensional metric and dilaton field are 
\begin{eqnarray}
ds_{10(st)}^2&=&
d\xi_{2}^2+z_0d\widetilde{s}_{CP^2}^2+\frac{1}{2gy}\left[
dy^2+y^2\left(
d\psi^2+\sin^2{\psi}\left(d\phi_1+\widetilde{A}\right)^2\right)
+\frac{1}{g^2}{d\phi_2^2}\right],\nonumber\\
e^{-2\Phi+2\Phi_0}&=&2gye^{2g^2z_0},
\end{eqnarray}
near the singularity.
The metric represents 
near horizon configuration of wrapped NS5 branes
smeared on a circle parameterized by $\phi_2$ direction 
We may interpret that this metric represents 
a slice of Coulomb branch spanned by
motions of the wrapped NS5 branes in directions
which are transverse both to $CY_3$
and wrapped NS5 branes \cite{GaKiMaWa1}.

An exact world-sheet model for the wrapped NS5 branes 
has been provided in \cite{HoKa}.
This model is a Landau-Ginzburg model with following superpotential
\begin{eqnarray}
W_{LG}=e^{-NZ}\left(e^{-Y_1}+e^{-Y_2}+e^{Y_1+Y_2}\right)+X^N,
\end{eqnarray}
where $X, Y_1,Y_2$ and $Z$ are chiral superfields.
Then, we are able to consider one-parameter deformation
for this model
\begin{eqnarray}
e^{\frac{t}{2}}e^{-NZ},
\end{eqnarray}
where $t$ is a complex deformation parameter.
Geometrically, this parameter $t$ can be identified with 
K\"ahler moduli of holomorphic $CP^2$ cycle.
Furthermore, supergravity background valid at large $t\to\infty$
has been given by string world-sheet arguments.  
Intuitively, this is  
because the ten-dimensional solution for wrapped NS5 branes
are regarded as a background with $SU(4)$ holonomy structure 
including non-trivial dilaton field and torsion structure \cite{GaKiMaWa2}.
We are able to relate the present ten-dimensional solution 
to a solution \cite{HoKa} by an identification of parameter
in each solution
\begin{eqnarray}k=\left(1-\frac{s}{N}\right)e^{\frac{s}{N}}, 
\qquad
s={\rm Re}\; t,
\end{eqnarray}
and a change of radial coordinate $z=2y/3$.
This changes the ranges of radial direction into $s\leq y<\infty$.
Here, we have a precise parameterization
of a volume for $CP^2$ by $s$ in the solution.
The solution is valid as a large volume description of
world-sheet model when $s\to\infty$.
Note that we have obtained the solutions 
from supersymmetric variations of fermions in the 
seven-dimensional $SO(4)$ gauged supergravity.
This construction ensures that the solutions preserve four supercharges.
This gives an explanation on a preserved supersymmetry 
for the solution in \cite{HoKa}.

Let us mention possible dual two-dimensional theory 
on the wrapped NS5 branes \cite{HoKa}.
Naively, we would have $D=2$ ${\cal N}=(2,2)$ $SU(N)$ 
super Yang-Mills theory. 
This theory is obtained as a Kaluza-Klein reduction of
$D=6$ ${\cal N}=2$ $SU(N)$ super Yang-Mills theory which is
a low energy effective theory of little string theory.
The two-dimensional super Yang-Mills theory 
has vector and axial $U(1)$ R symmetry.
We expect that these two R symmetries correspond to the translation 
invariance about a shift of $\phi_1$ and $\phi_2$ 
in the supergravity solution. 
The axial $U(1)$ R symmetry seems to break 
if we incorporate an effect of string world-sheet instantons 
which violate translation invariance under $\phi_2$ as in \cite{MaNu2}.
Here, it is likely that the effect of world-sheet instantons dumps in the
decoupling limit \cite{Lerche, HoKa}.  
Thus, we still detect a perturbative aspect of super Yang-Mills theory  
by the supergravity solution consistently.
In fact, we should go to scales much smaller than the energy scale of
little string theory in order to decouple the two-dimensional super
Yang-Mills theory.
In this limit, we need to adopt a $S$-dual background which
represents wrapped D5 branes around the same supersymmetric
cycle $CP^2$ in $CY_3$ \cite{ItMaSoYa}.
Then, we may do D5 probe calculation on the singular ten-dimensional
background \cite{GaKiMaWa1, BiCoZa, GoRu, GaKiMaWa2}.
This calculation will show that we have a 
smooth holomorphic moduli space with K\"ahler metric suitable for a
system with four supercharges as in \cite{KeWa, JoLoPa1, JoLoPa2}. 

Let us note that the solution for wrapped NS5 branes 
around holomorphic $CP^2$ inside $CY_3$ has a very similar
form to a solution for wrapped NS5 branes around holomorphic $CP^1$ 
inside non-compact $K3$ surfaces \cite{GaKiMaWa1, BiCoZa}.
The ten-dimensional solution is given by using
metric and K\"ahler potential for wrapped holomorphic $CP^1$ cycle
\begin{eqnarray}
d\widetilde{s}_{CP^1}^2=d\widetilde{\theta}^2+
\sin^2{\widetilde{\theta}}d\widetilde{\phi}^2,\qquad
\widetilde{A}=\cos{\widetilde{\theta}}d\widetilde{\phi},
\end{eqnarray}
where the range of angles are $0\leq\widetilde{\theta}<\pi$
and $0\leq\widetilde{\phi}<2\pi$, and
adopting slightly modified forms of $e^{-2x(z)}$ and the dilaton field
\begin{eqnarray}
e^{-2x(z)}&=&1-\frac{1+ke^{-2g^2z}}{2g^2z},\\
e^{-2\Phi+2\Phi_0}&=&e^{2g^2z}\left(1-\sin^2{\theta}
\frac{1+ke^{-2g^2z}}{2g^2z}\right).\nonumber
\end{eqnarray}
The solution shows a similar asymptotic behaviors to
that for $CP^2$ case.
Again there is an exact world-sheet description
for the wrapped NS5 branes \cite{HoKa}.
A supergravity approximation of this description 
is obtained by introducing a moduli $s$ parameterizing a volume of $CP^1$
\begin{eqnarray}
k=\left(\frac{s}{N}-1\right)e^{\frac{s}{N}},
\end{eqnarray}
and a change of radial coordinates $z=y/2$.
This identification is valid for $s\to\infty$.
Here, we have a relation $N=1/g^2$ among the number of NS5 branes
$N$ and the $SO(4)$ gauge coupling constant in $SO(4)$ gauged supergravity.
A similarity between $CP^1$ and $CP^2$ cases 
originates also from a universal string world-sheet construction
for $CP^n$. 
The supergravity solution have been interpreted 
as a dual background to perturbative Coulomb branch of 
$D=4$ ${\cal N}=2$ $SU(N)$ super Yang-Mills theory.
By a probe calculation, coordinates and parameters 
in the supergravity solution 
is mapped into those for the Seiberg-Witten solutions 
without instanton corrections \cite{GaKiMaWa1}.
This identification follows from
a form of effective actions required by eight supercharges.
Furthermore, the string world-sheet model 
precisely reproduces the Seiberg-Witten solution \cite{HoKa}.
In fact, it is known that 
world-sheet instantons are mapped into space-time gauge theoretical 
instantons in a context of geometric engineering \cite{KaKlVa}.

Let us discuss behaviors of wrapped NS5 branes around 
supersymmetric four cycles with negative constant curvature $(\ell=-1)$
inside non-compact Calabi-Yau threefolds.
We are able to construct a ten-dimensional solution 
for the wrapped NS5 branes in the same way as previous $CP^2$ case.
Here, we show behaviors of a function $e^{-2x(z)}$ and 
the ten-dimensional dilaton field.
The function $e^{-2x(z)}$ is given by
\begin{eqnarray}
e^{-2x(z)}=1+\frac{1}{g^2z}+\frac{1+ke^{2g^2z}}{2g^4z^2},
\end{eqnarray}
where $k$ is an integration constant.
We have only to change $z$ to $-z$ in order to obtain this
function from that for $CP^2$ (\ref{xbeh}).
This shows that only solutions with $k>-1$ are allowed.
Solutions with $k\leq-1$ 
always become negative for all positive values of $z$.   
We depict behaviors of this one-parameter function $e^{-2x(z)}$ in Figure 9.
\begin{figure}
\begin{center}
\includegraphics[height=6cm]{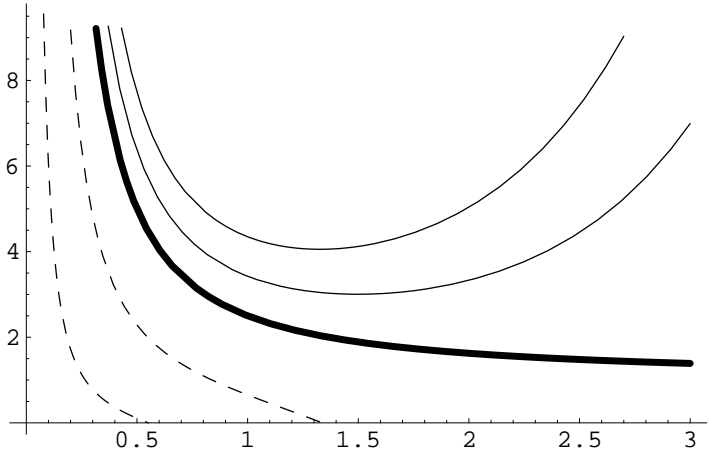}
\end{center}
\caption{Behaviors of $e^{-2x(z)}$ for $k=-0.9,-0.5,0,0.25$ and $0.5$.
We denote a plot of behavior for $k=0$ by bold lines,
plots for $k=-0.9, -0.5$ by dotted lines,
and plots for $k=0, 0.25$ by usual lines.}
\end{figure}
For $-1< k<0$, $e^{-2x(z)}$ goes to zero at finite $z=z_0$.
We see a divergence for $e^{-2x(z)}$ with $k>0$ at large $z$.
Ten-dimensional dilaton field $\Phi$ is given by 
\begin{eqnarray}
e^{-2\Phi+2\Phi_0}=e^{-2g^2z}\left[
1+\sin^2{\theta}\left(\frac{1}{g^2z}+\frac{1+ke^{2g^2z}}{2g^4z^2}\right)
\right],
\end{eqnarray}
where $\Phi_0$ is the expectation value of the dilaton field.
We show behaviors of this dilaton as a radial function of $z$ 
with several values of $k$ in Figure 10.
\begin{figure}
\begin{center}
\includegraphics[height=6cm]{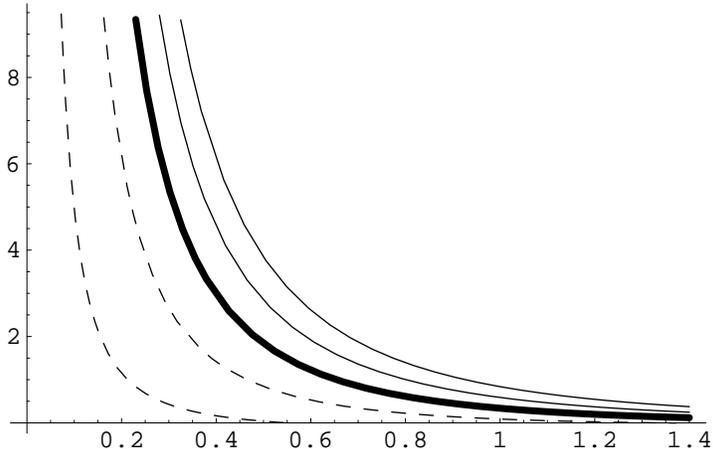}
\end{center}
\caption{Behaviors of $e^{-2\Phi(z)}$ for
$k=-0.9,-0.5,0,0.25$ and $0.5$.
Bold line corresponds the behavior for $k=0$.
We denote plots for $k=-0.9, -0.5$ by dotted lines
and plots for $k=0, 0.25$ by usual lines.
Plots for $k=-0.9$ and $-0.5$ become zero at finite positive $z$.
Other three plots go to zero at $z\to\infty$.
All plots diverge at $z\to0$.}
\end{figure}
Here, we are not able to compare
behaviors at $z\to \infty$ with a linear dilaton background 
from flat NS5 branes naively. 
We also have bad singularity at $z\to0$
due to a divergence in a $g_{00}$ component 
of each ten-dimensional metric in Einstein frame.
Thus, it might be possible that we are not able to investigate 
quantitatively dual two-dimensional theory
on the wrapped NS5 branes by the supergravity solution.
When we wrap type IIB NS5 branes around 
certain supersymmetric four cycles with negative constant curvature,
we may have several zero modes which give several hypermultiplets
in addition to a super Yang-Mills theory.
Then, dual two-dimensional theories could be non-renormalizable theories.
We need more detailed understanding of these dual theories 
in order to make a precise argument from the supergravity solution.
Similar situation is also seen in wrapped NS5 branes around
Riemann genus $g\geq 2$ Riemann surfaces inside $K3$ surfaces \cite{GaKiMaWa1}.

We wish to close this section with a remark.
We have two classes of singular supergravity
solutions for wrapped NS5 branes.
First, one class arises from solutions of wrapped NS5 branes 
with $q\not=0$ configurations.
These solutions are likely to be dual to Coulomb branches of 
worldvolume theories on wrapped NS5 branes. 
Motions transverse both to NS5 branes
and non-compact special holonomy manifolds give 
modes parameterizing Coulomb branch.
Coulomb branches from $1/4, 1/8$ supersymmetric $(d,p,q)=
(2,2,2), (4,2,2)$ configurations
have been solved by a string world-sheet language \cite{HoKa}. 
This shows that 
we need to explore a string theoretical argument
beyond the supergravity solutions
in order to resolve singularities and describe
dual Coulomb branches exactly.
Then generalized supergravity ansatz to resolve
singularities like \cite{GoRu} would be irrelevant 
in order to study dual worldvolume theories.
This picture seems to be advocated in context of enhancon singularities 
\cite{BuPePo}.
In this sense, we may expect that singularities
are often superficial due to absence of massive string modes.
As for solutions with four supercharges,
wrapped NS5 branes around supersymmetric four cycles inside
$G_2$ holonomy manifolds $(d=4, p=3, q=1)$
numerically show a similar behavior to 
that with three cycles in $CY_3$ $(d=3, p=3, q=1)$ \cite{GoRu, GaKiMaWa2}.
It should exist techniques to give a world-sheet description 
for these two wrapped NS5 branes.
Secondly,
the other class of solutions arises from wrapped NS5 branes with
$p=4, q=0$ configurations.
We expect no Coulomb branches from these configurations.
Singularities in $d=2,3$ configurations are resolved 
by assuming monopole or instanton configurations 
for gauge fields in the $SO(4)$ gauged supergravity\cite{MaNu2, MaNa}.
Then, smooth supergravity solutions are likely to correspond
to confining phases of dual worldvolume theories.
The other solutions with supersymmetric four cycles 
inside eight-dimensional special holonomy manifolds
would have a similar mechanism to resolve singularities.
A technique \cite{GaMaPaWa} would be useful in order to
construct these non-singular supergravity solutions.
Certainly, it is necessary to give a reason 
why we have these two different resolutions
in terms of string arguments or supergravity solutions.

\section{Non-BPS solutions in massive IIA supergravity}
\hspace{5mm}
In this section, we study non-supersymmetric
generalizations of wrapped branes.
We have studied first order differential
equations for supersymmetric wrapped branes.
In order to generalize them to non-supersymmetric situations,
we should use second order equations of motion.
We find that we are able to
claim about exact solutions which contain AdS space-time.
Here, we obtain such
solutions for wrapped D4-D8 systems in massive IIA
supergravity by using the same ansatz for
bosonic fields as supersymmetric cases.
One additional input is ansatz which assumes solutions with
AdS space-time found in supersymmetric wrapped D4-D8 branes \cite{NuPaScTr}.
Similar idea has been analyzed for wrapped M5 branes \cite{GaKiPaWa2}. 
We have checked that similar non-supersymmetric solutions
do not exist for wrapped D3 and M2 branes.
It might be possible that D2-D6 configurations studied in \cite{GuNuSc}
have non-supersymmetric solutions with AdS space-time.

\subsection{$D=6$ $N=2$ $SU(2)\times U(1)$ massive gauged supergravity}
\hspace{5mm}
Wrapped D4-D8 systems are constructed from
massive gauged supergravity in six-dimensions.
Known example with maximal amount of supersymmetry
is $D=6$ $N=2$ $SU(2)\times U(1)$ massive gauged supergravity \cite{Romans}.
Let us remind necessary data about this massive gauged supergravity.
This theory contains a graviton $e^{\alpha}_{\mu}$,
three $SU(2)$ gauge fields $A_{\mu}^I$
$(I=1,2,3)$, one Abelian
gauge field $a_{\mu}$, a two-index tensor gauge field $B_{\mu\nu}$,
one scalar field $\phi$, four gravitinos $\psi_{\mu i}$ and
four gauginos $\chi_i$.
Indices $\mu, \nu,\dots$ are curved indices, and $\alpha, \beta, \dots$
are local Lorentz indices.
Here, we set Abelian gauge fields $a_{\mu}$ and 
tensor fields $B_{\mu\nu}$ to be zero.
The bosonic Lagrangian is
\begin{eqnarray}
\label{d6lag}
L=\sqrt{-g}\left[
R-2\left(\partial_{\mu}\phi\right)^2
-e^{-\sqrt{2}\phi}F_{\mu\nu}^IF^{\mu\nu}_I
+\frac{1}{2}\left(g^2e^{\sqrt{2}\phi}
+4gme^{-\sqrt{2}\phi}-m^2e^{-3\sqrt{2}\phi}\right)
\right].
\end{eqnarray}
Here, we use signature $(- + + + + +)$ for a six-dimensional metric.
We denote $SU(2)$ gauge coupling constant by $g$,
and mass parameter by $m$. 
The $SU(2)$ gauge field strength $F_{\mu\nu}^I$ is defined by
$F_{\mu\nu}^I=\partial_{\mu}A_{\nu}^I-\partial_{\nu}A_{\mu}^I
+g \epsilon_{IJK}A_{\mu}^JA_{\nu}^K$.
Supersymmetry transformations for fermionic fields are
\begin{eqnarray}
\label{d6susy}
\delta\psi_{\mu i}&=&\partial_{\mu}\epsilon_i
+\frac{1}{4}\omega_{\mu}^{\nu\rho}\gamma_{\nu\rho}\epsilon_i
+gA_{\mu}^I\left(T^I\right)_i^j\epsilon_j\\
&&
+\frac{1}{8\sqrt{2}}\left(
ge^{\frac{\phi}{\sqrt{2}}}+me^{-\frac{3\phi}{\sqrt{2}}}
\right)\gamma_{\mu}\gamma_7\epsilon_i
-\frac{1}{4\sqrt{2}}\left(
\gamma_{\mu}^{\nu\rho}-6\delta_{\mu}^{\nu}\gamma^{\rho}
\right)
e^{-\frac{\phi}{\sqrt{2}}}
\gamma_7F_{\nu\rho}^I\left(T^I\right)_i^j\epsilon_j,\nonumber\\
\delta\chi_i&=&\frac{1}{\sqrt{2}}
\gamma^{\mu}\partial_{\mu}\phi\epsilon_i
-\frac{1}{4\sqrt{2}}\left(
ge^{\frac{\phi}{\sqrt{2}}}
-3me^{-\frac{3\phi}{\sqrt{2}}}
\right)\gamma_7\epsilon_i
+\frac{1}{2\sqrt{2}}\gamma^{\mu\nu}
e^{-\frac{\phi}{\sqrt{2}}}
\gamma_7F_{\mu\nu}^I\left(T^I\right)_i^j\epsilon_j.\nonumber
\end{eqnarray}
Here, $\omega_{\mu}^{\nu\rho}$ are spin connections of a metric.
We use $SU(2)$ generators defined by $T^I=-\frac{i}{2}\sigma^I$
where $\sigma^I (I=1,2,3)$ are $2\times 2$ Pauli matrices
whose matrix indices are $i,j=1,2$.
We denote $D=6$ gamma matrices by $\gamma_{\mu}$ satisfying
$\{\gamma_{\mu}, \gamma_{\nu}\}=2g_{\mu\nu}$ with a six-dimensional
metric tensor $g_{\mu\nu}$.
The gamma matrix $\gamma_7$ is defined as 
$\varepsilon_{\mu\nu\rho\sigma\tau\kappa}\gamma_7
=\gamma_{\mu\nu\rho\sigma\tau\kappa}
=\frac{1}{6!}\gamma_{[\mu}\dots\gamma_{\kappa]}$
where $\varepsilon_{\mu\nu\rho\sigma\tau\kappa}$
is a Levi-Civita tensor density satisfying
$\varepsilon_{123456}=1$.

Solutions in this gauged supergravity
can be embedded into those in massive type IIA supergravity
\cite{CvLuPo2}.
Ten-dimensional metric $ds_{10}$ is constructed from 
six-dimensional metric $ds_6$, scalar $\phi$ and gauge fields $A_{\mu}^I$
in six-dimensional gauged supergravity 
\begin{eqnarray}
\label{d6d10me}
ds_{10}^2&=&
\sin^{12}{\xi}\; X^{\frac{1}{8}}
\left[
\Delta^{\frac{3}{8}}ds_6^2
+
\frac{2\Delta^{\frac{3}{8}}X^2}{g^2}d\xi^2
+\frac{1}{2g^2\Delta^{\frac{5}{8}}X}\cos^2\xi
\sum_{I=1}^3\left(\Sigma^I-gA^I\right)^2
\right],
\end{eqnarray}
where $\Delta=X \cos^2\xi+\frac{1}{X^3}\sin^2\xi$, 
$X=e^{-\frac{1}{\sqrt{2}}\phi}$,
and
$\Sigma^I (I=1,2,3)$ are left invariant one-form of $SU(2)$
parameterizing $S^3$.
These one-forms satisfy
$d\Sigma^I=-\frac{1}{2}\epsilon_{IJK}\Sigma^J\wedge \Sigma^K$.
Here, squashed four sphere $S^4$ is parameterized 
by $\left(\xi, \Sigma^I\right)$.
The anti-symmetric four-form tensor fields are written down as
\begin{eqnarray}
F_4&=&
-\frac{\sqrt{2}\sin^{\frac{1}{3}}\xi\cos^3\xi \; U}{6g^3\Delta^2}d\xi
\wedge \prod_{I=1}^3\left(\Sigma^I-gA^I\right)
-\frac{\sqrt{2}\sin^{\frac{4}{3}}\xi\cos^4\xi}
{g^3\Delta^2X^3}dX\wedge\prod_{I=1}^3\left(\Sigma^I-gA^I\right)\nonumber
\\
&&+\frac{\sin^{\frac{1}{3}}\xi\cos^3\xi}{\sqrt{2}g}
\sum_{I=1}^3F^I\wedge\left(\Sigma^I-gA^I\right)
\wedge d\xi\nonumber\\
&&
-\frac{\sin^{\frac{4}{3}}\xi\cos^2\xi}{4\sqrt{2}g^2\Delta X^3}
\epsilon_{IJK}F^I\wedge \left(\Sigma^J-gA^J\right)
\wedge\left(\Sigma^K-gA^K\right),
\end{eqnarray}
where $U=\frac{\sin^2\xi}{X^6}-3X^2\cos^2{\xi}
+\frac{4\cos^2\xi}{X^2}-\frac{6}{X^2}$.
The dilaton field $\Phi$ is given by
\begin{eqnarray}
e^{\Phi}=
\frac{\Delta^{\frac{1}{4}}}{\sin^{\frac{5}{6}}\xi X^{\frac{5}{4}}}.
\end{eqnarray}
Note that supersymmetric extremum of a scalar potential in the 
six-dimensional massive gauged supergravity
gives a possible maximally supersymmetric background with $g=3m$.
This condition fixes a parameter of the six-dimensional gauged theory.
A relation among parameters in ten- and six-dimensional massive
supergravity is given by $m_{10D}=\frac{\sqrt{2}}{3}g$.
Note that we will have dilatonic solutions in ten-dimensions
even if we have solutions with AdS space-time in six-dimensions.
This situation in ten-dimensions can be seen 
from a supergravity solution of flat D4-D8 system \cite{Imamura}.

\subsection{BPS solutions}
\hspace{5mm}
Let us introduce supersymmetric examples for wrapped D4-D8
systems in \cite{NuPaScTr}.
We also mention two other examples in the end of this subsection.

We begin with wrapped D4-D8 systems around
holomorphic two-cycles inside non-compact $K3$ surfaces.
We use following ansatz on six-dimensional metric
\begin{eqnarray}
\label{d6k3me}
ds^2_6=e^{2f(r)}\left(d\xi_3^2+dr^2\right)+
e^{2g(r)}d\widetilde{s}^2_{\Sigma_2},
\end{eqnarray}
where $d\xi_3^2$ is a metric on unwrapped directions of D4 branes
$R^{1,2}$, and $d\widetilde{s}_{\Sigma_2}^2$
is a metric on curved supersymmetric two-cycles.
We normalize metric $d\widetilde{s}_{\Sigma_2}^2$ so that its Ricci tensor
satisfies $\widetilde{R}_{ab}=\ell \widetilde{g}_{ab}$
with $\ell=1,-1$. 
Let us introduce orthonormal basis of six-dimensional metric
\begin{eqnarray}
e^0=e^{f(r)}d\xi_0,
\quad
e^1=e^{f(r)}d\xi_1,
\quad
e^2=e^{f(r)}d\xi_2,
\quad
e^3=e^{f(r)}dr,\quad
e^{3+a}=e^{g(r)}\widetilde{e}^a,
\end{eqnarray}
where $\widetilde{e}^a \; (a=1,2)$ are orthonormal basis for 
supersymmetric two-cycles with respect to
a metric $d\widetilde{s}^2_{\Sigma_2}$.
Now let us consider about a twisting operation.
Maximally supersymmetric six-dimensional supergravity should have $SO(5)$ 
gauge group.
Then we extract $SO(2)$ symmetry from subgroup in $SO(3)$
which appears in a decomposition $SO(3)\times SO(2)\subset SO(5)$.
We are able to map this $SO(3)\times SO(2)$
into $SU(2)\times U(1)$ symmetry in the gauged theory.
We identify the $SO(2)=U(1)$ with a structure group $U(1)$
of spin connection for holomorphic 2-cycles.
This operation is implemented by following conditions on spinors and
ansatz on gauge fields  
\begin{eqnarray}
\label{d6k3ga}
&&\gamma_{45}T^3\epsilon=\frac{1}{2}\epsilon,
\qquad
\gamma_3\epsilon=\gamma_7\epsilon,\\
&& F_{45}^{3}=\frac{\ell}{g}e^{-2g(r)}.\nonumber
\end{eqnarray}
Solutions satisfying these conditions
will give $1/8$ supersymmetric solutions in six-dimensional gauged
supergravity and massive IIA supergravity.

We proceed to wrapped D4-D8 systems around 
special Lagrangian three cycles inside non-compact Calabi-Yau threefolds.
Ansatz on a six-dimensional metric is 
\begin{eqnarray}
\label{d6cyme}
ds^2_6=e^{2f(r)}\left(d\xi_2^2+dr^2\right)+
e^{2g(r)}d\widetilde{s}^2_{\Sigma_3},
\end{eqnarray}
where $d\xi_2^2$ is a metric on $R^{1,1}$, and 
$d\widetilde{s}^2_{\Sigma_3}$ is a metric on
supersymmetric three-cycles.
We normalize curved metric 
$d\widetilde{s}_{\Sigma_3}^2$ so that its Ricci tensor
satisfies $\widetilde{R}_{ab}=\ell \widetilde{g}_{ab}$
with $\ell=1,-1$. 
We introduce orthonormal basis of this six-dimensional metric by
\begin{eqnarray}
e^0=e^{f(r)}d\xi_0,
\quad
e^1=e^{f(r)}d\xi_1,
\quad
e^2=e^{f(r)}dr,\quad
e^{2+a}=e^{g(r)}\widetilde{e}^a,
\end{eqnarray}
where $\widetilde{e}^a \; (a=1,2,3)$ are orthonormal basis for 
metric $d\widetilde{s}^2_{\Sigma_3}$.
Let us specify a twisting operation.
We identify $SO(3)$ symmetry
which appears in decomposition $SO(3)\times SO(2)\subset SO(5)$
with structure group a $SO(3)$
of spin connection for special Lagrangian three-cycles.
This identification is realized by a projection on spinors
and ansatz on gauge fields  
\begin{eqnarray}
\label{d6cyga}
&&\gamma_{45}T^3\epsilon=\gamma_{53}T^2\epsilon=\gamma_{34}T^1\epsilon
=\frac{1}{2}\epsilon,
\qquad
\gamma_2\epsilon=\gamma_7\epsilon,\\
&&
F_{45}^3=F_{53}^2=F_{34}^1=\frac{\ell}{2g}e^{-2g(r)}.\nonumber
\end{eqnarray}
Solutions with these conditions preserves $1/16$ supersymmetry.

BPS equations obtained from supersymmetry transformation of fermions
(\ref{d6susy}) with above ansatz on bosonic fields are
\begin{eqnarray}
f'(r)e^{-f(r)}&=&-\frac{1}{4\sqrt{2}}\left(
ge^{\frac{\phi(r)}{\sqrt{2}}}+me^{-\frac{3\phi(r)}{\sqrt{2}}}
\right)-\frac{d\ell}{4\sqrt{2}g}e^{-\frac{\phi(r)}{\sqrt{2}}-2g(r)},
\nonumber\\
g'(r)e^{-f(r)}&=&-\frac{1}{4\sqrt{2}}\left(
ge^{\frac{\phi(r)}{\sqrt{2}}}+me^{-\frac{3\phi(r)}{\sqrt{2}}}
\right)+\frac{(8-d)}{4\sqrt{2}g}e^{-\frac{\phi(r)}{\sqrt{2}}-2g(r)},\\
\frac{1}{\sqrt{2}}\phi'(r)e^{-f(r)}&=&
\frac{1}{4\sqrt{2}}\left(ge^{\frac{\phi(r)}{\sqrt{2}}}
-3me^{-\frac{3\phi(r)}{\sqrt{2}}}\right)
+\frac{d}{4\sqrt{2}g}e^{-\frac{\phi(r)}{\sqrt{2}}-2g(r)},\nonumber
\end{eqnarray}
where $d=2, 3$ denotes a dimension of supersymmetric cycles
under consideration.
Solutions always have spinor fields of a form 
$\epsilon_i(r)=e^{\frac{f(r)}{2}} \epsilon_{0i}$ 
with constant spinor fields $\epsilon_{0i}$.

We concentrate on solutions with AdS space-time for BPS equations.
It has been shown that each configuration has such a solution.
Solution with AdS space-time for holomorphic two-cycles is given by  
\begin{eqnarray}
\label{d6k3ads}
e^{f(r)}=\frac{2\sqrt{2}}{g}e^{-\frac{1}{\sqrt{2}}\phi}\frac{1}{r},
\;\qquad
e^{g(r)}=\frac{2}{g}e^{-\frac{1}{\sqrt{2}}\phi},
\;\;\;\qquad
e^{-2\sqrt{2}\phi(r)}=\frac{g}{2m},
\qquad \ell=-1.
\end{eqnarray}
On the other hand, we have following values when we obtain 
solution with AdS space-time 
for wrapped D4-D8 systems around special Lagrangian three-cycles
\begin{eqnarray}
\label{d6cyads}
e^{f(r)}=\frac{3}{\sqrt{2}g}e^{-\frac{1}{\sqrt{2}}\phi}\frac{1}{r},
\qquad
e^{g(r)}=\frac{\sqrt{3}}{g}e^{-\frac{1}{\sqrt{2}}\phi},
\qquad
e^{-2\sqrt{2}\phi(r)}=\frac{2g}{3m},
\qquad
\ell=-1.
\end{eqnarray}
Behavior of ten-dimensional metrics for these solutions 
are obtained from embedding formula (\ref{d6d10me}).

Note that it is straightforward to derive other possible solutions  
by the six-dimensional $SU(2)\times U(1)$ gauged supergravity.
Possible configurations are wrapped D4-D8 systems around
K\"ahler four cycles in non-compact Calabi-Yau threefolds,
and coassociative four cycles in non-compact $G_2$ holonomy manifolds.
These cases turn out to give the same form for a six-dimensional metric.
Solution with AdS space-time is given by
\begin{eqnarray}
e^{f(r)}=\frac{\sqrt{2}}{g}e^{-\frac{1}{\sqrt{2}}\phi}\frac{1}{r},
\qquad
e^{g(r)}=\frac{\sqrt{2}}{g}e^{-\frac{1}{\sqrt{2}}\phi},
\;\qquad
e^{-2\sqrt{2}\phi(r)}=\frac{g}{m},
\qquad
\;\ell=-1.
\end{eqnarray}

\subsection{Non-BPS solutions}
\hspace{5mm}
We proceed to a study of non-supersymmetric solutions for
wrapped D4-D8 systems.
We have discussed first order equations of motion from supersymmetry
variation of fermions for supersymmetric solutions in previous subsection.
Now, we should consider second order equations of motion in the six-dimensional
massive gauged supergravity.
Then, we start with Lagrangian (\ref{d6lag}) of
the gauged supergravity in Einstein frame
\begin{eqnarray}
\label{d6lagde}
L&=&\sqrt{-g}\left(R-\sqrt{2}\left(\partial_{\mu}\phi\right)^2
+\frac{1}{2}\left(g^2e^{\sqrt{2}\phi}+4gme^{-\sqrt{2}\phi}
-m^2e^{-3\sqrt{2}\phi}\right)-e^{-\sqrt{2}\phi}
F_{\mu\nu}^IF_{I}^{\mu\nu}\right),\nonumber\\
&=&L_G+L_{\phi}+L_{V}+L_{F},
\end{eqnarray}
where we denote Einstein-Hilbert term, kinetic term of scalar field,
scalar potential and kinetic term of gauge fields by
$L_{G}, L_{\phi}, L_{V}$ and $L_F$.

We first consider non-supersymmetric generalization of
wrapped D4-D8 system around two cycles within non-compact $K3$ manifolds. 
Here, we choose the same ansatz for metric and gauge fields as 
supersymmetric solution (\ref{d6k3me}), (\ref{d6k3ga}). 
We substitute the ansatz with three radial functions $f(r), g(r)$
and $\phi(r)$
into each term in Lagrangian (\ref{d6lagde}). 
Then, we have following expressions of Lagrangian
\begin{eqnarray}
L_G&=&2e^{2f}\left[\ell
e^{2f}+e^{2g}\left(3f'^2+6f'g'+g'^2\right)\right],\nonumber\\
L_{\phi}&=&-2e^{4f+2g}\phi'^2,\nonumber\\
L_{V}&=&-\frac{1}{2}e^{4f+2g}\left(
g^2e^{\sqrt{2}\phi}+4gme^{-\sqrt{2}\phi}
-m^2e^{-3\sqrt{2}\phi}\right),\\
L_{F}&=&\frac{2}{g^2}e^{4f-2g-\sqrt{2}\phi}\nonumber,
\end{eqnarray}
where we have dropped two derivative terms in $L_G$ by partial integrations.
Now we analyze a radial evolution specified by this 
one-dimensional effective Lagrangian. 
We arrange this Lagrangian as a 
combination of kinetic term $T$ and potential term $V$
\begin{eqnarray}
\label{lag}
L=T-V.
\end{eqnarray}
Here, we also impose the zero-energy condition specified by Hamiltonian
\begin{eqnarray}
\label{ham}
H=T+V=0.
\end{eqnarray}
Let us introduce another radial functions $F(r), G(r)$ and $\Lambda(r)$ 
in order to obtain simple expressions 
for equations of motion 
\begin{eqnarray}
\label{nonasz}
F(r)=e^{f(r)},\qquad G(r)=e^{g(r)},\qquad \Lambda(r)=e^{\sqrt{2}\phi(r)}.
\end{eqnarray}
We have following kinetic and potential term
in Lagrangian with new radial functions
\begin{eqnarray}
T&=&6F'^2G^2+12F'G'FG+2F^2G'^2-2F^4G^2\frac{\Lambda'^2}{\Lambda^2},\\
V&=&-2\ell F^4+\frac{g^2}{2}F^4G^2\Lambda
-2gm\frac{F^4G^2}{\Lambda}+\frac{m^2}{2}\frac{F^4G^2}{\Lambda^3}
+\frac{2}{g^2}\frac{F^4}{G^2\Lambda}.
\end{eqnarray}
Now, we further 
assume following form of radial functions $F(r), G(r)$ and $\Lambda(r)$
motivated by supersymmetric solution with AdS space-time (\ref{d6k3ads})
\begin{eqnarray}
F(r)=\frac{\alpha}{g}e^{-\frac{1}{\sqrt{2}}\phi}\frac{1}{r},
\qquad
G(r)=\frac{\beta}{g}e^{-\frac{1}{\sqrt{2}}\phi},
\qquad
\Lambda(r)=e^{-\sqrt{2}\phi},
\qquad
e^{-2\sqrt{2}\phi(r)}=\gamma,
\end{eqnarray}
where $\alpha,\beta$ and $\gamma$ are unknown constants to be determined.
This assumption makes it easy to solve the second order equations of motion.
Equations of motion with this ansatz are obtained by  
Euler-Lagrange equations of Lagrangian $L$ (\ref{lag})
\begin{eqnarray}
\frac{\partial L}{\partial F}-\left(
\frac{\partial L}{\partial F'}\right)'=0,
\qquad
\frac{\partial L}{\partial G}-\left(
\frac{\partial L}{\partial G'}\right)'=0,
\qquad
\frac{\partial L}{\partial \Lambda}-\left(
\frac{\partial L}{\partial \Lambda'}\right)'=0,
\end{eqnarray}
and a constraint on Hamiltonian $H$ (\ref{ham}).
We write down resulting independent equations for unknown constants
$\alpha, \beta$ and $\gamma$ by
\begin{eqnarray}
&&2\alpha^2+6\beta^4-2\ell\alpha^2\beta^2-\frac{1}{2}\alpha^2\beta^4-
\frac{2m}{g}\alpha^2\beta^4\gamma+\frac{m^2}{2g^2}\alpha^2\beta^4\gamma^2=0,
\nonumber\\
&&\beta^4\left(\frac{3m}{g}\gamma-1\right)\left(\frac{m}{g}\gamma-1\right)
=-4,\\
&&4\alpha^2-24\beta^4+\alpha^2\beta^4+\frac{4m}{g}\alpha^2\beta^4\gamma-
\frac{m^2}{g^2}\alpha^2\beta^4\gamma^2=0.\nonumber
\end{eqnarray}
These equations have two solutions both with $\ell=-1$ for wrapped cycles
\begin{eqnarray}
{\rm supersymmetric}&&\alpha=2\sqrt{2},\qquad \beta=2,\qquad\quad
 \gamma=\frac{g}{2m},\qquad\quad \ell=-1,\\
{\rm non}\;{\rm supersymmetric}
&&\alpha=2.98,\qquad \beta=1.84,\qquad \gamma=\frac{0.68 g}{m},\qquad \ell=-1,
\end{eqnarray}
where we have written numerical approximate values for the second solution.
Note that the first one is an original supersymmetric solution (\ref{d6k3ads}).
The second solution gives new non-supersymmetric solution with AdS space-time.
These solutions likely give dual backgrounds to $D=3$ 
conformal field theories realized on wrapped D4 branes at 
low energies compared with inverse size of wrapped two cycles.
It will be interesting to make a precise study
about these conformal field theories. 
Let us consider a relation of these two exact AdS solutions.
Values of $\alpha$ is proportional to radius of AdS$_4$ space-times in 
six-dimensional metrics.
Here, we have found that the value of non-supersymmetric solution is
larger than one of supersymmetric solution.
These values can be interpreted as central charges for dual conformal field
theories on wrapped D4 branes because their central charge are 
proportional to square values of radius of AdS$_4$ space-times.
It would be interesting to see whether we can
interpret a present situation from dual field theories on D4 branes.

Similarly we proceed to non-supersymmetric wrapped D4-D8 system
around special Lagrangian three-cycles in Calabi-Yau threefolds.
We use the same ansatz on metric and the gauge fields as
supersymmetric solution (\ref{d6cyme}), (\ref{d6cyga}).
We are able to calculate in the same manner as a previous example.
We again obtain two solutions with AdS$_3$ space-time
for three-cycles with negative curvature $\ell=-1$  
\begin{eqnarray}
{\rm supersymmertic} &&\alpha=\frac{3}{\sqrt{2}},\;\qquad 
\beta=\sqrt{3},\;\qquad \gamma=\frac{2g}{3m},\qquad\;\;\; \ell=-1,\\
{\rm non}\;{\rm supersymmetric}
&&\alpha=2.21,\qquad \beta=1.79,\qquad \gamma=\frac{0.54g}{m},
\qquad \ell=-1.
\end{eqnarray}
The former solution gives a supersymmetric solution (\ref{d6cyads}).
We have new latter numerical solution which gives non-supersymmetric
solution with AdS$_3$ space-time.
The ten-dimensional metrics can be obtained from the map (\ref{d6d10me}).
These solutions would be dual to $D=2$ 
conformal field theories on wrapped D4 branes.
Central charges of $D=2$ conformal field theories are proportional to
the values of radius of AdS$_3$ space-time.
Note that the value of radius for AdS$_3$ space-time
for non-supersymmetric solution is again larger than one
for supersymmetric solution.

In present examples, we always have
larger values of radius of AdS space-times for non-supersymmetric 
solutions than those for supersymmetric solutions.
This suggests a picture that unstable non-supersymmetric 
solutions with AdS space-times
flow into stable supersymmetric solutions with AdS space-times.
This picture is expected from a holographic $c$-theorem 
\cite{FrGuPiWa} within the six-dimensional massive gauged supergravity.
On the other hand, by assuming a $c$-theorem,
inverse flows from supersymmetric solutions to non-supersymmetric
solutions have observed from wrapped M5 brane solutions \cite{GaKiPaWa2}.
We would be able to give a precise statement about a direction of flows 
by using truncated effective actions with two scalar fields used 
in \cite{GaKiPaWa2}.
Then, we would be able to solve
kink solutions interpolating two AdS space-times 
numerically along an analysis in \cite{FrGuPiWa}.
We will also need to do an intensive analysis 
on a stability of non-supersymmetric solutions with AdS space-times
about various fluctuations of six-dimensional metric, scalar field
and gauge fields by generalizing a study \cite{DeFrGuHoMi}.
It would be further involved to study a stability 
about fields in massive type IIA supergravity.
Eventually, we should clarify a relation between a stability of
non-supersymmetric AdS solutions and a direction of flows interpolating 
two solutions with AdS space-times.

%%%%%%%%%%%%%%%%%%%%%%%%%%%%%%%%%%%%%
\section{Conclusions and discussions}
%%%%%%%%%%%%%%%%%%%%%%%%%%%%%%%%%%%%%
\hspace{5mm}
We have given supersymmetric and non-supersymmetric 
wrapped branes by using possible maximally
supersymmetric gauged supergravities and
suitable embedding ansatz of their solutions into 
those of ten- and eleven-dimensional supergravities.
A construction is fully explored  
for the simplest twisting procedure in \cite{MaNu1}.
More general twisting such as discussed in \cite{HeSf}
would provide another systematic study of wrapped branes in general.
As a next problem, we should study resolutions
of singularities for wrapped brane solutions
from a viewpoint of supergravity. 
Present understandings are limited to 
singularities of wrapped NS5-branes \cite{MaNu2,MaNa, HoKa}.
We will have to clarify whether singularities in wrapped branes can be 
resolved by modes of supergravity.
For this purpose, it would be useful to 
continue a study of non-supersymmetric solutions.
Non-supersymmetric curved space resolution of 
singularities in \cite{BuTs} is likely to be applicable to wrapped branes.  
Here, we will include event horizons to hide singularities by hand.
We also think that various truncated effective theories 
with metric and scalar fields \cite{GaKiPaWa2} play some roles.
On the other hand, we need to discuss a detailed
interpretation of results in supergravity in terms of 
worldvolume theories on wrapped branes.
Various solutions with AdS space-time give a class of examples for 
AdS/CFT correspondence with less amount of supersymmetry.
Here, we need a check about R symmetry in both theories.
We always expect superconformal symmetries in both sides.
One may wish to apply usual AdS supergroups \cite{Nahm} for 
candidate R symmetries.
We will have to understand a role of Kaluza-Klein modes 
arising from wrapped cycles in order to give a precise statement.
In principle, we should also be able to study 
correlation functions in dual theories by assuming holographic ideas. 

Finally, let us close with future problems along recent discussions in
string theory.
At first, it will be very interesting to construct de Sitter solutions 
combining with an idea of wrapped branes.
After a discussion on dS/CFT
correspondence in \cite{Strominger}, such an idea has been applied
into certain supergravity backgrounds \cite{Buchel}.
It might be interesting to test whether singularities 
in wrapped branes can be resolved by de Sitter like deformations.
As a related direction, supersymmetric
de Sitter massive $D=6$ $N=2$ gauged supergravity \cite{auva} 
would give a rigid starting point for wrapped D4-D8 branes.
Secondly, it would be interesting to consider 
Penrose limit \cite{BlFiHuPa1, BeMaNa, BlFiHuPa2}
of supergravity solutions for wrapped branes.
We have many examples of explicit solutions including
AdS space-time with lower amount of supersymmetry.
As for wrapped type IIB NS5 branes, 
Maldacena-Nunez background \cite{MaNu2} for resolved wrapped NS5 branes
around holomorphic $CP^1$ in Calabi-Yau threefold turns into a generalized 
Nappi-Witten type background with maximal amount of supersymmetry 
in Penrose limit \cite{GoOo}.
There are related discussions on Penrose limit of
supergravity backgrounds with varying fluxes
\cite{CoHaKeWa, GiPaSo, BrJoLoMy}.
We intend to give a systematic study preserved supersymmetry for 
Penrose limits of solutions for various wrapped branes.
In some cases, it would be possible to solve string theory on the backgrounds.
We will return these problems in future work.

%%%%%%%%%%%%%%%%%%%%%%%%%%%%%%%%%%%%%%%%%%%%%%%%%%%%%%%%%%%%%%%%%%%%%%%%%%%%%%
%%%%%%%%%%%%%%%%%%%%%%%%%%%%%%%%%%%%%%%%%%%%%%%%%%%%%%%%%%%%%%%%%%%%%%%%%%%%%%
\section*{Acknowledgements}
%%%%%%%%%%%%%%%%%%%%%%%%%%%%%%%%%%%%%%%%%%%%%%%%%%%%%%%%%%%%%%%%%%%%%%%%%%%%%
%%%%%%%%%%%%%%%%%%%%%%%%%%%%%%%%%%%%%%%%%%%%%%%%%%%%%%%%%%%%%%%%%%%%%%%%%%%%%%
\hspace{5mm}
We are very grateful to J. Gauntlett and Y. Tanii for valuable advise.
We also thank 
C. Hull, Y. Imamura, N. Kim, Y. Konishi, D. Martelli, S. Matsuura,
M. Nishimura, H. Nishino and D. Waldram for comments.
We are also grateful to members
at particle theory group in 
University of Tokyo, Komaba and
Yukawa Institute of Theoretical Physics.  
We thank hospitality at Queen Mary and Westfields college
in University of London while a part of this work was done.
This work is supported by 
Japanese Society for the Promotion of Science under Post-doctorial
Research Program $(\sharp 0206911)$.

%%%%%%%%%%%%%%%%%%%%%%%%%%%%%%%%%%%%%%%%%%%%%%%%%%%%%%%%%%%%%%%%%%%%%%%%%%%%%

\end{document}